\documentclass{pasa}%

\usepackage{graphicx}
\usepackage{tablefootnote}

\title[Disk masses]{The formation of planetary systems with SPICA}

\author[Kamp et al.]{Kamp, I.$^1$, 
Honda, M.$^2$, 
Nomura, H.$^3$, 
Audard, M.$^4$, 
Fedele, D.$^5$, 
Waters, L.B.F.M.$^{6,7}$, 
Aikawa, Y.$^8$, 
Banzatti, A.$^9$, 
Bowey, J.E.$^{10}$,
Bradford, M.$^{11}$, 
Dominik, C.$^{12}$,
Furuya, K.$^{13}$, 
Habart, E.$^{14}$, 
Ishihara, D.$^{15}$,
Johnstone, D.$^{16}$,
Kennedy, G.$^{17}$,
Kim, M.$^{18,39}$,
Kral, Q.$^{19}$,
Lai, S.-P.$^{20,36}$,
Larsson, B.$^{21}$,
McClure, M.$^{22}$,
Miotello, A.$^{23}$
Momose, M.$^{24}$,
Nakagawa, T.$^{15}$,
Naylor, D.$^{25}$,
Nisini, B.$^{26}$,
Notsu, S.$^{27}$,
Onaka, T.$^{28, 8}$,
Pantin, E.$^{29}$,
Podio, L.$^{5}$,
Riviere Marichalar, P.$^{30}$,
Rocha, W. R. M.$^{31}$,
Roelfsema, P.$^{32,1}$,
Santos, F.$^{33}$,
Shimonishi, T.$^{34}$,
Tang, Y.-W.$^{35}$,
Takami, M.$^{35}$,
Tazaki, R.$^{12}$,
Wolf, S.$^{18}$,
Wyatt, M.$^{37}$,
Ysard, N.$^{14}$\\

\affil{$^1$Kapteyn Astronomical Institute, University of Groningen, PO Box 800, 9700 AV Groningen,The Netherlands}%
\affil{$^2$Department of Biosphere-Geosphere Science, Okayama University of Science, 1-1 Ridai-cho, Kita-ku, Okayama, Okayama 700-0005, Japan}
\affil{$^3$ Division of Science, National Astronomical Observatory of Japan, 2-21-1 Osawa, Mitaka, Tokyo 181-8588, Japan}
\affil{$^4$Department of Astronomy, University of Geneva, Ch.\ d'Ecogia 16, 1290 Versoix, Switzerland}
\affil{$^5$INAF Osservatorio Astrofisico di Arcetri, L.go Fermi 5, 50126 Firenze, Italy}
\affil{$^6$ Institute for Mathematics, Astrophysics \& Particle Physics, Department of Astrophysics, Radboud University, P.O. Box 9010, NL-6500 GL Nijmegen, The Netherlands}
\affil{$^7$SRON Netherlands Institute for Space Research, Sorbonnelaan 2, NL-3584 CA Utrecht, the Netherlands}
\affil{... Please find the remaining affiliations at the end of the paper}
}%

\jid{PASA}
\doi{10.1017/pas.\the\year.xxx}
\jyear{\the\year}

\usepackage{aas_macros}
\usepackage{hyperref} 
\hypersetup{colorlinks,citecolor=blue,linkcolor=blue,urlcolor=blue}

\hypersetup{draft}

\begin{document}

\begin{frontmatter}
\maketitle

\begin{abstract}
In this era of spatially resolved observations of planet forming disks with ALMA and large ground-based telescopes such as the VLT, Keck and Subaru, we still lack statistically relevant information on the quantity and composition of the material that is building the planets, such as the total disk gas mass, the ice content of dust, and the state of water in planetesimals. {\it SPICA} is an infrared space mission concept developed jointly by JAXA and ESA to address these questions. The key unique capabilities of {\it SPICA} that enable this research are (1) the wide spectral coverage $10-220$\,$\mu$m, (2) the high line detection sensitivity of $(1-2) \times 10^{-19}$\,W\,m$^{-2}$ with $R\!\sim\!2000-5000$ in the far-IR (SAFARI) and $10^{-20}$\,W\,m$^{-2}$ with $R\!\sim\!29000$ in the mid-IR (SMI, spectrally resolving line profiles), (3) the high far-IR continuum sensitivity of 0.45\,mJy (SAFARI), and (4) the observing efficiency for point source surveys. This 
paper details how mid- to far-IR infrared spectra will be 
unique in measuring the gas masses and water/ice content of disks and how these quantities evolve during the planet forming period. These observations will clarify the crucial transition when disks exhaust their primordial gas and further planet formation requires secondary gas produced from planetesimals. The high spectral resolution mid-IR is also unique for determining the location of the snowline dividing the rocky and icy mass reservoirs within the disk and how the divide evolves during the build-up of planetary systems. Infrared spectroscopy (mid- to far-IR) of key solid state bands is crucial for assessing whether extensive radial mixing, which is part of our Solar System history, is a general process occurring in 
most planetary systems and whether extrasolar planetesimals are similar to our Solar System comets/asteroids. We demonstrate that the SPICA mission concept would allow us to achieve the above ambitious science goals through large surveys of several hundred disks within $\sim\!2.5$\,months of observing time.
\end{abstract}

\end{frontmatter}

\section*{PREFACE} 
\label{sec:preface}

The articles of this special issue focus on some of the major scientific questions that a future IR observatory will be able to address. We adopt the {\it SPace Infrared telescope for Cosmology and Astrophysics (SPICA)} design \citep{PRR2018, PRR2020a} as a baseline to demonstrate how to achieve the major scientific goals in the fields of galaxy evolution, Galactic star formation and protoplanetary disks formation and evolution. The studies developed for the \textit{SPICA} mission serve as a reference for future work in the field, even though the mission proposal has been cancelled by ESA from its M5 competition.

The mission concept of {\it SPICA} employs a 2.5\,m telescope, actively cooled to below $\sim\!8$\,K, and a suite of mid- to far-IR spectrometers and photometric cameras, equipped with state of the art detectors \citep{PRR2018}. In particular the {\it SPICA} Far-infrared Instrument (\mbox{SAFARI}) is a grating spectrograph with low ($R\!\sim\!200 - 300$) and medium ($R\!\sim\!3000 - 11\,000$) resolution observing modes instantaneously covering the $35$--$210\,\mu$m wavelength range. The {\it SPICA} Mid-infrared Instrument (SMI) has three operating modes: a large field of view ($10' \times 12'$) low-resolution $17 - 36\,\mu$m imaging spectroscopic ($R\!\sim\!50$--$120$) mode and photometric camera at $34\,\mu$m (SMI-LR), a medium resolution ($R\!\sim\!1300 - 2300$) grating spectrometer covering wavelengths of $18 - 36\,\mu$m (SMI-MR) and a high-resolution echelle module ($R\!\sim\!29\,000$) for the $10 - 18\,\mu$m domain (SMI-HR). Finally, B-BOP, a large field of view ($2'.6 \times 2'.6$), three channel ($70\,\mu$m, $200\,\mu$m and $350\,\mu$m) polarimetric camera complements the science payload.

\section{INTRODUCTION}
\label{sec:intro}

\begin{figure*}[th]
    \centering
	\includegraphics[width=14cm]{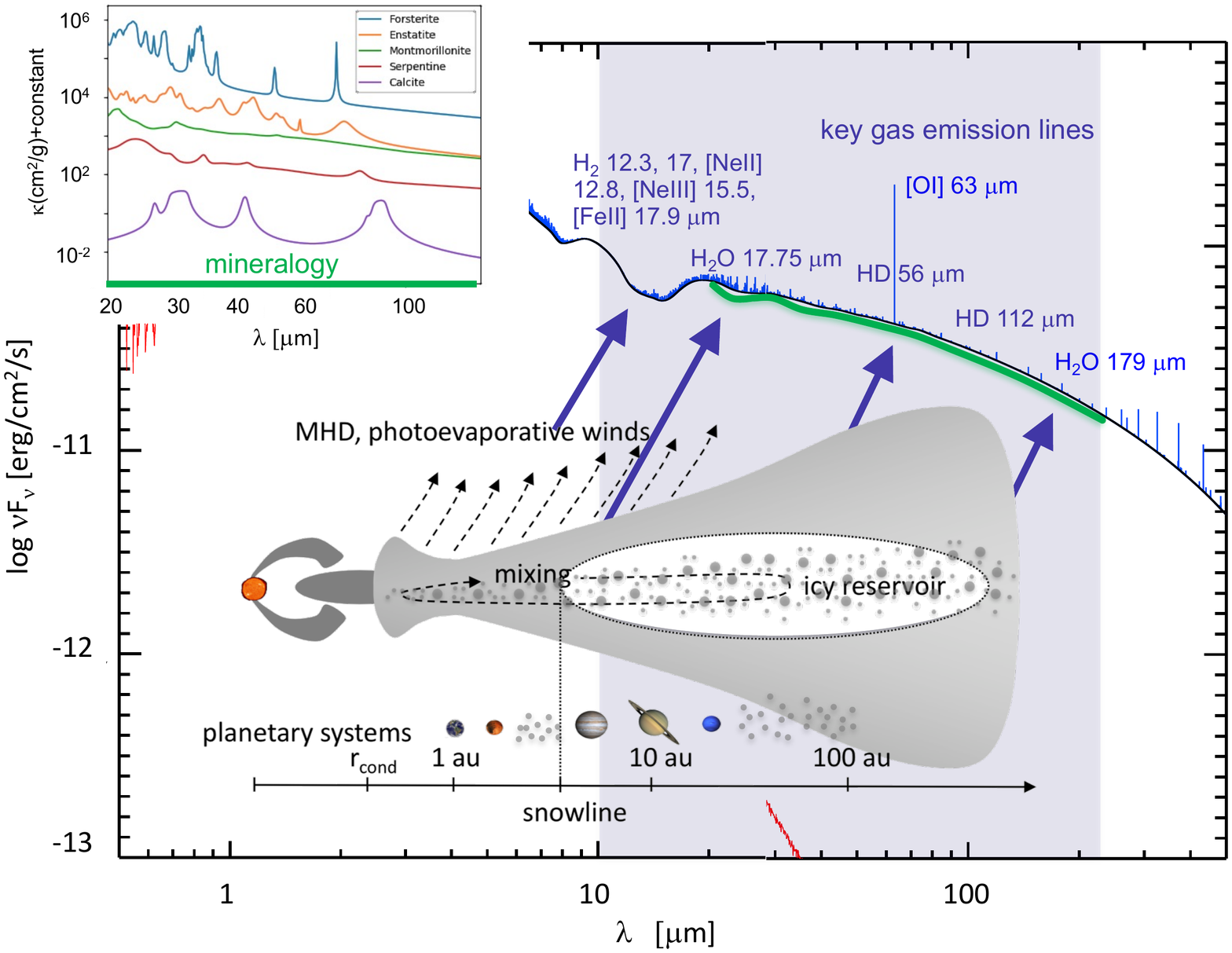}
\vspace*{-4mm}
	\caption{Sketch summarizing where {\it SPICA} provides unique insight into how planetary systems form.}\label{fig:sketch-disks-SPICA}
\end{figure*}

A key challenge to understanding planet formation is that we must explain both our own Solar System, with all its planets and minor bodies, as well as extrasolar planetary systems, which appear to differ vastly from our own. In the past five years, high spatial resolution imaging of planet forming disks with ALMA, VLT/SPHERE and/or Keck/GPI \citep[e.g.][]{Perez2014, ALMA2015, Benisty2015, Rapson2015} has revolutionised our view and our understanding of how disks evolve into planetary systems. These instruments reveal a wealth of substructures within disks, for both targeted \citep[e.g.][]{Andrews2018, Huang2018} and unbiased surveys \citep{Long2018}, as well as in the case of protoplanetary disk candidates \citep[e.g.][]{Ginski2018, Keppler2018, Pinte2018b} which likely indicate that planet formation is not only starting well before the disks appear as isolated objects (class\,II phase without strong envelope/jet components), but furthermore that planet formation must also be a very efficient process. Indeed, the Kepler mission has found that the probability of a low-mass star hosting at least one planet is close to 100\% \citep[e.g.][]{Tuomi2014}. 

In this era of spatially resolved observations of planet forming disks, we still lack statistically relevant information on the quantity and composition of the material that is building the planets and how it evolves during and beyond the era of planet formation. Key open questions are:\\[-6mm]
\begin{enumerate}
    \item How does the disk gas reservoir which is driving planet formation evolve? --- Sects.~\ref{sec:gasmass},\ref{sec:dispersal}
    \item When does the transition from primordial to secondary gas occur? --- Sects.~\ref{sec:gasmass},\ref{sec:dispersal}
    \item Are the extrasolar planetesimals similar to our Solar System comets/asteroids? --- Sect.~\ref{sec:debrisdisks}
    \item How do water and ice abundances within the disk evolve during the planet forming era and beyond? --- Sect.~\ref{sec:water-trail}
    \item How does mineral/ice mixing proceed in planet forming disks? --- Sect.~\ref{sec:mineralogy}\\[-5mm]
\end{enumerate}

In order to answer these questions we need measurements that can only be done using very sensitive infrared (IR) spectral observations due to the unique diagnostic lines and features that trace the planet building material and its evolution (HD and H$_2$O lines, emission bands from water ice, and large dust grains with a wide range of mineral composition). Available spatially resolved data capture only a single tracer at once (e.g. mm-sized dust, micron-sized dust, CO) and lack crucial calibration measurements to quantify the total gas reservoir. For example, dust grains grow from submicron particles into km-sized planetesimals in a gas-rich disk \citep[e.g.][]{Birnstiel2016}; during these steps, the solid material composition (mineralogy, ices) is altered, transported, and mixed through the disk. Our Solar System comets show the presence of highly processed dust, a strong indication for large scale mixing from the inner Solar System \citep{Nittler2016}. While ALMA clearly shows dust concentrations \citep[e.g.][]{vanderMarel2013, Casassus2013}, we cannot test our understanding of radial migration, settling, and grain growth without quantifying the gas reservoir in which this happens and the relevant dispersal timescales \citep[see][for a recent review]{Ercolano2017}.  Evidence from hydrodynamical simulations \citep[e.g.][]{Paardekooper2006b, Dipierro2015, Zhang2018} suggest that the gas surface density, in tandem with forming proto-planets, shapes the remaining dust in the disk into intricate substructures such as rings, spirals and vortices. Thus, the quantitative interpretation of the wealth of observed dust disk substructures requires firm knowledge of gas surface densities and hence gas masses in disks.

The snowline, where water condenses onto the refractory dust, plays a crucial role in planet formation since the solid mass reservoir available to build planets is significantly enhanced beyond it \citep{Hayashi1981} and the sticking properties of icy dust 
could be conducive to planet formation \citep{Okuzumi2012}. Several planet formation scenarios consider the water snowline as the prime location for material to overcome growth barriers in planet formation, enabling the formation of planetary cores \citep[e.g.][]{Schoonenberg2017}. Also at the snowline, the C/O ratio of the gas changes, with direct implications for planet formation \citep[e.g.][]{Oeberg2011,Helling2014,Eistrup2016,Notsu2020}.
None of the present facilities providing spatial resolution, however, has the sensitivity or diagnostic power to measure the snowline location across large samples of disks. 

The manner in which planetary cores subsequently grow into planets, both rocky and gas/ice giants, depends again on the total gas reservoir of the disk: the total gas mass controls processes such as migration of planetary cores \citep{Baruteau2014}, growth of gaseous envelopes and atmospheres \citep{Lissauer2007,Lammer2018}, and the circularization of planetary orbits \citep{Muto2011,Kikuchi2014}. The specific water vapour/ice content of the disk is crucial to quantifying the degree of hydration of refractory dust \citep{DAngelo2019,Thi2020}, the aqueous alteration of larger km-sized bodies \citep{Beck2014}, and the delivery of water to planets at late stages – for the build-up of oceans as well as primary/secondary atmospheres \citep[e.g.][]{Massol2016, Kral2018}. Eventually, the water content is key to assess the origin of life and habitability of the rocky planets forming in these disks.

During the process of planet formation, the primordial gas mass is decreasing as it is accreted onto the star (building it up to its final mass), incorporated into planets (gas giants), and dispersed through winds and jets (magnetically driven or photoevaporative). Late evolutionary stages of young stars (class\,III) have disks with very low dust masses \citep[$<\!0.3$~M$_{\rm Earth}$,][]{Hardy2015}, potentially they already formed planetesimals and planetary cores. Regular planetesimal collisions will replenish small dust grains and produce a secondary gas component that reflects the volatile content (ices) of the parent planetsimals \citep{Wyatt2015, Lovell2020}. Once a planet forming disk has lost its primordial gas, giant planet formation will come to a halt and also the primordial atmospheres of rocky planets will be lost \citep{Lammer2014, Massol2016}.

A missing link within the wealth of data characterizing the composition of exoplanets and their atmospheres is the evolution of 
dust with time, from young ($1-10$\,Myr) planet-forming disks to debris disks with planetesimals (tens of Myr). Existing and upcoming instruments operate in the near- to mid-IR (e.g.\ VLT MATISSE/VISIR/CRIRES, JWST, and ELT METIS) and therefore will not be able to trace dust grains larger than a few micron, which emit at longer far-IR wavelengths. In addition, thermal emission from water vapour and ices are uniquely observed at far-IR wavelengths, revealing a needed window into the evolution of water, the key element for life (as we know it), during the planet build-up.

This overview paper on the formation of planetary systems argues for the uniqueness of a cooled infrared space mission such as {\it SPICA} in order to\\[-6mm]
\begin{itemize} 
\item establish an absolute calibration of disk gas mass estimates using HD,
\item measure the gas dissipation timescale,
\item determine when the transition from 'primordial' to 'secondary' gas occurs,
\item quantify the evolution of the water vapour and ice reservoirs,
\item characterize the evolving mineralogy as dust grains grow into planetesimals,
\item characterize the volatile content in planetesimals in late stages of planetary system formation.\\[-6mm]
\end{itemize} 
These important results will require large statistical surveys of many thousands of systems throughout the planet formation phase (1-500\,Myr). Figure~\ref{fig:sketch-disks-SPICA} shows that {\it SPICA's} wide spectral coverage $10-220$\,$\mu$m, high line detection sensitivity of $(1-2) \times 10^{-19}$\,W\,m$^{-2}$ with $R\!\sim\!2000-5000$ in the far-IR (SAFARI) and $10^{-20}$\,W\,m$^{-2}$ with $R\!\sim\!29000$ in the mid-IR (SMI, spectrally resolving line profiles), and high far-IR continuum sensitivity of 0.45\,mJy (SAFARI) are key to achieve the above outlined scientific goals.

\section{DISK GAS MASSES}
\label{sec:gasmass}

Over the past three decades, many different observational techniques have been employed to measure disk masses (roughly ordered in time): mm-continuum fluxes, CO line fluxes, mm-continuum and sub-mm line interferometry, CO isotopologue line ratios, [O\,I]\,63\,$\mu$m+CO sub-mm lines, HD line(s) and wavelength dependent dust outer radii. 
Important steps in the disk mass determination were (1) the use of disk models in the interpretation of interferometric data, (2) the recent revision of dust opacities, and (3) the combined interpretation of multi-wavelength datasets. We discuss below the most recently employed techniques to estimate disk masses as well as the advantages and disadvantages of each method. 

{\bf Continuum mm/sub-mm fluxes:} 
Thermal continuum emission can be used to determine the dust mass within a disk under the assumption of a dust opacity (typically uncertain by at least a factor of two) and a dust temperature (better modeled as varying with radial location in the disk). Reliable estimates of the dust mass also require attention to the optical depth \citep{Beckwith1990}. Recent ALMA observations raise again the possibility that the inner parts of planet forming disks are optically thick \citep{Ballering2019, Zhu2019}. \citet{Woitke2016} show how the assumptions of a single representative dust temperature and an optically thick contribution to the emission affect the dust mass derivation. Once the dust mass is determined, conversion to gas mass requires an estimate of the dust to gas ratio which is usually assumed to be 100 (based on the canonical interstellar medium value).

{\bf CO sub-mm and [O\,I]\,63~$\mu$m:} Herschel was able to detect the strongest gas cooling line from planet forming disks, the [O\,I]\,63\,$\mu$m line, in a large fraction of young planet forming disks \citep{Dent2013, Riviere2016}. The [O\,I]\,63~$\mu$m line emitting region shifts, radially and vertically, with disk mass, making it an indirect tracer in combination with a temperature indicator. Hence, in combination with a CO sub-mm line, these fine structure line fluxes can be used to estimate disk masses within a factor three \citep{Kamp2011, Meeus2012}. However, for disks with gas masses above $\sim\!10^{-3}$\,M$_\odot$, or for disks with non-solar carbon and/or oxygen abundances, this method becomes highly uncertain.

\begin{figure}[!t]
\begin{center}
\vspace*{5mm}
\hbox{\hspace{-1.5cm}\includegraphics[width=1.0\columnwidth, angle=-90]{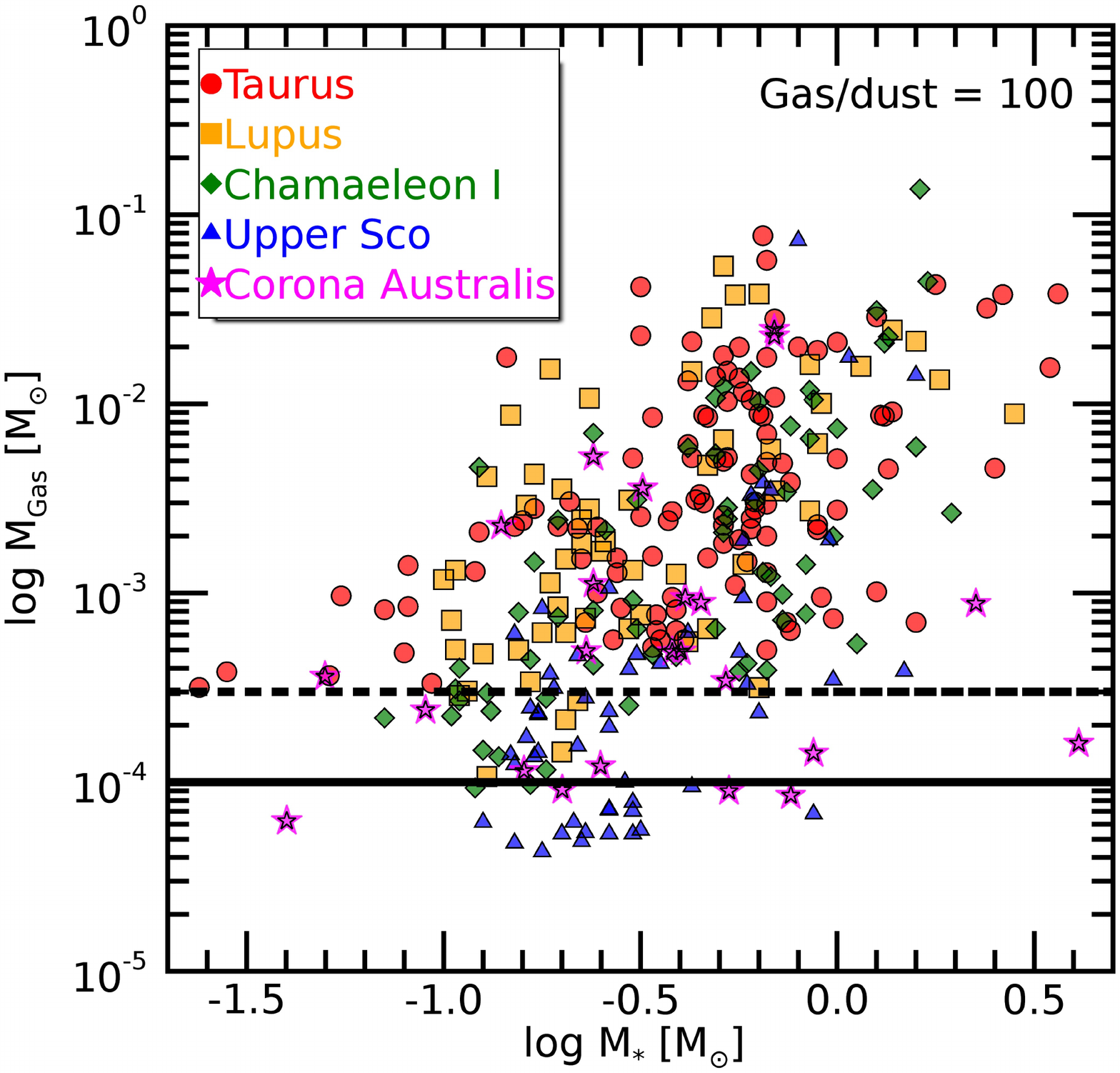}}
\vspace*{-6mm}
\caption{Distribution of disk masses as a function of the stellar mass for nearby star forming regions, based on recent millimeter dust continuum observations with the SMA and ALMA \citep{Andrews2013, Ansdell2016, Barenfeld2016, Pascucci2016, Cieza2019, Cazzoletti2019} and assuming a global gas-to-dust mass ratio of 100. The two horizontal lines indicate the detection limits that will be achieved with SAFARI observations of the HD $J\!=\!1-0$ line fluxes in the case of the shallow (1\,h on-source, dashed line) and deep (10\,h on-source, dotted) surveys, following the HD line flux predictions from \citet{Trapman2017}.}
\label{fig:diskmass}
\end{center}
\end{figure}

{\bf CO isotopologue sub-mm lines:} A more direct estimate of disk gas mass comes from observations of the rotational lines of CO and its rarer isotopologues\footnote{The rare isotopologue $^{13}$C$^{17}$O has only recently been detected in disks with ALMA by \citet{Booth2019} and \citet{BoothIlee2020}} ($^{13}$CO, C$^{18}$O, $^{13}$C$^{18}$O, $^{13}$C$^{17}$O) using (sub)-millimeter telescopes.
The rarer isotopologues are optically thinner 
and hence trace disk regions down to the CO iceline. The uncertainty in the gas-phase abundances of these species relative to H$_2$ molecules, however, limits the measurement reliability. In addition, CO rotational lines predominantly measure the cold gas mass, typically residing beyond 50\,au. Processes such as CO freeze-out \citep{Panic2009} and CO isotope selective photodissociation, by stellar and interstellar photons \citep{Miotello2014, Miotello2016}, as well as the gas phase carbon and oxygen abundances \citep{Bruderer2012, Miotello2017} affect these gas mass estimates. \citet{Williams2014} consider use of the fluxes of the two rarer isotopologues $^{13}$CO and C$^{18}$O to derive disk gas mass estimates. The method was refined by \citet{Miotello2016}, accounting self-consistently for isotope selective dissociation. Both grids of models, however, neglect the gas-to-dust mass ratio as a confounding parameter in their study; \citet{Woitke2016} show that the gas-to-dust parameter can strongly affect the CO isotopologue line fluxes; the local gas-to-dust mass ratio determines the disk temperature and affects the location of the CO iceline as well as the gas temperature of the emitting CO isotopologues. Thus, we require an
iterative approach, relying on successive dust and the gas mass determinations. And again, chemical effects changing the canonical CO-to-H$_2$ conversion factor continue to present a significant source of uncertainty \citep{Yu2017,Molyarova2017}. 

{\bf HD:} The major constituent of the disk gas is molecular hydrogen, H$_2$, but its rotational lines are not sensitive to main disk temperature conditions, i.e., these lines are not effective disk mass tracers. The deuterated variant, HD, has a very simple chemistry and is expected to have a constant abundance of $2 \pm 0.1 \times 10^{-5}$ \citep[the average local ISM value,][]{Prodanovic2010} throughout the disk. This constancy makes the lowest two HD rotational lines --- $J\!=\!1-0$ at 112\,$\mu$m ($E_{\rm up}\!=\!128.5$\,K) and HD $J\!=\!2-1$ at 56\,$\mu$m ($E_{\rm up}\!=\!384.6$\,K) --- reliable tracers of the warm gas mass in the inner disk (inside $\sim\!100$\,au). These inner disk regions are of prime importance since the majority of planets are thought to be formed there, rather than in the cold outer disk beyond $\sim\!100$\,au 
where CO sub-mm lines become good tracers of mass. \citet{Bergin2013} and \citet{McClure2016} detected the lower excitation HD line in three planet forming disks with Herschel/PACS and \citet{Kama2020} derived upper limits to the gas masses for disks around 15 objects from Herschel upper limits on both HD lines.

\subsection{DISK GAS MASSES FROM HD LINES WITH SPICA}
\label{subsec:HDmasses}

Over the next decade, we anticipate ALMA to have undertaken complete disk surveys (dust and CO isotopologues) in all relevant nearby star forming regions (within $500$\,pc). For large, $>\!100$\,au,  disks in low mass star forming regions, most of the disk gas mass is expected to be well below 50\,K; with ALMA, we will thus have a good understanding of the cold outer disk component. Planet formation similar to our Solar System, however, happens in the inner regions of the disk, inside 100\,au, and \citet{Trapman2017} showed that most of the HD emission is originating along the inner warm disk surfaces ($50-200$\,K). Furthermore, ALMA surveys have shown recently that the majority of disks are likely smaller than 100\,au \citep{Ansdell2018, Long2019}, making HD a direct tracer of the bulk gas mass for those objects. On the other hand, in high mass star forming regions, the external radiation field of nearby O/B stars could warm the entire disk surface well above 50\,K, enhancing the importance of HD measurements.

Both HD lines ($J\!=\!1-0$ and $J\!=\!2-1$ at 112 and 56\,$\mu$m respectively) are only accessible from space/stratosphere and SAFARI is uniquely equipped in terms of spectral coverage and sensitivity to measure them across statistical samples of disks. For an observational estimate of the gas mass in the planet forming region, within 100\,au, the HD line ratio will be a proxy of the `disk gas temperature' and the line fluxes will be used to to derive disk mass estimates using model grids, similar to what has been done with ALMA for the CO isotopologues \citep{Miotello2017, Long2018}. These model grids can be easily refined, taking the disk substructure known from the ALMA surveys into account. Since the HD abundance is not affected by element abundances/chemistry/freeze-out, this remains the most robust technique to estimate disk gas masses. 

\begin{figure}
\begin{center}
\includegraphics[width=1.0\columnwidth]{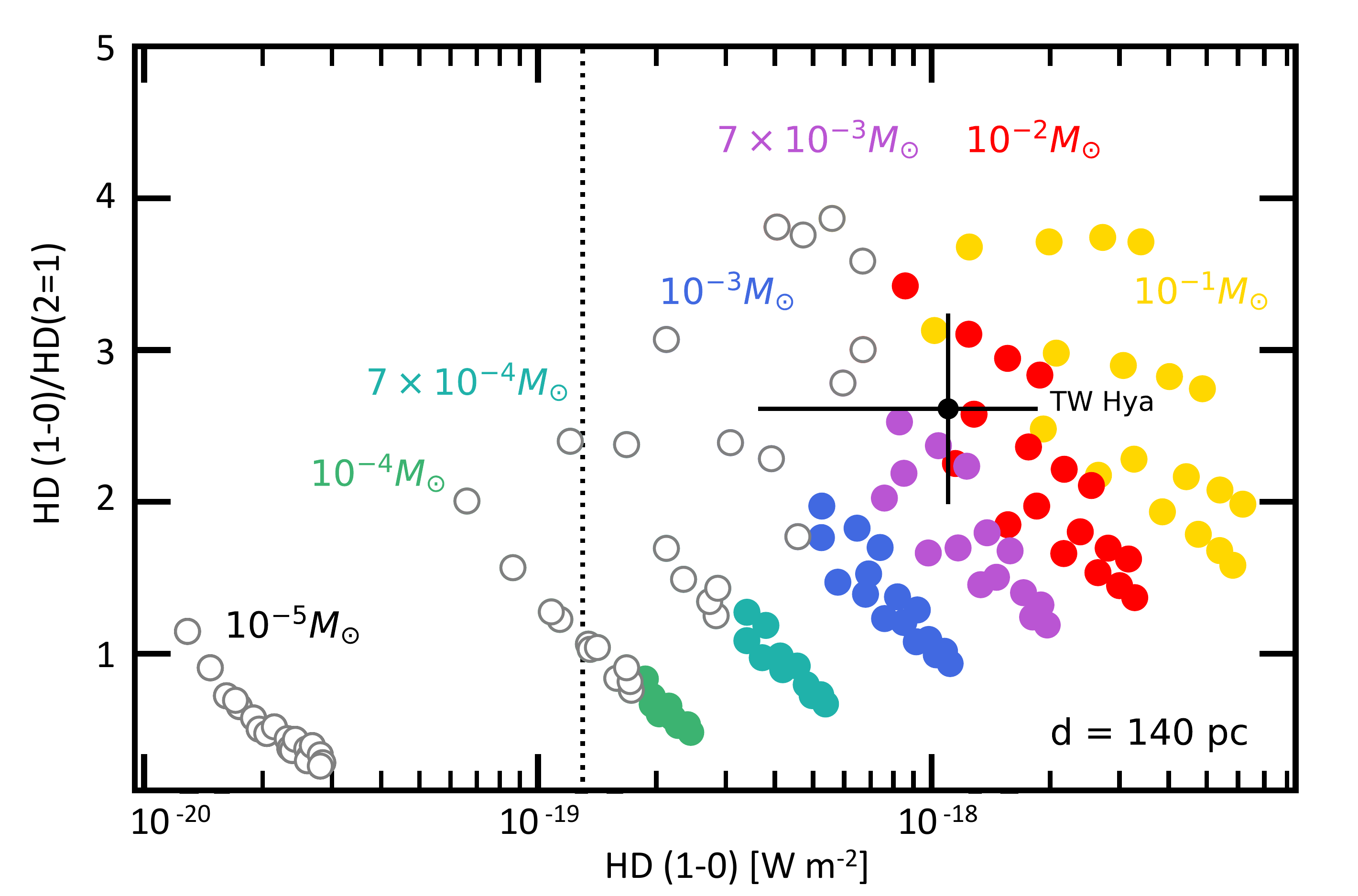}
\vspace*{-5mm}
\caption{HD diagnostic diagram: Disk gas masses can be derived from a combination of the two HD lines, based on a grid of T\,Tauri disk models from \citet{Trapman2017}. Open symbols denote disk model predictions for a distance of 140\,pc (e.g.\ the Taurus star forming region), where {\it SPICA} will not be able to detect both HD lines in a SAFARI/HR 1\,h integration ($<\!5\,\sigma$) due to the lower sensitivity at $56\,\mu$m (the higher continuum lowers the line sensitivity). The black dot with error bars shows the only object, TW~Hya ($60.1$\,pc, put here at a distance of 140\,pc), for which both HD lines have been measured with Herschel/PACS. The vertical dotted line indicates the 1\,h sensitivity limit ($5\,\sigma$) for the HD $J\!=\!1-0$ line.}
\label{fig:HD-to-diskmass}
\end{center}
\end{figure}

Figure~\ref{fig:diskmass} shows the ALMA disk samples for which dust masses have been measured in five star forming regions \citep[Taurus, Lupus, Chameleon\,I, Upper Sco and Corona Australis, from][]{Andrews2013, Ansdell2016, Barenfeld2016, Pascucci2016, Cieza2019, Cazzoletti2019}. We assume here a gas-to-dust mass ratio of 100 and indicate the SAFARI sensitivity limits based on a 1 and 10\,h per source deep survey. If all the disks in these surveys are T\,Tauri disks with disk/stellar properties similar to those modeled by \citet{Trapman2017}, the HD line flux predictions reveal that we can detect the gaseous counterpart from all ALMA dust detected disks. Both HD lines are optically thin for disks as massive as a few $10^{-2}\,M_{\odot}$ 
\citep[i.e.\ a few times the value of the minimum mass Solar Nebula,][]{Trapman2017}. The vast majority of the ALMA surveyed disk population has a total gas mass below this limit, utilizing a gas-to-dust ratio of 100. Therefore, HD line flux measurements, together with the constant HD abundance, provide a temperature and gas column density estimate which can be converted into a precise mass estimate of the gas. Figure~\ref{fig:HD-to-diskmass} demonstrates the feasibility of this approach across a grid of T\,Tauri disk models \citep{Trapman2017}. Combined with ALMA dust continuum fluxes, the SAFARI measurements of the two HD lines will provide the most complete survey for bulk gas mass estimates and gas-to-dust mass ratios for disk planet forming regions at a sensitivity comparable to that of the current ALMA dust surveys \citep{Ansdell2016, Barenfeld2016, Pascucci2016, Long2018, Cieza2019, Cazzoletti2019}. Having HD gas masses for statistically relevant disk samples allows also to calibrate and understand the issues around the CO isotopologue mass estimate.

To mitigate the temperature dependence of the HD gas mass estimates, we will use the $^{12}$CO and $^{13}$CO rotational ladders to derive the 2D temperature structure (see Sect.~\ref{subsec:COladder}), providing an improvement over the simple HD line ratio `disk gas temperature' estimate. We will thus measure the {\it total} gas mass from a simultaneous fit of the two HD line fluxes and the $^{12}$CO and $^{13}$CO rotational ladders \citep[e.g.][]{Fedele2016}.

Disk flaring is key for the gas heating and disk temperature structure \citep[e.g.]{Kamp2011}. Hence, to further refine the disk gas mass estimates, the [O\,I] fine structure lines have been shown to be sensitive tracers of disk flaring out to $\sim\!100$\,au \citep[regions where HD is also predicted to emit,][]{Trapman2017} and good probes of disk gas temperature \citep{Woitke2010, Kamp2011}; the [O\,I]\,63\,$\mu$m line is the brightest cooling line in disks, regularly detected with Herschel, and hence will be detected routinely with {\it SPICA}.
 
ALMA CO surveys can provide radial surface gas density profiles. The combination of ALMA CO and far-IR [O\,{\sc i}] together with modeling lines can break degeneracies in the determination of disk masses. Ancillary and routine detection of the [O\,I]\,63, 145\,$\mu$m and [C\,II]\,158\,$\mu$m lines with SAFARI, can also address the frequently discussed question of the disk C/O abundances being different from typical ISM values, possibly due to planet forming processes, migration of icy planetesimals, and filtering of sizes due to dust traps \citep[see e.g., ][]{Cleeves2018, Miotello2019}. 

\begin{figure*}[t]
\begin{center}
\vspace*{-1.5cm}
\includegraphics[width=1.3\columnwidth, angle=-90]{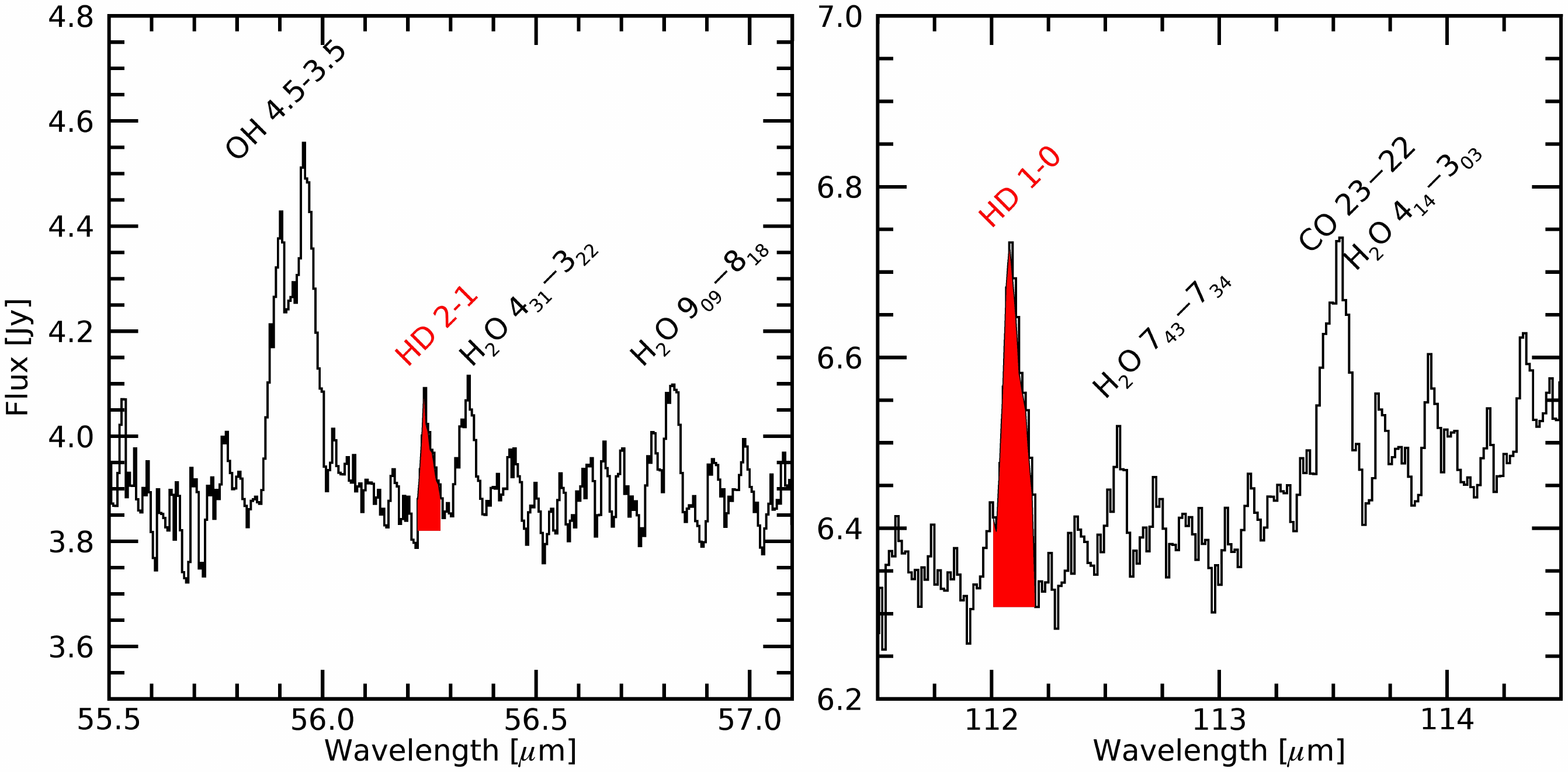}
\vspace*{-22mm}
\caption{Herschel/PACS spectrum of TW Hya showing the detection of the lowest two HD rotational lines at 56\,$\mu$m and 112\,$\mu$m (spectra re-reduced by D.\,Fedele). }
\label{fig:pacs_hd}
\end{center}
\end{figure*}

The HD lines do not need to be spectrally resolved in the SAFARI wavelength range. However, since molecular lines have a very low line-to-continuum ratio ($\lesssim$ 0.05), i.e.\ weak narrow gas lines (unresolved) on a strong continuum, we require a spectral resolution of at least a few 1000 to maximize the line detection rate \citep[see][for a detailed discussion]{Trapman2017}. The low excitation HD lines, $J\!=\!1-0$, $J\!=\!2-1$, are expected to have typical widths of $5-20$ km/s, dependent also on inclination. Figure~\ref{fig:pacs_hd} shows that even with the Herschel/PACS resolution of $R\!\sim\!1500$, the HD lines were not blended. Given the higher spectral resolution of SAFARI/FTS ($R\!\sim\!7300$ and $3700$ at 56 and 112\,$\mu$m respectively) line fluxes can be extracted in a straightforward manner.

From thermo-chemical disk models by \citet{Trapman2017} for T\,Tauri disks spanning a typical range of disk properties (flaring, disk size, scale heights), we find HD $J\!=\!1-0$ line fluxes ranging from $7\!\times\!10^{-20}$\,W\,m$^{-2}$ for $10^{-4}$\,M$_\odot$ disks (low mass, optically thin) to $2.3 \times 10^{-18}$\,W\,m$^{-2}$ for $10^{-1}$\,M$_\odot$ disks (massive, optically thick). The HD $J\!=\!2-1$ line is stronger for low mass optically thin disks ($10^{-19}$\,W\,m$^{-2}$) and fainter for optically thick massive disks ($10^{-18}$\,W\,m$^{-2}$).

Since the line fluxes scale with the continuum (see Fig.~\ref{fig:HDline-cont} in Appendix~\ref{app:HDsurvey}), we propose a survey based on the continuum brightness of the sources.
The HD $J\!=\!1-0$ and $J\!=\!2-1$ lines fall into the medium (MW) and long wavelength band (LW) with a spectral resolution of $R\!\sim\!3700$ and $R\!\sim\!7300$ respectively. For sources with a continuum flux $<\!0.1$\,Jy (MW) and $<\!1$\,Jy (LW) respectively, SAFARI/HR reaches a line sensitivity of $\sim\!1.3 \times 10^{-19}$\,W\,m$^{-2}$ in 1\,h (5\,$\sigma$);
this corresponds to disk gas masses down to $\sim\!10^{-3}$\,M$_\odot$. We will set our integration times in the HD disk gas mass survey such that we can push down to $2 \times 10^{-4}$\,M$_\odot$ while still detecting both HD lines. Fig.~\ref{fig:diskmass} shows that we would be able to provide, in a survey with $1-10$ h integration times, reliable gas mass estimates for all disks that are currently detected in dust surveys of star forming regions with ALMA. The sensitivity limit also overlaps with the upper end of the gas masses detected via CO in debris disks 
\citep[see Sect.~\ref{sec:debrisdisks}]{Moor2019}. Should those disks harbor remnant primordial gas from the younger phase (instead of secondary H$_2$-poor gas), we would detect that gas in the HD lines. Including in the survey a statistically significant number of class\,III disks, potentially young debris disks, we will be able to determine for the first time the transition from primordial to secondary gas.

\subsection{RADIAL TEMPERATURE PROFILES - THE CO-LADDER}
\label{subsec:COladder}

Disks are not isothermal structures. Rather, they are characterised by a large gradient in the gas temperature, which can go from nearly 10$^3$\,K in the proximity of the star down to $\sim\!10\,$K (or even lower) in the outer part of the disk midplane. Knowledge of the gas temperature gradient is important for the interpretation of both molecular and atomic spectra, as $T_{\rm gas}$ regulates line intensities. The vertical gas temperature gradient is, for example, key to understanding the emission from CO isotopologues which probe successively deeper layers in the disk \citep{Dartois2003}. Moreover, the local gas temperature also controls the gas chemistry, hence the formation of molecules and the resultant chemical enrichment. For example, at warm temperatures ($>\!200$\,K) neutral-neutral chemistry can proceed to efficiently form water via the radiative association H+OH \citep{Glassgold2009,Kamp2013}.

\begin{figure}
\begin{center}
\hbox{\hspace{-1.5cm}\includegraphics[width=0.95\columnwidth, angle=-90]{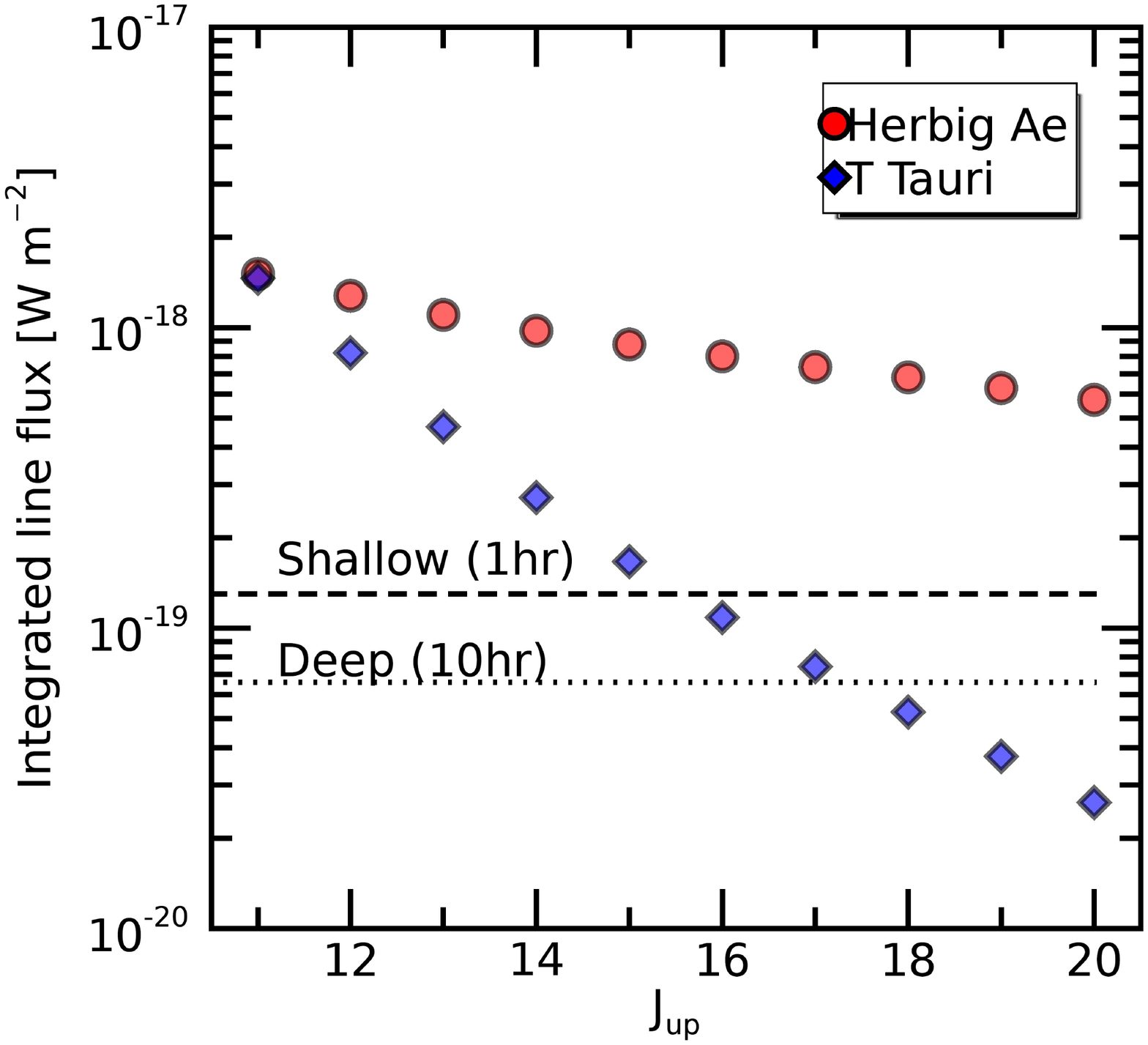}}
\vspace*{-4mm}
\caption{Predicted line flux of high-$J$ CO rotational transitions in protoplanetary disks to be measured with SAFARI. The two models refer to the case of a flat disk with low scale height around a Herbig Ae star (spectral type A0, 20\,$L_{\odot}$) and to a cold T\,Tauri disk (K5, 1.5\,$L_{\odot}$), both located at a distance of 140\,pc from the Sun \citep[DALI models,][]{Bruderer2012, Bruderer2013}. The two horizontal lines indicate the limiting line flux ($5\,\sigma$ detection limit) for a shallow and deep survey, respectively.}
\label{fig:co-ladder}
\end{center}
\end{figure}

An ideal thermometer of disk surface layers is the CO rotational ladder: the CO rotational transitions are optically thick and, due to their low critical density, are quickly thermalized in disks. Moreover, the CO transitions span a large range of upper energy levels, from 5.5\,K ($J\!=\!1-0$) up to a few 1000\,K (for $J_{\rm up}\!>\!20$), see Table~\ref{tab:co}. Rotational lines up to $J\!=\!10-9$ can be obtained from the ground, but the higher rotational lines are uniquely measured from space. As such the CO rotational ladder provides direct information about the kinetic temperature throughout the entire disk surface. In particular, a simultaneous model fit to the various CO transitions allows us to constrain the 2D temperature structure, not only the radial gradient but also the vertical gradient in the surface. Additional constraints to the vertical gradient reaching deeper into the disk are provided by the $^{13}$CO ladder. Due to the lower optical depth with respect to the main isotopologue, the $^{13}$CO transitions probe different density and temperature regions inside the disk. Figure~\ref{fig:co-ladder} shows that we can routinely detect the CO ladder in our HD survey ($1-10$\,h exposure times) toward the warm disks around Herbig stars, and typically up to $J\!=\!17-16$ for the colder T\,Tauri disks.

\begin{table}[!t]
    \caption{Wavelengths and upper level energies ($E_{\rm up}$) of CO lines to be observed with SAFARI to constrain the 2D gas temperature structure.}
    \centering
    \begin{tabular}{llll}
    \hline\hline
         & Transition & Wavelength [$\mu$m] & $E_{\rm up}$ [K] \\
\hline
         & 23-22 & 113.46 & 1524  \\
         & 22-21 & 118.58 & 1397  \\
         & 21-20 & 124.19 & 1276  \\
         & 20-19 & 130.37 & 1160  \\
         & 19-18 & 137.20 & 1050  \\
         & 18-17 & 144.78 & 945 \\
         & 17-16 & 153.27 & 846  \\
         & 16-15 & 162.81 & 752  \\
         & 15-14 & 173.63 & 663  \\
         & 14-13 & 186.00 & 581  \\
         & 13-12 & 200.27 & 503  \\
         & 12-11 & 216.93 & 431  \\
         \hline\hline
    \end{tabular}
    \label{tab:co}
\end{table}

\subsection{DISENTANGLING DISK AND SHOCK ORIGIN FOR HD}
\label{subsec:disk-shock}

\begin{figure}[h]
    \vspace*{-47mm}
	\hbox{\hspace{-1.5cm}\includegraphics[width=11.cm]{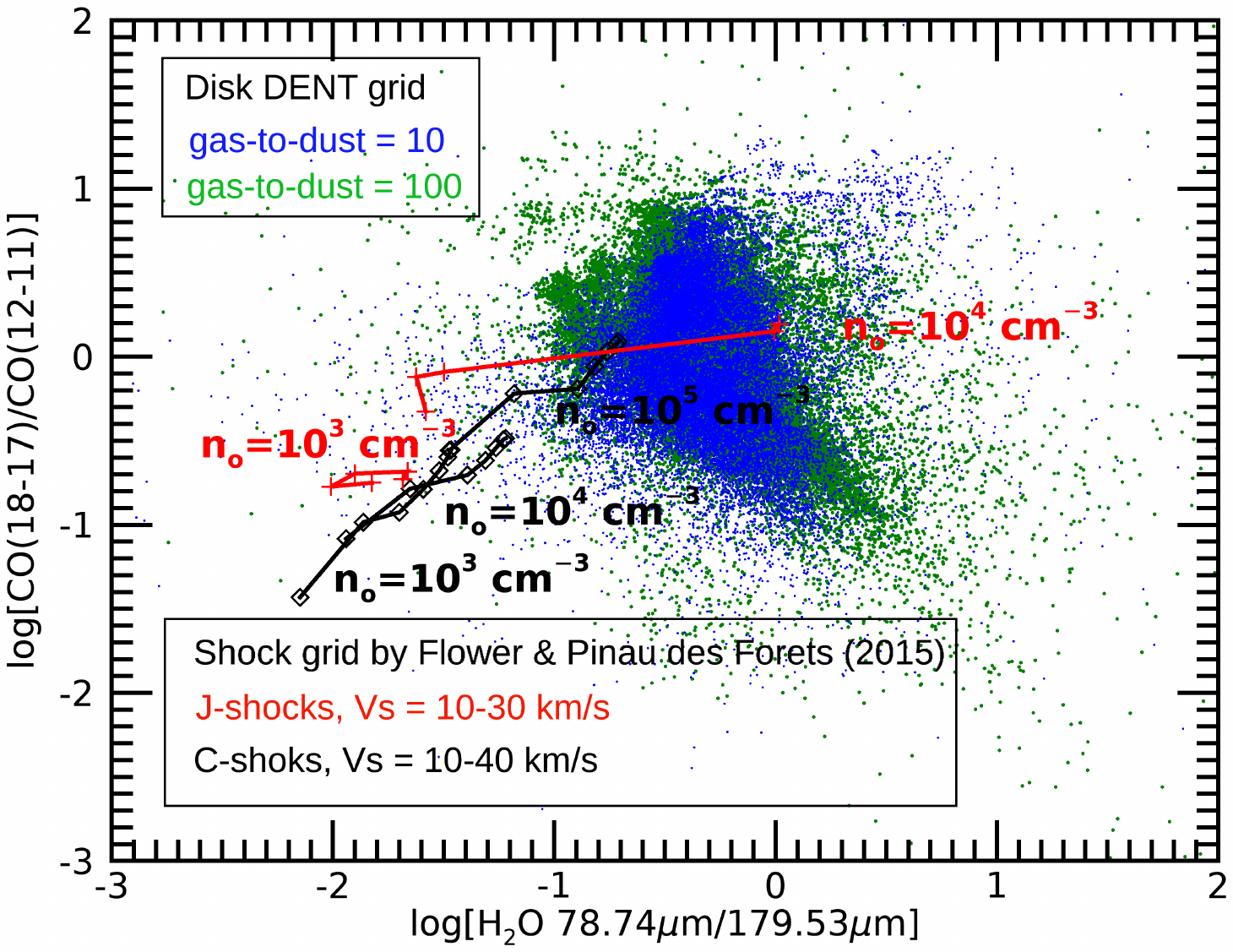}}
    \vspace*{-46mm}
	\caption{Line intensity ratio of CO(18-17)/CO(12-11) vs H$_2$O 78\,$\mu$m/179\,$\mu$m line ratios predicted by the DENT grid of disk models compared with the predictions of J- and C-type shock models of Flower \& Pineau des Forets (2015) for different pre-shock densities and pre-shock velocities.}
	\label{fig:shocks_disks}
    \vspace*{-2mm}
\end{figure}

Often disks and shocks produce the same spectral lines in star forming regions. In the absence of spatial resolution, spectral resolution can be used to disentangle these components. However, {\it SPICA} has only limited spectral resolution. We can overcome this by using diagnostic line ratios for disentangling disk and shocks. This is demonstrated in Fig.~\ref{fig:shocks_disks}, where we show the line ratio of CO\,(18-17)/CO\,(12-11) versus the line ratio of H$_2$O\,4$_{23}$-3$_{12}$ (78.74$\,\mu$m)/2$_{12}$-1$_{01}$(179.53\,$\mu$m) for the disk models explored by the DENT (Disk Evolution with Neat Theory) grid \citep{Woitke2010,Kamp2011} and for a non-dissociative shock models grid by \citet{Flower2015}. The figure shows that shocks and disks populate relatively separated areas of the line ratio plane. A larger contamination occurs only for dissociative shock models at the highest pre-shock density of 10$^4$\,cm$^{-3}$. 

\section{DISK DISPERSAL - SETTING THE CLOCK FOR PLANET FORMATION}
\label{sec:dispersal}

One important aspect of star and planet formation is the timescale for the dispersal of the protoplanetary disk. Indeed, it is crucial to determine how fast gas evaporates or gets accreted onto the central star, as the amount of gas available in the disk will set the clock for planet formation (see \citealt{Alexander2014}). ALMA provides now increasing evidence that class\,{\sc ii} disks may already host giant protoplanets \citep[e.g.][]{Andrews2018, Pinte2018b}. However, the gas dispersal timescale ends the runaway gas accretion onto giant protoplanets and halts their migration.

Broad-band near- to mid-IR excess (optically thick continuum emission from the dust) above the photospheric emission is an easy means to determine if an inner dust disk exists. The mean dispersal time is estimated from the fraction of such disks still observed as a function of the region's age \citep[e.g.,][]{Haisch01, Hernandez08,Mamajek09,Richert18}. Studies across many star forming regions have allowed for the determination of a mean disk dispersal timescale of a few Myr; however, this method only traces the optically thick inner dust disk component, which comprises about 1\% of the disk mass, and not the gas disk component, which drives the dynamics. \citet{Ohsawa2015} showed that this method also leads to a bias in the estimate of the disk dissipation timescale if the disk mass function is not properly taken into account. Alternative studies using H$\alpha$ emission, as a tracer for ongoing mass accretion and thus of the continued presence of gas in the disk, provide a similar, albeit possibly slightly shorter, timescale \citep{Fedele10}, suggesting a coevolving dust and gas disk with a somewhat longer-lived dust component. The few Myr timescale is comparable to the timescale expected for planet formation via the core accretion process, as well as to the timescale for giant planets to migrate (see \citealt{Ercolano2017} for a detailed review).  The exact mechanism of disk dispersal --- photoevaporation and/or MHD winds --- and the amount of mass loss through these processes will therefore have important implications for gas giant planet formation, helping to constrain the available time for orbital migration or circularization \citep{Tanaka2004, Kikuchi2014}. Inner disk evolution and dispersal, thus, play a crucial role in the final configuration of planetary systems, and there is a clear need for unbiased gas disk dispersal timescale studies.  

One of the challenges of gas dispersal studies is the difficulty to spatially resolve the inner gas disks. ALMA can measure the gas content in the outer disk regions, however, most planets form in the inner disk, i.e.\ inner 50\,au.
Therefore, we need to directly quantify the dissipation processes in the inner region of the disk. {\em Spectrally} resolved molecular line profiles can provide an ideal substitute for the lack of spatial resolution. These gas line profiles reveal both the spatial and physical components contributing to disk gas dispersal, namely fast jets and photoevaporative, or magneto-hydrodynamic (MHD), slow disk winds. 

Disk winds can be generated by MHD processes, that can also explain jets and outflows, or they can be launched due to photoevaporation \citep[e.g.][]{Shu1993, Yorke1996, Font2004}.  Hydrodynamical simulations show that EUV radiation can result in photoevaporative mass loss rates of order $10^{-10}$\,M$_\odot$/yr \citep[see][for a review]{Alexander2014}, while including X-rays can increase the mass loss by a factor of one hundred, assuming common X-ray luminosities. The mass-loss rate in the photoevaporative case is directly proportional to $L_\mathrm{x}$ (the power found in simulations is 1.14). Though observations show a wide scatter for young stars, there exists a strong correlation with the bolometric luminosity, or stellar mass. Line profiles produced by X-ray induced winds are more extended than those produced by EUV radiation.  

Disk winds around T\,Tauri stars have been extensively studied from the ground via high resolution optical/IR spectroscopy. In particular, observations of the [O\,I]\,630\,nm line profile have shown that components at different velocities likely trace distinct types of winds. In particular, the high velocity component at 
$>\!30$\,km/s testifies to the presence of a collimated jet, while components at lower velocity originate in disk winds \citep[e.g.,][]{Hartigan95,Natta14,Nisini18}. The low velocity component further reveals a broad, blueshifted, component due to emission within the inner 0.5\,au of the disk, and associated with MHD winds. In addition, there is also a low velocity narrower component that has been suggested to trace winds from more distant gas at 0.5-5\,au, indicating winds that might be photoevaporative or MHD in nature \citep{Rigliaco13,Simon16,Banzatti19}. 

The information that can be obtained from the [O\,I] emission is, however, limited since [O\,I], and other optical lines, are (1) excited at very high temperatures ($>\!5000$\,K) and therefore unable to probe colder disk winds (lower than 2000\,K) that might significantly contribute to the mass loss and (2) modeled wind mechanisms do not lead to different [O\,I] profiles so that the interpretation of those lines is often limited. 

\begin{figure}
\begin{center}
\vspace{-38mm}
\hbox{\hspace{-1.3cm}\includegraphics[width=1.2\columnwidth]{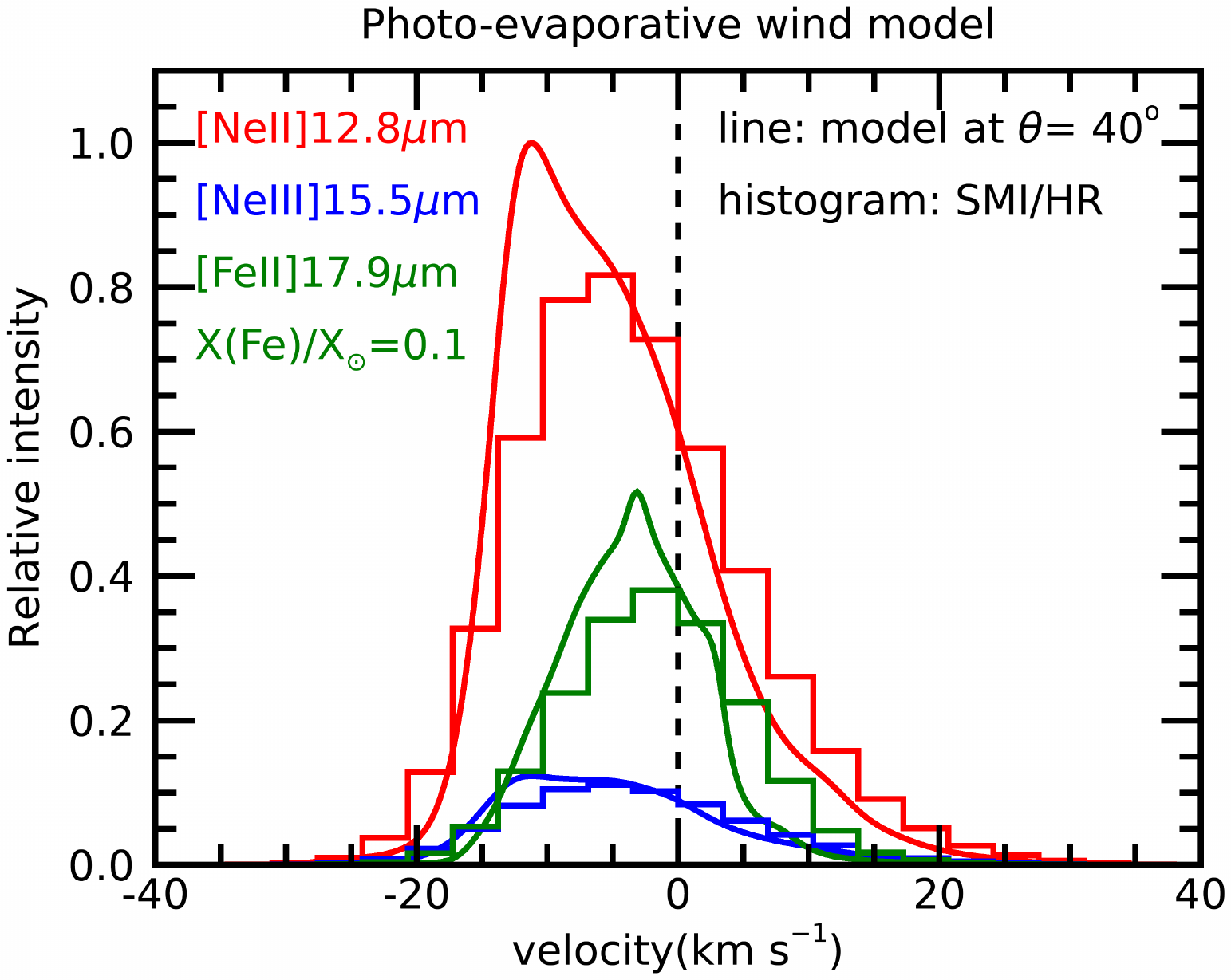}}
\vspace{-43mm}
\caption{Expected profiles of the [Ne II]12.8$\mu$m (red), [Ne III] 15.5$\mu$m (blue) and [Fe II]17.9$\mu$m (green) lines excited in a X-ray induced photoevaporative wind, based on the models of \citet{Picogna2019} and the calculations of \citet{weber2020}. Histograms show how the lines are seen by the SMI instrument. An iron abundance ten times smaller than solar has been assumed, taking into account the expected Fe depletion onto dust grains.}
\label{fig:visir_ne}
\end{center}
\end{figure}

Fundamental information necessary to clarify this picture will come from observations in the mid-IR. In particular, over the range covered by SMI/HR many gas tracers are found, i.e. {\em molecular} emission, such as H$_2$ rotational lines at 12.3 and 17$\,\mu$m, HD, H$_2$O and OH lines, together with diagnostic {\em atomic} lines, such as [Ne\,II]\,12.8, [Ne\,III]\,15.6, [Fe\,II]\,17.9 and [S\,I]\,17.43\,$\mu$m tracing atomic and ionized material at lower excitation temperatures (i.e.\ below 2000\,K) than probed by optical forbidden lines. 
SMI, with its high spectral resolution and sensitivity, will break new ground in the study of gas dispersal in protoplanetary disks. Figure~\ref{fig:visir_ne} shows that SMI/HR can spectrally resolve the the expected photoevaporative wind line profiles as well as detect shifts in the peak and line asymmetries.

Pioneering studies with ground observations at high ($R\!=\!17\,000$) resolution with VLT/VISIR have revealed the potential of [Ne\,II], at 12.8\,$\mu$m, as a key diagnostic tracer of photoevaporative winds, where Ne ionization occurs by the action of both X-ray and UV photons from the central star and the accretion shocks \citep[e.g.,][]{Pascucci2007,Pascucci2011,BaldovinSaavedra2012}. The limited observations performed with VISIR have shown that this line usually comes as a single component, and that high-resolution ($R\!>\!20\,000$) is needed to discriminate its origin in slow disk winds \mbox{($v\!\sim\!5-10$\,km/s)}, in the high velocity jet \mbox{($v\!>\!20-50$\,km/s)}, or in the inner gaseous disk. 

Molecular line diagnostics originate from regions located at greater distances from the star, i.e., $1-10$\,au. Near-IR ro-vibrational lines trace gas temperatures $>\!1000$~K and probe regions in the disk $\sim\!0.5-5$\,au (gas and dust can thermally de-couple where these lines originate). Spectrally and spatially resolved observations of H$_2$ in the near-IR have shown that the emission most often comes from wide-angle low velocity winds; however, evidence for emission from bound gas in the inner disk has been also found \citep[e.g.,][]{Bary2003,Beck2012}. H$_2$ mid-IR lines originate from a region further out, of the order of $5-10$\,au, at $T\!=\!150-1000$\,K at the disk surface and thus have the unique potential of tracing the gas within the giant planet formation region. 
Emission from H$_2$ mid-IR rotational lines have been searched for in many T\,Tauri and Herbig sources (from ground-based telescopes and from space); however, positive detections have been obtained only for an handful of objects, mostly via the space-based Spitzer-IRS instrument at low spectral resolution \citep[e.g.,][]{Lahuis2007} and thus without the possibility to separate the disk wind contribution from the overall emission. 
Indeed, models of H$_2$ emission from disks \citep[e.g.,][]{Nomura2007} predict flux levels below the detection limit of present instrumentation, but well above the sensitivity limit of SMI. Significantly, a detailed modeling of theses molecular lines using a thermo-chemical treatment for optically thin disk winds is still lacking.
Combined, spectrally resolved observations of both near-IR and mid-IR molecular lines should have the potential of tracing the gas temperature stratification of the inner disk. 

In the infrared, high-resolution spectra (e.g., VLT/VISIR and CRIRES+, and in the future \mbox{E-ELT/METIS} and TMT/MICHI) can provide very high spectral resolution, but with poorer flux sensitivity than SMI, mainly due to water vapour in the Earth atmosphere and the high thermal background due to the Earth's atmosphere.
Thus, SMI/HR fills a niche ideal to probe, at high spectral resolution, both molecular and atomic lines that trace the disk wind. Furthermore, SMI will probe disk winds across a very large sample of many hundred planet forming disks, Ideally, the same sample will be observed as undertaken by the disk gas mass survey with SAFARI, providing a census of both the disk conditions and the disk wind properties that cannot be obtained with ground-based observations.

\section{TRACING THE GAS IN DEBRIS DISKS - VOLATILE CONTENT OF PLANETESIMALS}
\label{sec:debrisdisks}

Gas that is found in debris disk systems is generally thought to be secondary, released from the debris produced in collisions between planetesimals, rather than a direct remnant from the initial star-forming nebula \citep[e.g.][]{Roberge2006, Matra2015, Kral2016a}. As such, this secondary gas provides a unique opportunity to study indirectly the composition of the debris (i.e.\ the volatile/ice content), and hence the planetesimals, the building blocks of planets. In our Solar System, meteorites (carbonaceous chondrites) contain up to 15\,weight\% water\footnote{Note that this contains possible contamination from the terrestrial atmosphere and the intrinsic value should be much less.} and water dominates the volatile content in comets \citep[$>\!50$\,\%,][]{Alexander2018}. Hydrous silicates have also been confirmed in C-complex asteroids by AKARI \citep{Usui2019} and by the sample return missions for Ryugu \citep{Kitazato2019} and Bennu \citep{Hamilton2019}, both of which are C-type asteroids. It is unclear whether extrasolar systems contain similar levels of water and volatiles or whether they are generally drier than the Solar System.

Planetesimals that were formed far enough from the star, in regions of the protoplanetary disk that are sufficiently cold,
are expected to have a significant volatile component. For example, comets in the Solar System are made up by mass of $\sim\!50$\,\% H$_2$O and $3-10$\,\% CO \citep{Bockelee-Morvan2017}. The volatiles locked inside planetesimals may play a key role in the development of life on exoplanets, since these volatiles can be released from the planetesimals and then migrate to the planets and be captured into their atmospheres \citep{Kral2020a} or be released upon direct impact with the planet, potentially delivering water, a key ingredient in the origin of life as we know it, and replenishing a secondary atmosphere \citep{Chyba1990, Kral2018, Wyatt2020}.
Understanding this volatile content can also help inform how and where in planet forming disks planetesimals are able to form; it is for example suggested that planetesimal belts form preferentially at ice-lines \citep{Matra2018} which should be reflected in a relatively high volatile content in that localized region. Of particular interest is the C/O abundance ratio \citep{Oeberg2011, Helling2014, Moriarty2014}, since the formation of solids and ultimately planets is believed to be sensitively dependent on this ratio. This is directly due to the fact that the condensation sequence forming solids changes dramatically at C/O\,$\sim\!0.98$ \citep{Larimer1975}. Also, this ratio determines the C/O in primary atmospheres which originate from the accretion of remnant disk material \citep{Massol2016}. The C/O ratio also has an impact on secondary atmospheres, as these depend on the bulk planet composition.  

A new technique for probing the volatile content of planetesimals has recently been developed by studying gas in the debris disks found around nearby stars (these are the exo-Kuiper belts found around $\sim\!20$\,\% of stars). The planetesimals collide and are ground down into dust, observable through infrared emission from the dust itself.
This process has been studied extensively over the last few decades \citep[e.g.][]{Wyatt2008}. In this process, the volatiles are also released as gas. The [C\,I]\,610\,$\mu$m fine structure line and the CO sub-mm lines have been detected by ALMA from debris disks, revealing the spatial distribution of the gas \citep[e.g.][]{Cataldi2018, Moor2019} and in some cases strong asymmetries possibly attributed to recent collisions \citep[e.g.][for the case of $\beta$\,Pic]{Dent2014}. 

\begin{figure}
    \centering
    \includegraphics[width=8cm]{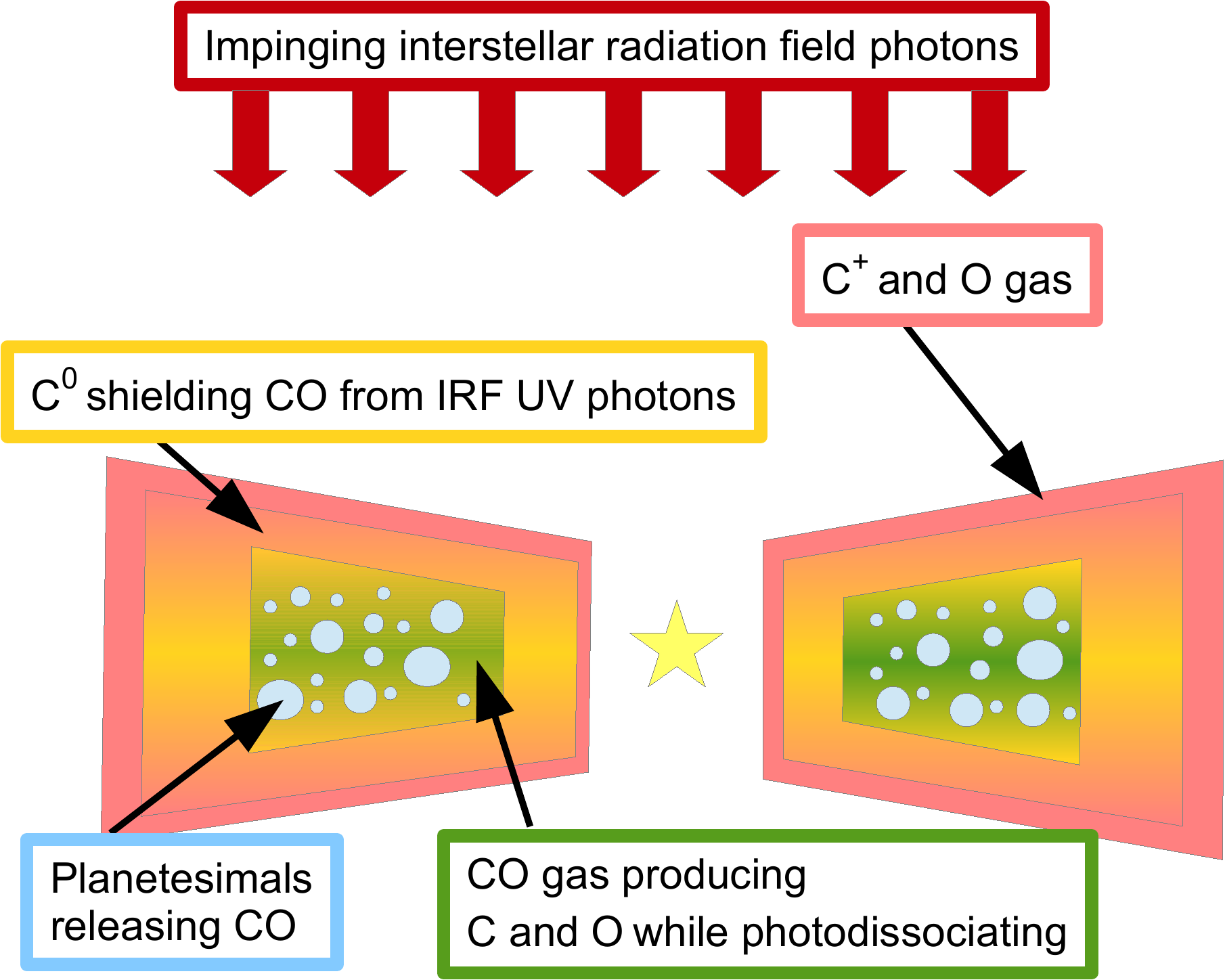}
    \caption{Sketch of gas distribution in a debris disk based on a detailed dissociation/ionisation model including an interstellar radiation field (IRF). This illustrates the complementarity of ALMA and {\it SPICA} gas observations in debris disks. Figure modified from \citet{Kral2019}.}
    \label{fig:cartoonALMA-SPICA}
\end{figure}

\begin{figure*}[ht]
	\includegraphics[width=8.8cm]{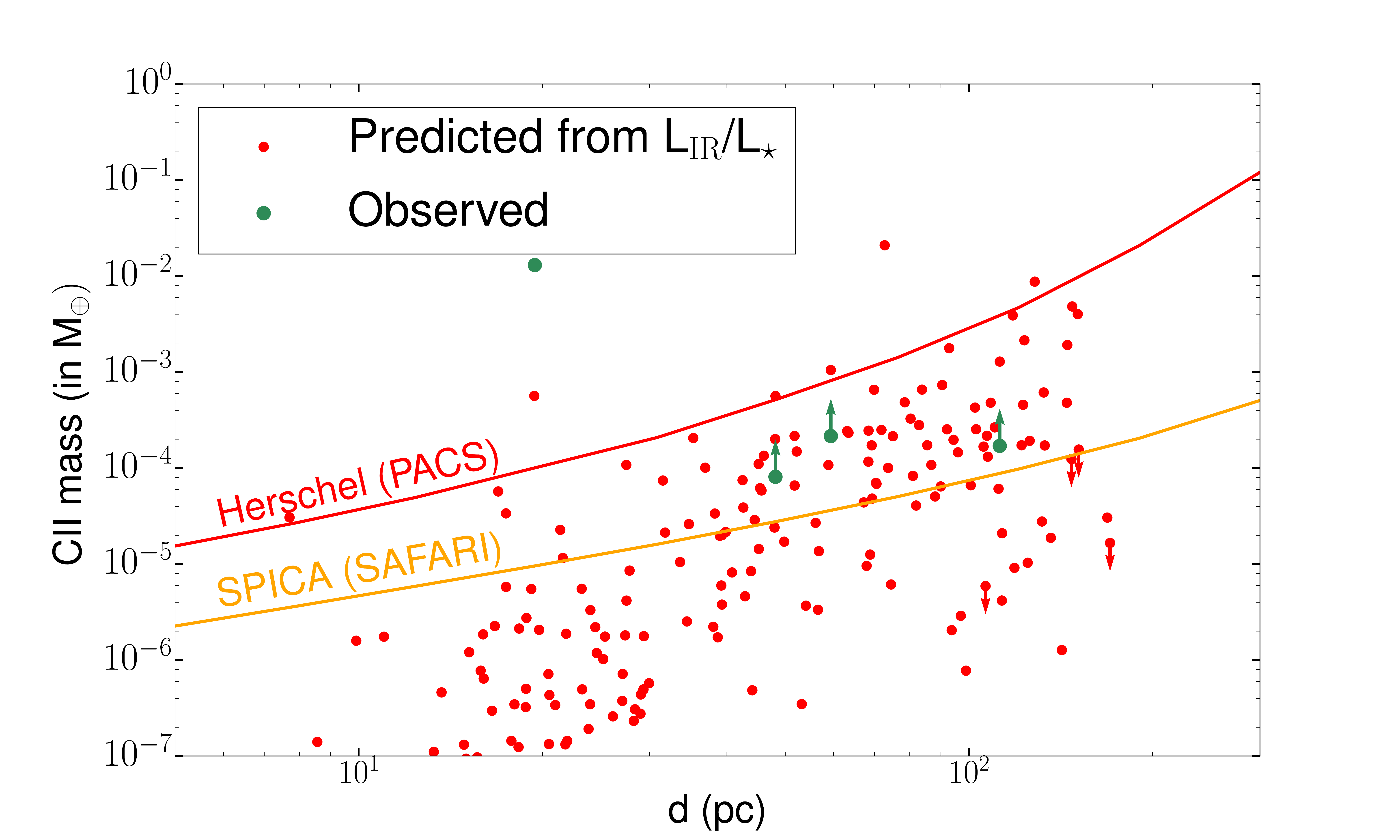}
	\includegraphics[width=8.8cm]{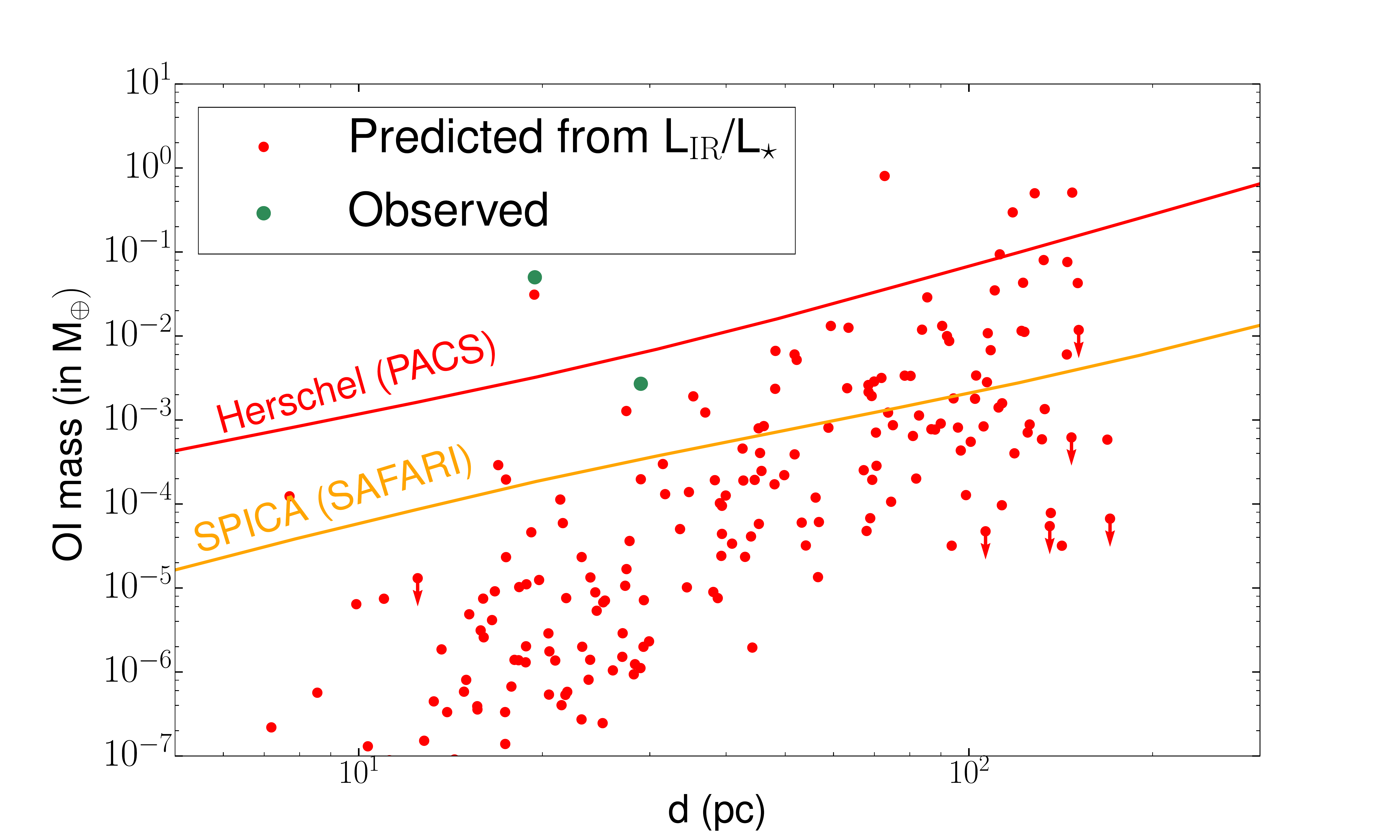}
\caption{Predicted [C\,II] (left) and [O\,I] (right) masses for the sample of 192 debris disks in \citet{Kral2017}. Approximate detection thresholds for surveys with Herschel/PACS (red, 1\,h, 5\,$\sigma$) and with SAFARI/HR (yellow, 1\,h, 5\,$\sigma$) are shown; the sensitivity of Herschel was only sufficient to detect (green dots) the most extreme disks in the population in deep integrations (longer than 1\,h), whereas with {\it SPICA} we can build statistically representative samples of debris disk gas detections. The mass estimates for the few objects that have detections from Herschel are indicated by green dots. Green arrows indicate lower limits of carbon masses related to the large uncertainty on the excitation temperature.}
\label{fig:kral2017}
\end{figure*}

Assuming the gas is released in a steady state process, the production rates of CO and dust can then provide their relative quantities in the planetesimals. The systems studied to date, e.g.\ HD\,181327 and Fomalhaut, have CO-to-dust compositions that are comparable with Solar System comets \citep[e.g.][]{Marino2016, Matra2017}. However, \citet{Kral2020b} report two systems that may have less CO than Solar System comets. While the CO content of the planetesimals can be probed in the above way with ALMA, this technique relies on an assumption that the gas is produced in steady state and that the lifetime of CO is set by photodissociation via the interstellar radiation field with, in some cases, CO self-shielding \citep{Visser2009, Heays2017}. Yet we expect that once CO is destroyed the C and O atoms will accumulate until they spread through viscous processes to eventually accrete onto the star. Observations of [C\,I] with ALMA show that in some cases there is sufficient carbon to shield the CO from photodissociating \citep[see Fig.~\ref{fig:cartoonALMA-SPICA},][]{Kral2019}. Moreover, the ALMA [C\,I] observations call into question the steady state nature of the gas production \citep{Cataldi2019} and probably do not probe the bulk of the carbon which is likely to be predominantly ionised, at least for the disks that are not shielded. 
\citet{Cataldi2020} show that a complete knowledge of the gas composition (e.g. C$^+$, C, CO) in connection with a detailed photodissociation and ionisation model allows us to time potential large collision events, and determine the time since a collisional cascade was initiated.
In addition to CO, there may be other volatiles released, in particular H$_2$O which photodissociation turns into OH, H and O \citep{Kral2016c}. Also here, ALMA observations alone cannot constrain the H$_2$O content of the planetesimals since the ``daughter'' products are only detectable in the far-IR with, for example, {\it SPICA}. 

Even though it lacks spectral and spatial resolution, SAFARI is the only instrument that will be able to detect the [O\,I] lines at 63 and 145\,$\mu$m  and use them to measure the atomic oxygen abundance. Another unique line, [C\,II]\,158\,$\mu$m, is crucial to measure the carbon budget together with ALMA [C\,I]\,610\,$\mu$m and CO sub-mm line measurements. This information is not only relevant for debris disks, but also for planet forming disks (see Sect.~\ref{subsec:HDmasses}). The [C\,II] line, in conjunction with CO and [C\,I] observations with ALMA, will allow measurement of the carbon mass, the carbon ionisation fraction, and the electron density in debris disks, all of which constrain both the CO content of the planetesimals and the steady state versus stochastic production process \citep[as well as the mechanism controlling the viscous evolution, e.g., MRI,][]{Kral2016b}. With this information the [O\,I]\,63\,$\mu$m line flux can be used to determine the H$_2$O content of the planetesimals \citep[see][]{Kral2016c}. In addition, the H$_2$O (179.7\,$\mu$m) and OH (119.3+119.5\,$\mu$m) lines can be used to probe the water content directly, though detection requires a high production rate due to the short photodissociation time for these molecules \citep{Matra2018}.

Previous observations with Herschel \citep{Riviere2014} only had the sensitivity to detect [O\,I] and [C\,II] for a few debris disks, indicated in Fig.~\ref{fig:kral2017}. 
SAFARI (yellow line) improves significantly on Herschel/PACS\footnote{We assume a Herschel/PACS line sensitivity of $8 \times 10^{-18}$\,W\,m$^{-2}$ and $6 \times 10^{-18}$\,W\,m$^{-2}$ ($5\,\sigma$ in 1\,h) for the [C\,II] and [O\,I] lines respectively.} (red line), and opens the possibility to detect the gas component in a statistically significant sample of debris disks (about 100 potential detections), thereby measuring the volatile content of planetesimals. In addition, these observations characterize the nature of the gas production process, steady versus stochastic, thus informing how the late stages of planet assembly and atmosphere build-up proceed in extrasolar systems \citep{Kral2020a}.

\begin{figure*}[t]
\begin{center}
\includegraphics[width=0.7\columnwidth, angle=-90]{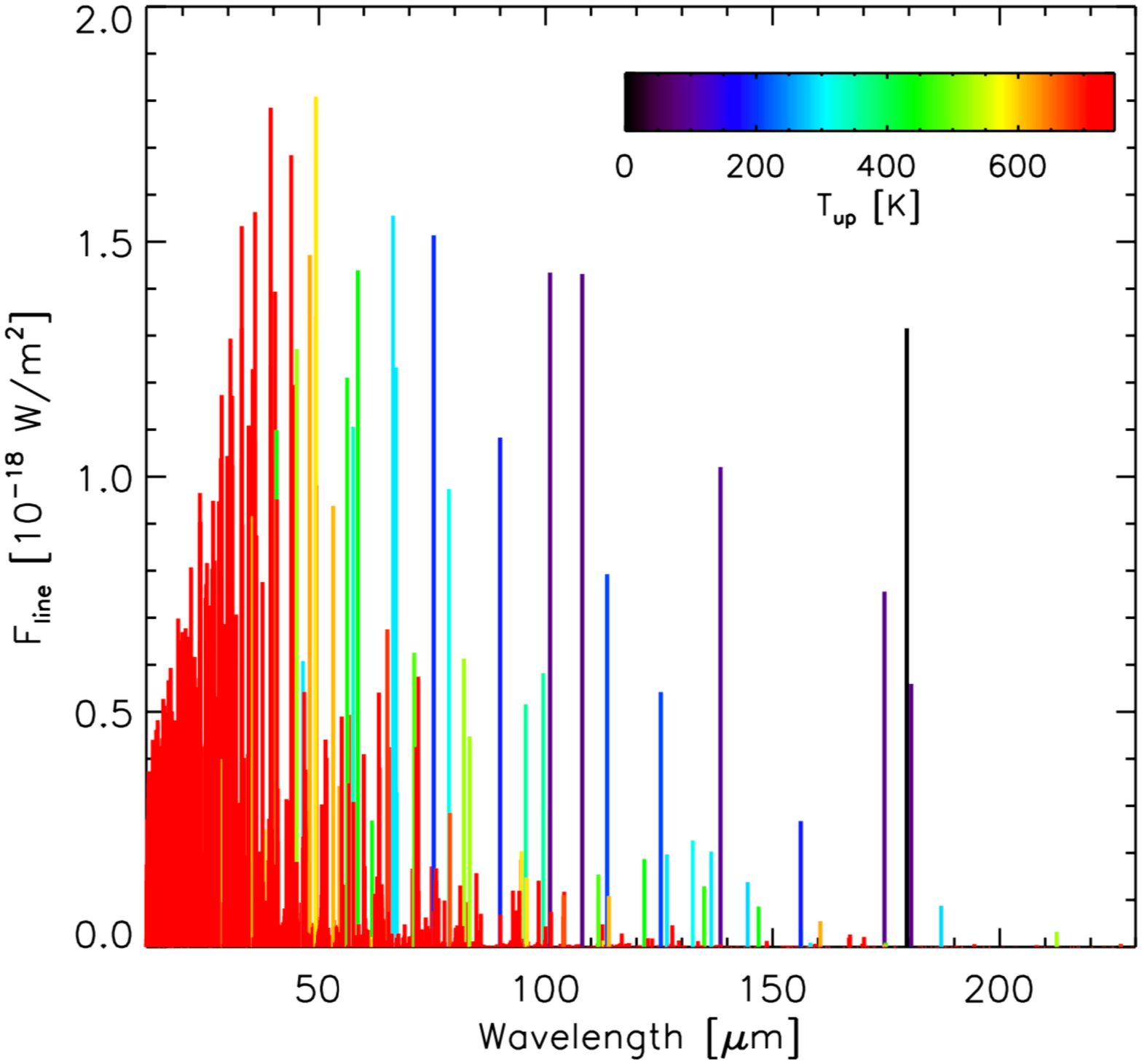}
\includegraphics[width=0.7\columnwidth, angle=-90]{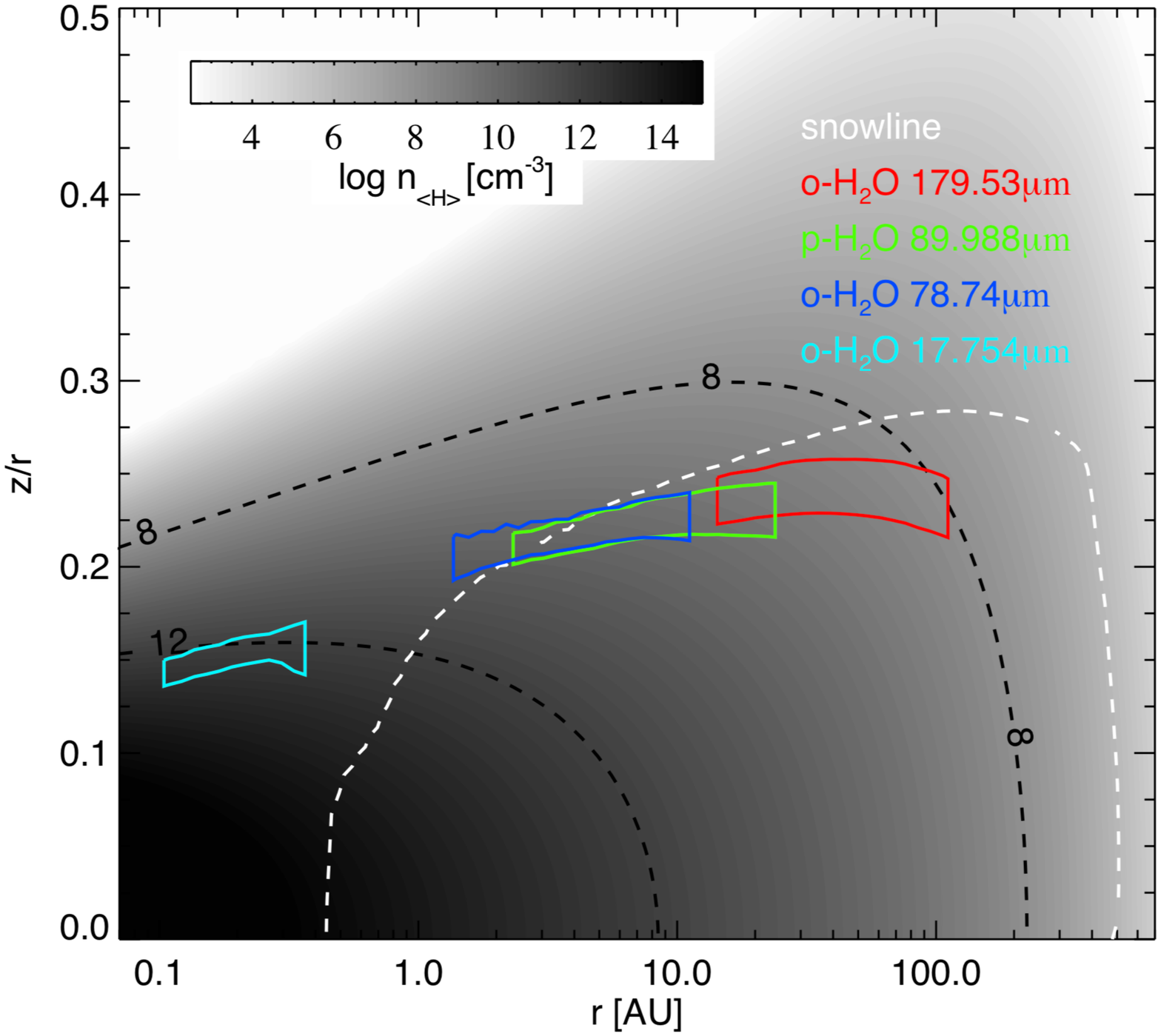}
\vspace*{-2mm}
\caption{Left: The water vapor line spectra after subtracting the dust continuum, at SMI and SAFARI wavelengths, color-coded by the energy of the upper levels of the line. Right: The line emitting regions of various water vapour transitions in a disk around a 1\,$M_{\odot}$ T\,Tauri star for the model in \citet{Kamp2017}. The white dashed line is the water snowline indicating where water vapor starts to freeze out.}
\vspace*{-6mm}
\label{fig:fig_5.1}
\end{center}
\end{figure*}

With SAFARI, a $\beta$~Pic-equivalent will be detectable with 5\,$\sigma$ in 1\,h out to 150\,pc in [O\,I]\,63\,$\mu$m and 400\,pc in [C\,II]\,158\,$\mu$m. This increases the volume in which $\beta$~Pic-like gas-rich debris disks can be detected 400- to 1200-fold, compared to Herschel/PACS. In this volume, there are hundreds to thousands of debris systems already known \citep{Cotten2016}, so a careful target selection is possible. To answer the broader question of how the chemical make-up correlates with the stellar spectral type and the presence of planets in the system, linking chemical composition to planet formation, a sample size on the order of 100 systems is required. There are already $\sim\!20$ debris disks with CO detected by ALMA, and this number is predicted to continue to rise steadily over the next decade. The detection thresholds for [O\,I] and [C\,II] predicted by \citet{Kral2017} are comparable to those of ALMA observations of CO and [C\,I]; this should result in detections of $\sim\!100$ debris disks with both ALMA and {\it SPICA}. 

\subsection{OTHER GAS TRACERS IN DEBRIS DISKS}
\label{sec:gastracersDD}

Other species of importance for planet formation and uniquely available to {\it SPICA} are silicon, through the strong [Si\,II]\,35\,$\mu$m line (using SMI), and molecules such as H$_2$O (ortho line at 179.7\,$\mu$m), and OH (doublet at 119.3, 119.5\,$\mu$m). Since molecules are quickly dissociated on timescales of days to hundreds of years \citep[see][]{Kral2019}, they will only be found in systems with significant ongoing gas production such as $\beta$~Pic, where CO is observed \citep[][]{Dent2014}. Since the molecular lines are likely optically thick, the emission will depend sensitively on the excitation temperature. Assuming a molecular gas distribution similar to the observed CO distribution in the $\beta$~Pic disk (that is, a solid angle of $\sim\!1$\,arcsec$^2$ at 20\,pc), at a velocity dispersion of 4\,km/s and excitation temperatures in the range $20 - 85$\,K, we find a predicted line flux of $\sim\!10^{-18} - 10^{-17}$\,W\,m$^{-2}$, sufficient to detect H$_2$O/OH with 5\,$\sigma$ in 20 minutes for a $\beta$\,Pic-equivalent out to 30-150\,pc.

\section{THE WATER TRAIL DURING PLANET FORMATION}
\label{sec:water-trail}

Understanding the water vapour and ice distribution in planet-forming disks will provide information on the origin of water in planetary systems, including our Solar System. Water can be delivered to rocky planets, including Earth, through icy pebbles, icy planetesimals, comets and asteroids, supporting the emergence of life on these planets. 

Water vapor line emission has 
been detected from protoplanetary disks using ground-based telescopes at near-infrared wavelength (e.g., \citealt{Carr2004}). {Resolved water line profiles confirmed that they originate in the inner few au of T\,Tauri disks \citep{Salyk2011b,Banzatti2017}.} The Spitzer Space Telescope detected water vapor lines towards dozens of T\,Tauri disks (e.g., \citealt{Carr2008, Pontoppidan2010, Salyk2011}). {More recently, the Herschel Space Observatory detected warm ($\sim\!1000$\,K) and cold water vapor ($\sim\!50$\,K) in a few disks around T\,Tauri and Herbig stars} (e.g., \citealt{Hogerheijde2011, Meeus2012, Riviere-Marichalar2012, Fedele2012}). Meanwhile, near-infrared absorption features of water ice have been detected towards some edge-on disks (e.g., \citealt{Pontoppidan2005, Aikawa2012, Terada2017}) and in scattered light towards some face-on disks (\citealt{Honda2009, Honda2016}) using ground-based telescopes. In addition, far-infrared emission features of water ice have been detected by the ISO and the Herschel Space Observatory {towards T\,Tauri and Herbig disks} (e.g., \citealt{Malfait1999, McClure2012, McClure2015,Min2016b}). 

Water vapor observations from the ground are intrinsically difficult due to the Earth's atmosphere and thus limited to very narrow windows of high excitation water lines and low abundance water isotopologues.
The Earth’s atmosphere is opaque to the rotational transitions of water, especially the low-$J$ transitions which trace the bulk of the cold gas reservoir in disks. Water remains mostly inaccessible even to air-borne facilities, like SOFIA.

{\it SPICA} will characterize the main reservoirs of warm and cold water vapour (see Fig.~\ref{fig:fig_5.1}) and the far-infrared crystalline and amorphous water ice features, giving unique access to the time evolution of the water throughout the entire disk. {\it SPICA} will trace the time evolution of the water snowlines in the planet-forming disks which divide the regions between rocky and gas giant planet formation \citep{Notsu2016,Notsu2017}. {\it SPICA} will characterize how the water vapour and ice reservoirs change in the presence of strong accretion events and substructures such as the inner cavities, gaps/rings, and spiral arms, which are related to the disk evolution and planet formation. {\it SPICA} will enable the water trail to be traced into the phase of mature planetary systems.

\subsection{TRACING THE WATER VAPOR AND ICE RESERVOIR}

Together, SAFARI and SMI cover a wide range of wavelength and many water vapor emission lines from the mid- to far-IR, as well as water ice features at far-IR wavelengths {(45 and 63\,$\mu$m)}. In particular, the wavelength range of 30-50\,$\mu$m opens new spectroscopic windows with {\it SPICA}. 
SAFARI and SMI provide access to more than a thousand ro-vibrational water vapour transitions spanning a wide range of excitation energies from $E_{\rm up} \sim\!100$\,K to several 1000\,K (Fig.~\ref{fig:fig_5.1}, left). Given this wide range of the energy levels, these transitions are ideal to trace the gaseous reservoirs in a variety of regions within {planet forming disks}. Figure~\ref{fig:fig_5.1} (right) shows the dominant emitting regions for four H$_2$O transitions in a {planet forming disk} around a 1\,$M_{\odot}$ T\,Tauri star: the H$_2$O transitions {in the SMI/SAFARI wavelength range} originate at different radial location in the disk. 

\begin{figure}
\begin{center}
	\hbox{\hspace{-0.8cm}\includegraphics[width=7.cm, angle=-90]{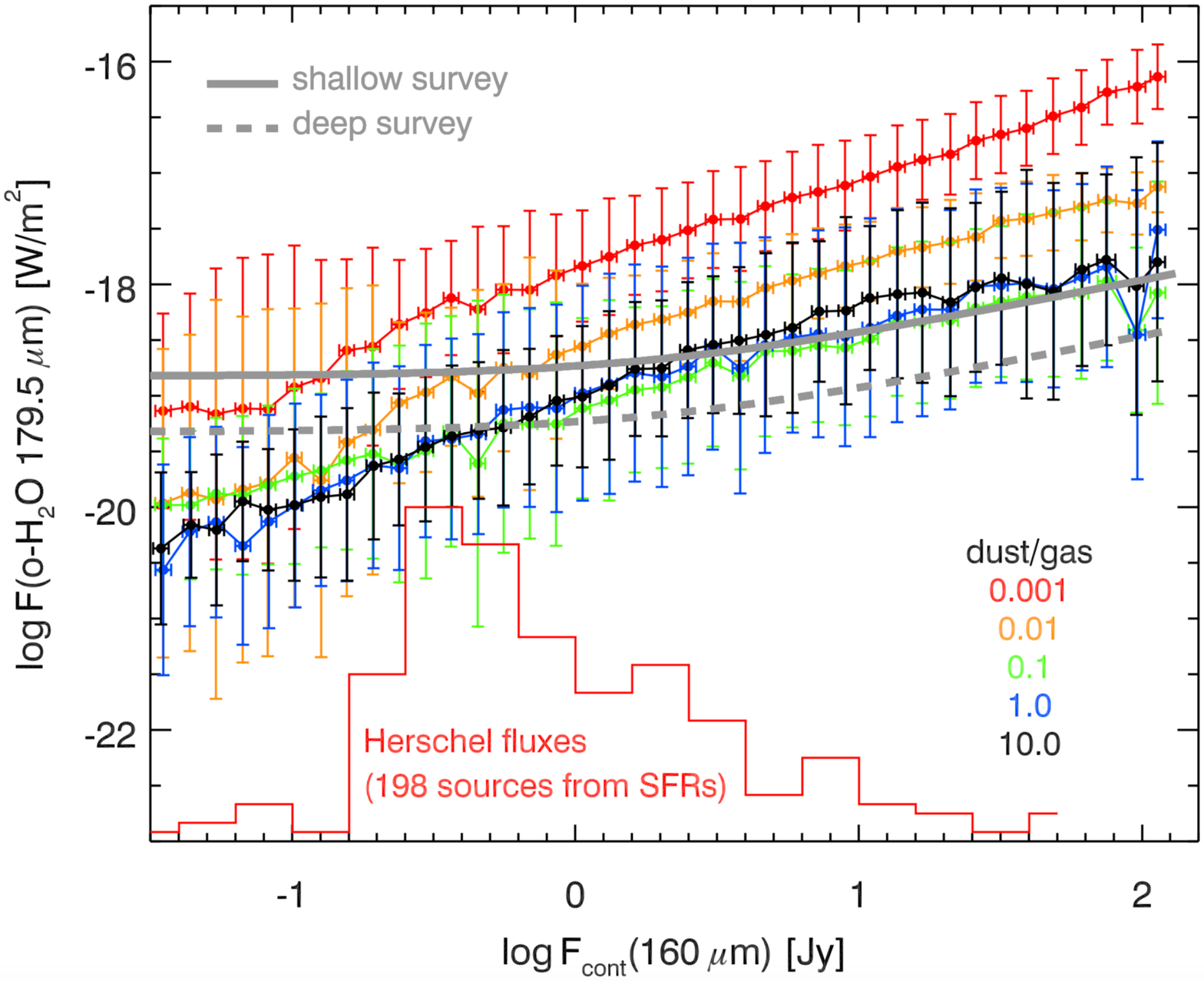}}
\vspace*{-0.2cm}
\caption{The 179.5\,$\mu$m water vapour line fluxes as a function of the dust continuum fluxes for the DENT grid of disk models with a variety of parameters \citep{Woitke2010,Kamp2011}. The sensitivity levels of SAFARI for a 1\,h (shallow) and 10\,h (deep) survey and the histogram of observed Herschel dust continuum fluxes (red) from selected star forming regions are also shown.}
\label{fig:fig_5.2}
\end{center}
\end{figure}

{\it SPICA} has a sensitivity which is a factor 50 higher than that of either the Spitzer Space Telescope or the Herschel Space Observatory. This enables the detection of both water vapour and ice with disks across a  variety of evolutionary stages. Figure~\ref{fig:fig_5.2} shows the flux from the 179.5\,$\mu$m water vapour line as a function of the underlying dust continuum flux, for models with a variety of parameters such as the gas-to-dust mass ratio, the stellar luminosity, and the disk mass \citep[DENT grid,][]{Woitke2010,Kamp2011}. The sensitivity levels of SPICA/SAFARI are shown for shallow and deep surveys, revealing that the water vapour line emission from  disks with dust continuum fluxes above $1$\,Jy are detectable. 
The embedded histogram of dust continuum fluxes {for a sample of} disks observed by the Herschel Space Observatory {reveals} that {there exists sufficient detectable targets in nearby star forming regions}. The fluxes of various water vapour lines are plotted in Fig.~\ref{fig:fig_5.1} (left), showing that many of the lines will be detectable together with the 179.5\,$\mu$m line. Thus, SAFARI and SMI have the potential to trace the evolution of the distribution of water vapor in disks.

\begin{figure*}
\begin{center}
\includegraphics[width=5.5cm]{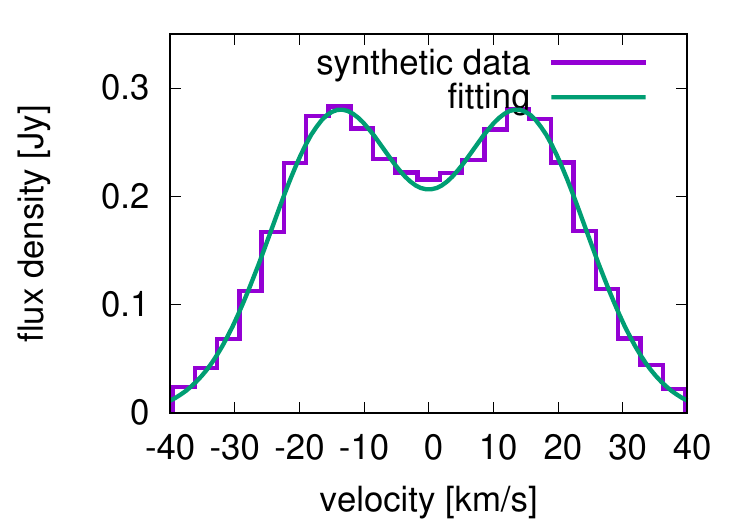}
\includegraphics[width=5.5cm]{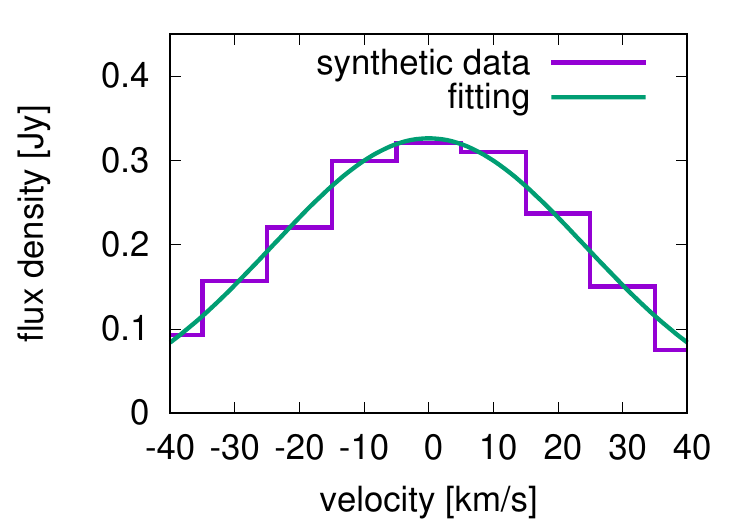}
\includegraphics[width=6.2cm]{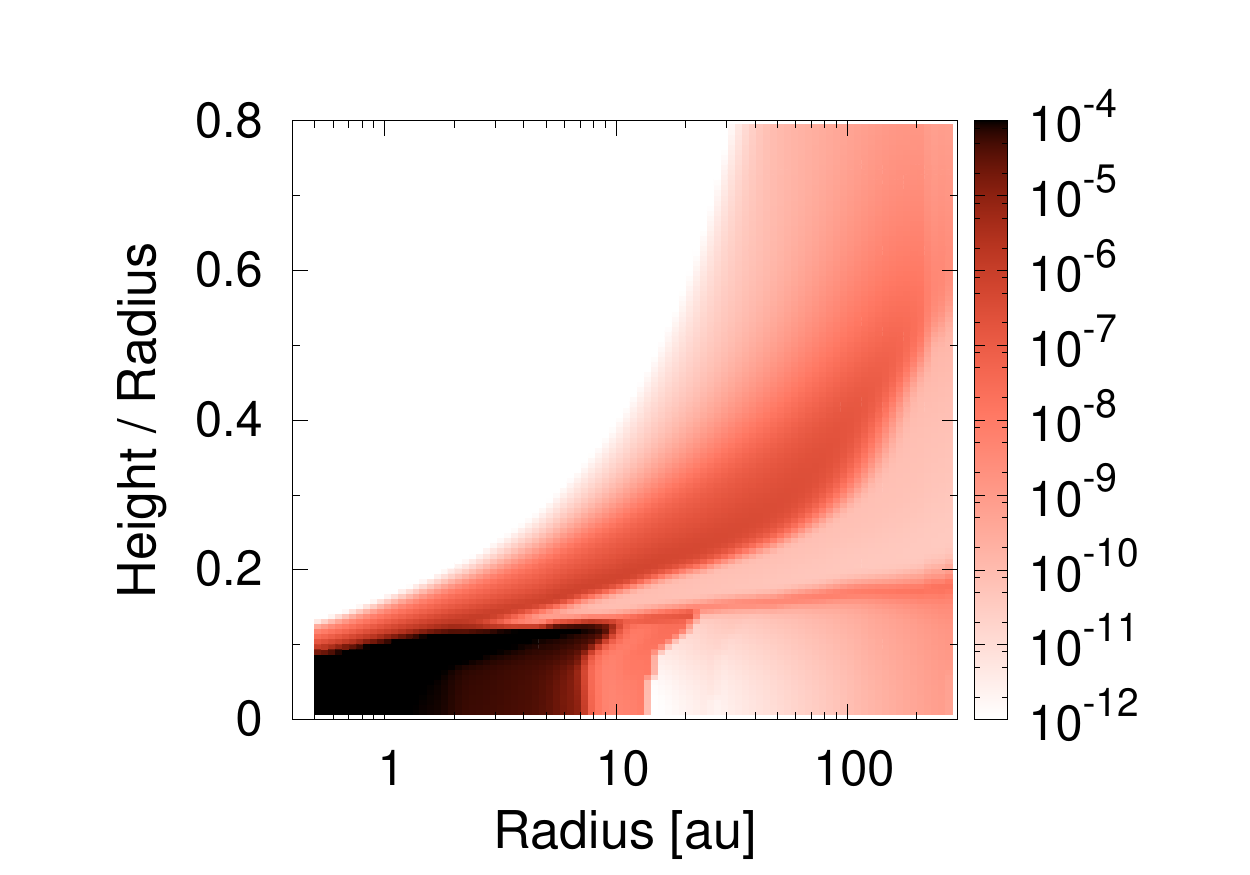}
\caption{Water profiles of the 17.75\,$\mu$m line (left) convolved with $R\!=\!29,000$ ($\Delta v\!\sim\!10$\,km/s) and the 37.98\,$\mu$m line (middle) convolved with $R\!=\!10,000$ ($\Delta v\!\sim\!30$\,km/s), emitted from a Herbig Ae disk with inclination angle of 45 degree at a distance of 140\,pc. An integration time of $\sim\!10$ minutes is assumed. Right: The water vapour abundance distribution ($n_{\rm H_2O}/n_{\rm H}$) in a disk around a 2.5 M$_{\odot}$ Herbig Ae star \citep{Notsu2017}.} 
\label{fig:fig_5.3}
\end{center}
\end{figure*}

\subsection{TRACING THE WATER SNOWLINES}

The high-resolution mode of SMI is unique for locating the water snowline near the midplane in planet forming disks. 
Since the line emission is Doppler-shifted due to the Keplerian rotation, the velocity profiles of the water lines are sensitive to the radial distribution of the water line-emitting region within the disks. SMI/HR ($R\!\sim\!30,000$) 
spectra enable us to spectroscopically resolve the water line profiles originating in the inner few au of the disk and thus locate kinematically the line-emitting regions, that is, the water snowlines in disks (Fig.~\ref{fig:fig_5.3}, {see also Fig.~\ref{fig:fig_5.1} showing the 17.75\,$\mu$m line emitting region is located inside the white dashed snowline}). 
Since water vapour is distributed not only inside the water snowline, but also in the {warm (a few 100\,K)} surface layer and the cold outer regions (Fig.~\ref{fig:fig_5.3}, {see also Fig.~\ref{fig:fig_5.1} showing the 78.74 and 179.53\,$\mu$m lines trace the outer disk}), we need to {select} specific transitions in order to only trace the water vapour inside the water snowline. 

Sophisticated physical and chemical disk models show that the transitions with low emissivity, {i.e.\ low} Einstein A coefficient of $10^{-6}-10^{-3}$ s$^{-1}$, and {an excitation energy} of $E_{\rm up}\sim\!1\,000$\,K are {suitable} to trace the water snowline (\citealt{Notsu2016,Notsu2017}). 
These transitions exist at wavelengths ranging from mid-IR to sub-millimeter, and the 17.75\,$\mu$m line, which satisfies the condition to trace the water snowline, {is the best candidate transition} in the wavelength range of SMI/HR ($12-18\,\mu$m). {The} 37.98\,$\mu$m line in the wavelength range of SAFARI is {similarly suited, but the spectral resolution even with the FTS is lower ($R\!\sim\!10\,000$)}. {Thus SAFARI will not be able to} resolve the double peak structure of {this emission line,} but will enable us to measure the line width. {This will} give us {additional} information on the location of the snowline (Fig.~\ref{fig:fig_5.3}). 

These {water lines cannot be accessed} from ground-based telescopes due to telluric absorption. The sensitivity of the Spitzer Space Telescope was not high enough to detect {these} lines {(\citealt{Antonellini2015,Antonellini2016b, Antonellini2016})}. JWST will be able to detect the 17.75\,$\mu$m line, but its highest spectral resolution {($R\sim\!3,000$) is} not high enough to resolve the line profiles. {However,} JWST observations will be {key in measuring} line strengths, {a key input for the planning} of future observations by {\it SPICA}. {Suitable water lines to trace snowlines with ALMA} require long integration times to detect the {line profiles} (\citealt{Carr2018, Notsu2018,Notsu2019}) and {thus} it will not be possible to study the time evolution of the location of water snowlines {across large samples of disks}. 

{The detection of the 17.75\,$\mu$m water line profiles towards Herbig Ae stars with enough sensitivity to locate the snowline} requires only short integration times ($\sim\!30$\,min) with SMI/HR even if the disks are located at the distance of the Orion Molecular Clouds ($\sim\!420$\,pc). T\,Tauri disks in nearby star forming regions ($<\!300$\,pc) will require {typically} {$\sim\!1$\,h} exposure times to achieve the {required} sensitivity {for a reliable snowline estimate}.

\begin{figure}
\begin{center}
\includegraphics[width=1.0\columnwidth]{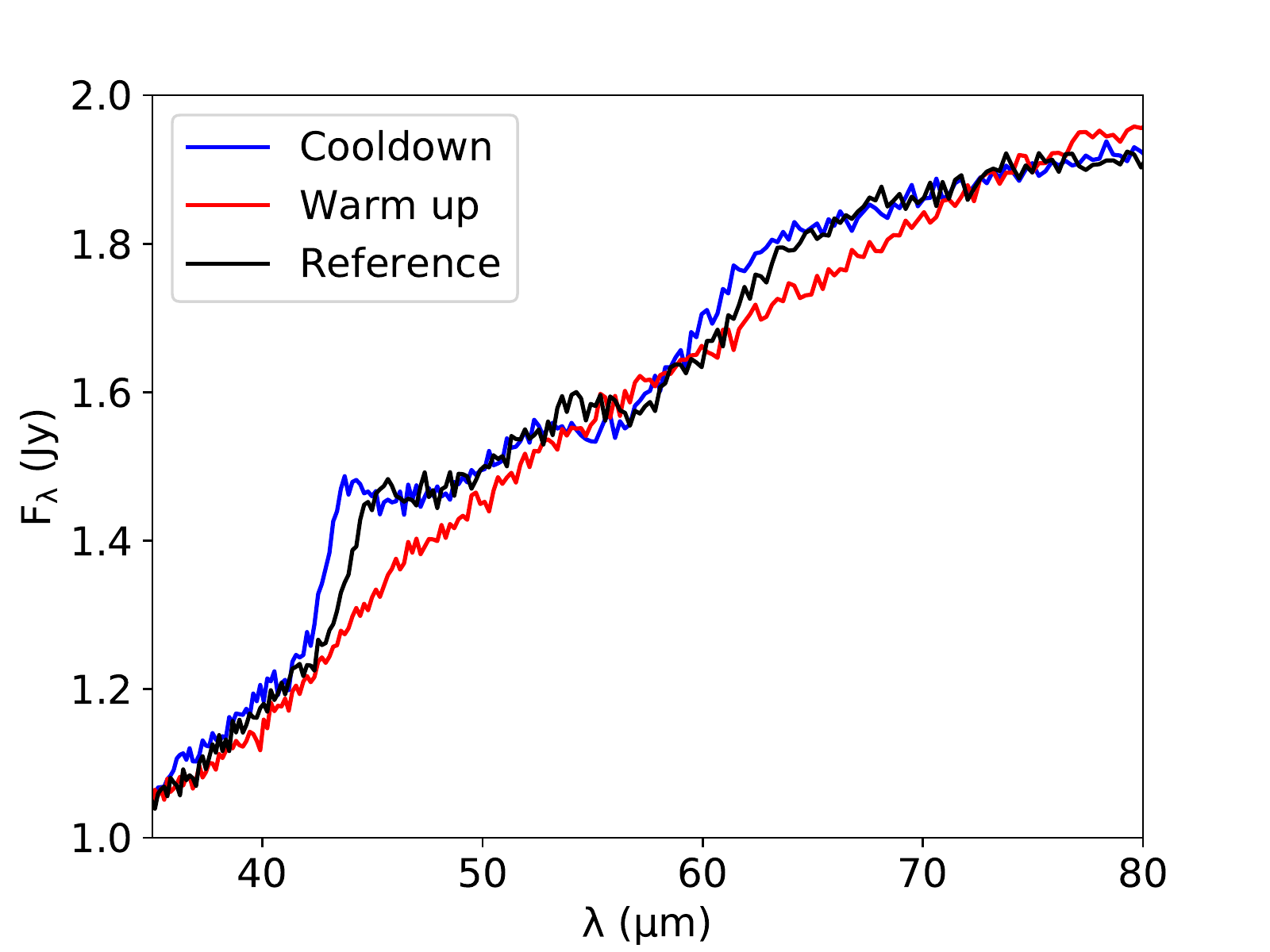}
\caption{Simulated water ice spectra at 45\,$\mu$m with different thermal histories. Using a 10\,min integration (SAFARI noise level) and spectral resolution of $R\!=\!250$, these are distinguishable. The "Reference" case is for using constant temperature crystalline ice opacities (140\,K).}
\vspace*{-0.5cm}
\label{fig:fig_5.4}
\end{center}
\end{figure}

\subsection{TRACING THE THERMAL HISTORY OF WATER ICE}

Water ice has multiple bands in the infrared, each of which are sensitive to the temperature of the ice. In the far-IR, two prominent features exist around 45\,$\mu$m and 63\,$\mu$m. The thermal history of water ice is imprinted in the peak strength and shape of the 45\,$\mu$m feature that is narrower in the crystalline phase compared to the amorphous form. {\it In situ} formation and mass transport from the inner disk has been discussed by \citet{Min2016a} as potential origin of crystalline water ice in planet forming disks. Such mixing of material between hot inner regions and cold outer regions of the disks has been also discussed based on the presence of crystalline silicates in comets \citep[e.g.,][]{Watson2009}.

Observing water bands in the far-IR also allows to trace deeper regions within planet forming disk due to the lower opacity at longer wavelengths. The mid-IR ice bands at 3, 6 and 13\,$\mu$m are mostly optically thick under typical disk conditions, and therefore do not probe the reservoir where most of the ice is located \citep{Rocha2020}. Hence, the 45\,$\mu$m feature is an excellent tracer of the formation of ice and transport of icy material within disks. Water ice formed in the warm region near the snowline and transported to the cold outer region (cooldown) has a different structure (and thus spectral appearance) compared to water ice formed in cold regions and transported into warm regions (warmup, e.g.\ \citealt{Smith1994}). The model calculations around a 0.7\,$M_{\odot}$ T~Tauri star show that ice features with these different thermal histories {are} distinguishable through observations with a spectral resolution of $R\!=\!250$ (Fig.~\ref{fig:fig_5.4}). SOFIA/HIRMES also will be able to detect this ice feature; however, SOFIA will require long exposure times, $\sim\!1$~hr for the most nearby star-forming regions. SPICA/SAFARI can achieve the same sensitivity with much shorter integration times ($\sim$\,10~min) and {thus will be a} unique instrument to perform {large surveys} of water ice features towards young objects {in} various evolutionary stages (\citealt{Kamp2018}). 

\section{FROM PRISTINE DUST TO PLANETESIMALS/COMETS}
\label{sec:mineralogy}

\begin{figure*}[ht!]
	\includegraphics[width=16cm]{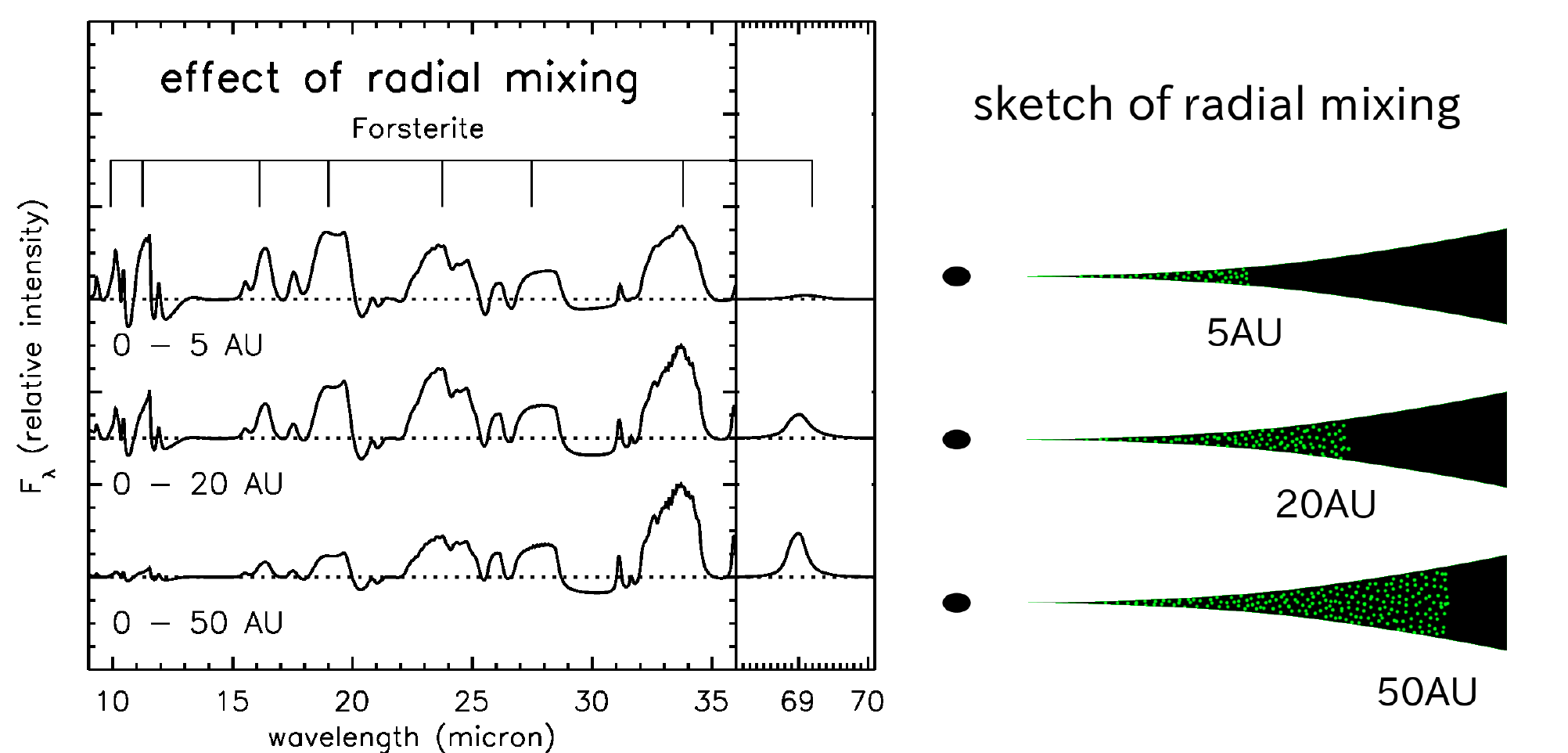}
\caption{
The effect of radial mixing of forsterite (crystalline silicate) grains in the disk on the infrared spectra. When fosterite resides only within 5 au, the 69\,$\mu$m forsterite feature is very weak. On the other hand, when forsterite appears within the 0-50 au radius of the disk, astrong 69\,$\mu$m feature appears \citep{Maaskant2015}.}
\label{fig:Maaskant2015}
\end{figure*}

While only one per cent of the matter in disks is solid, dust grains play a crucial role in the formation of planetary systems. In order to understand the properties of planets, it is important to constrain the chemical composition of their building blocks: these are small dust particles that enter the disk from the molecular cloud and grow as they are transported inwards. In the inner disk, annealing of the amorphous interstellar grains causes the lattice structure to become ordered, forming crystalline silicates. 

Crystalline grains can also be 
produced
by gas phase condensation. Radial mixing transports these hot grains from the inner disk outwards, and they become part of planetesimals and eventually planets. In the inner disk, collisions between dust particles result in a constant cycling of grains between larger and smaller particles, including larger parent bodies. Collisions dominate in the gas-poor debris disk environment, allowing a unique view on the composition of parent bodies. All these processes alter the chemical composition of dust. Infrared spectroscopy can be used to constrain the chemical composition of dust particles. 

ISO-SWS, Spitzer-IRS, and JWST-MIRI have studied or will study the dust emission from the inner, warmer disk regions, probing temperatures down to about 100-150 K. The ISO and Spitzer observations show clear evidence for grain growth and the presence of both Mg-rich olivines (Mg$_{2x}$Fe$_{(2-2x)}$SiO$_4$) and pyroxenes (Mg$_{x}$Fe$_{1-x}$SiO$_3$), with up to 10 per cent of Fe contained in some of the crystalline pyroxenes   \citep[see,][]{Juhasz2010,Sargent2009}. Spitzer spectra also reveal that olivines are more abundant in the \emph{outer disk}, and pyroxenes in the inner disk \citep{Juhasz2010}. Radial mixing models (e.g. \citealt{gail2004A&A}) suggest an opposite trend, with olivines (specifically forsterite Mg$_2$SiO$_4$) more abundant in the \emph{inner disk}. 

\begin{figure}[ht]
        \vspace*{-2cm}
	\hbox{\hspace{-1.5cm}\includegraphics[width=8cm, angle=-90]{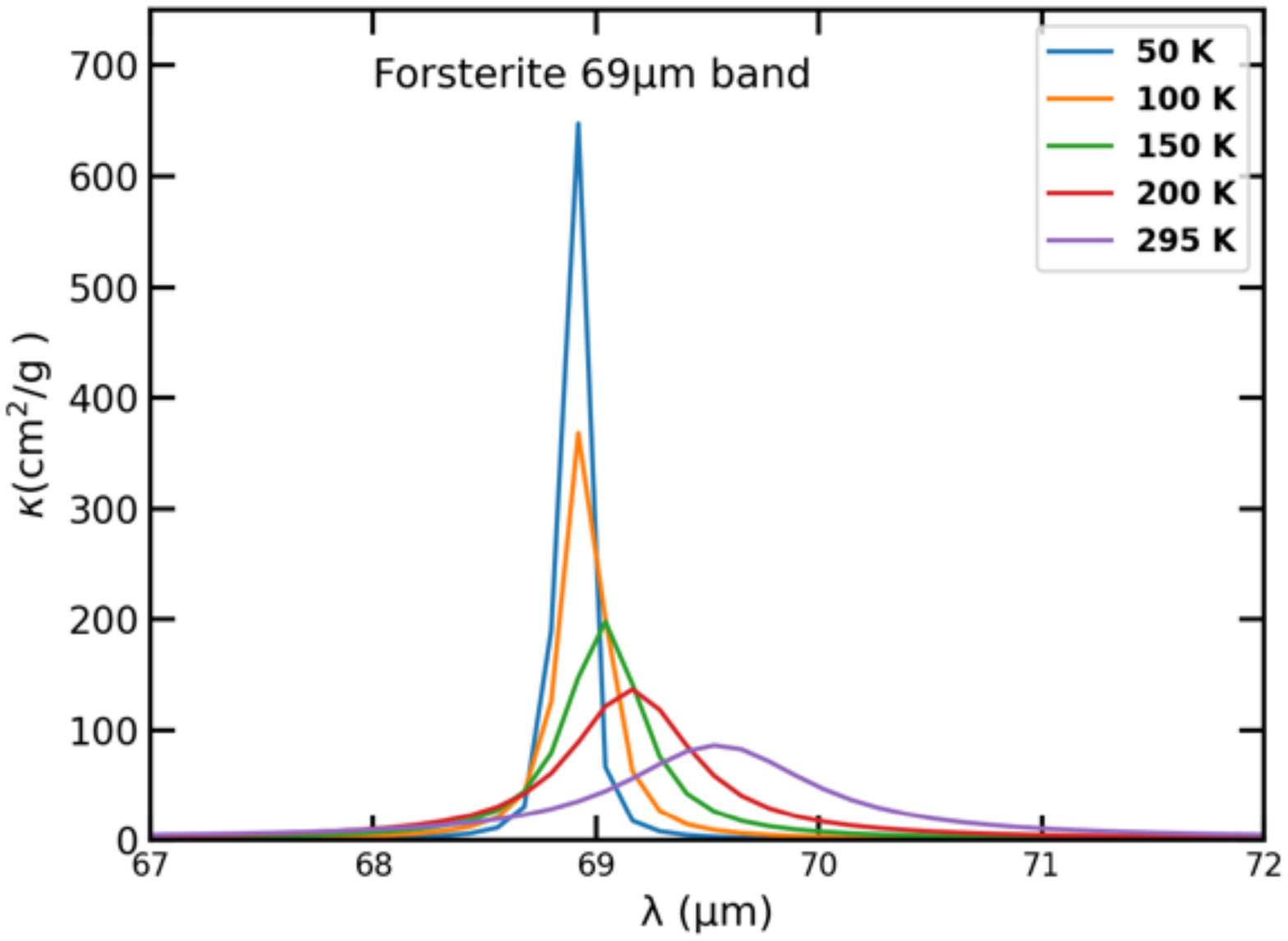}}
	\vspace*{-14.5cm}
	\hbox{\hspace{-8.cm}\includegraphics[width=24cm]{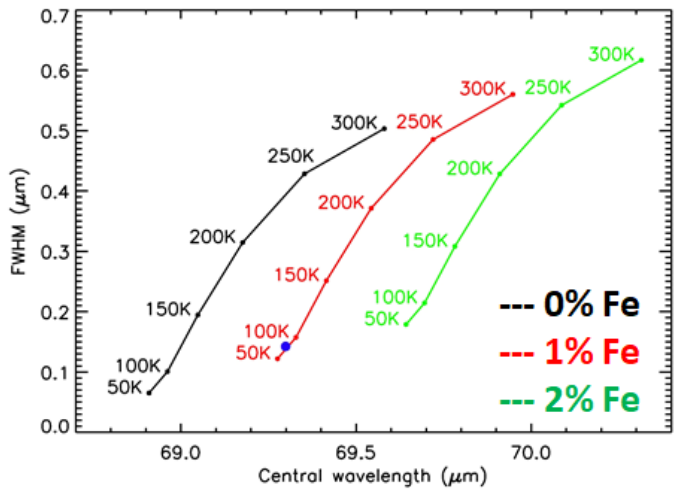}}
	\vspace*{-14cm}
\caption{(Top) The 69 $\mu$m resonance of Mg$_2$SiO$_4$ for a range of temperatures. The band position broadens and shifts redwards with increasing temperature. Note that the band also shifts redwards with increasing Fe content. Optical constants from \citealt{suto2006}, using a DHS grain model (\citealt{Min:2005cq}) with $f\!=\!0.7$ and a grain size of 1\,$\mu$m. 
(B) Forsterite 69\,$\mu$m feature peak and FWHM dependence on Fe content in olivine (Mg$_{2x}$Fe$_{(2-2x)}$SiO$_4$) \& temperature. Only 1\,\% inclusion results in $\sim\!0.3\,\mu$m shift to longer wavelengths \citep{devries2012}.
}
\label{fig:fo69TempDependance}
\end{figure}

Much less is known about the composition of the cold dust reservoir. Solar system comets that were formed in the outer regions of the proto-solar nebula contain significant amounts of refractory materials (crystalline silicates, FeS) that require high formation temperatures, that are more typical for the inner disk. (e.g. \citealt{Crovisier1997, Brownlee:2014gv,  Wooden:2017ev}). In contrast, the molecular and volatile content of comets provides strong evidence that these objects never experienced high temperatures (e.g. \citealt{Calmonte:2016iv}). The outer regions of the proto-solar nebula may have experienced substantial mixing of solids from high temperature regions in the disk. However, it is not clear such radial mixing is a common process in the disk during planet formation.     

Both ISO-LWS and Herschel have searched for solid-state features in the far-IR spectra of Herbig Ae/Be disks, and they were successful in detecting the 69 $\mu$m crystalline olivine band \citep[][see Figs.~\ref{fig:Maaskant2015} and \ref{fig:fo69TempDependance}]{malfait1998,bowey2002,sturm20102010A&A,sturm2013, Maaskant2015}. For Herbig Ae/Be proto-planetary disks, \citet{sturm2013} found eight detections of the 69 $\mu$m band of, almost pure, forsterite (Mg$_2$SiO$_4$) in a sample of 32 Herbig Ae/Be disk observations. For debris disks only one detection of the 69 $\mu$m band was found out of a set of eight observations \citep{devries2012}. The observations so far tell us about the diversity of the crystalline silicate abundance in the cold disk region, implying a variety of disk mixing histories. To directly compare with our Solar System radial mixing history, observations of T\,Tauri stars are required. However, such observations have been limited due to the sensitivity of previous instruments and issues with the pointing and base-line stability over a broad wavelength range.

\begin{figure}[ht]
\vspace*{-2cm}
	\hbox{\hspace{-1.5cm}\includegraphics[width=8cm, angle=-90]{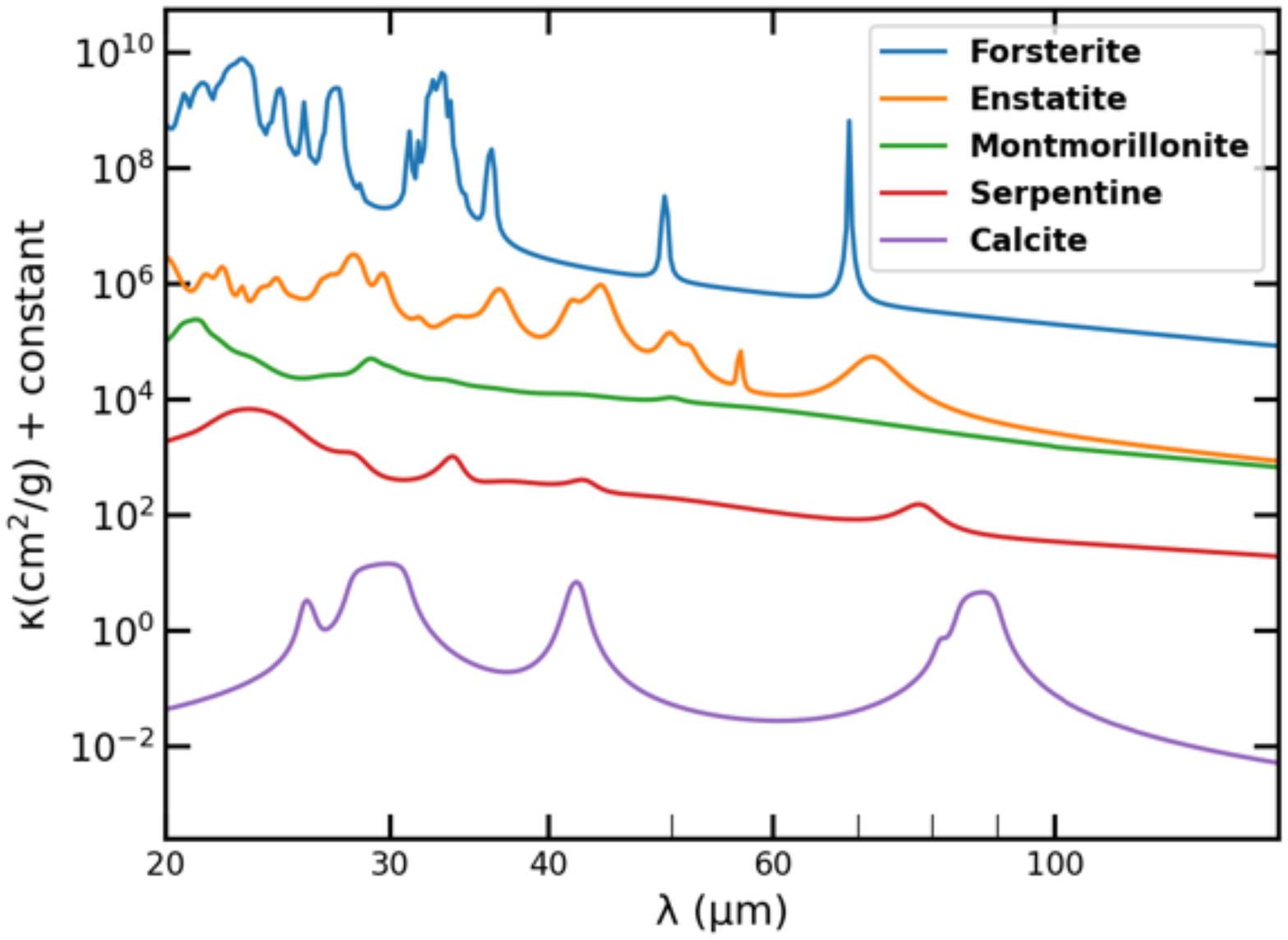}}
\vspace*{-1cm}
\caption{Opacities of forsterite, enstatite, calcite, and the hydrosilicates montmorillonite and serpentine, showing the potential of {\it SPICA} to detect these dust species over the full SMI and SAFARI wavelength range. Optical constants from \citet{suto2006}, \citet{Zeidler:2015fu}, and \citet{koike1990}  using a ``distribution of hollow spheres'' grain model (\citealt{Min:2005cq}) with a vacuum fraction of $f\!=\!0.7$ and a grain size of $0.1\,\mu$m.
}
\label{fig:dustopa}
\end{figure}

{\it SPICA} 
will be able to observe solid-state resonances from dust particles in the far-IR with {\it unprecedented sensitivity}, and will have the full mid- to far-IR spectral coverage to identify and characterize minerals that show vibrational resonances over a wide wavelength range (Fig.~\ref{fig:dustopa}). Far-IR resonances of minerals are invaluable since they allow the determination of mineral properties within the low temperature regions where comets are formed. 
By using the many resonances of these dust species, the spatial distribution of olivines and pyroxenes can be derived. Figure~\ref{fig:Maaskant2015} illustrates how FIR is important to understand the radial mixing history. The relative abundance of olivines to pyroxenes as a function of radius in the disk can be used as a probe of the relative importance of different processes in the disk that produce crystalline silicates, such as chemical equilibrium gas-phase condensation, thermal annealing of molecular cloud amorphous silicates, and (local) flash heating events caused by activities such as stellar outbursts, lightning, and planet formation. 

Far-infrared resonances also allow us to measure the Fe/Mg ratio in crystalline olivine and pyroxene with very high precision (down to an error of better than a percent). For example, the wavelength position and width of the 69\,$\mu$m band of olivine (Mg$_{2x}$Fe$_{(2-2x)}$SiO$_4$), is strongly dependent on the Fe/Mg ratio and grain temperature(see Fig.~\ref{fig:fo69TempDependance}). This allows a precise determination of the composition of olivine and pyroxene grains.
The Fe/Mg ratio of (crystalline) olivine and pyroxene is sensitive to parent body processing and can be used to constrain the size and formation timescale of parent bodies. In the case of a small parent body ($<10$ km) the composition of the mineral is un-altered since its formation in the proto-planetary disk and we expect a very low Fe content  (Fe/(Fe+Mg)\,$\sim\!0$). In slightly larger parent bodies ($\sim\!10$\,km to several hundreds of kilometres) the minerals undergo equilibration (Fe/(Fe+Mg)\,$\sim\!0.3$) and for large planetesimals and planets ($>\!200$\,km) igneous processes will strongly impact the minerals (Fe/(Fe+Mg)\,$>\!0.5$.
As in gas-rich disks, the mineralogy of $\mu$m-sized dust particles in debris disks directly probes the composition of their parent-bodies: extra-solar asteroids and comets. 

Furthermore, a search for carbon bearing solids such as calcite and dolomite will be interesting. Carbon bearing minerals like calcite (CaCO$_3$) and dolomite (CaMg)(CO$_3$)$_2$ only form under specific conditions, on Earth requiring the presence of liquid water or biogenic processes. Carbonates have been detected in evolved stars \citep{kemper2002} and protostars \citep{ceccarelli2002A&A}. Several formation mechanisms have been proposed (e.g.\ \citealt{toppani2005}), such as gas-grain chemical reactions and non-equilibrium gas-phase condensation, all requiring a CO$_2$ and H$_2$O-rich environment. The detection of such minerals in planetary systems would strongly impact our current view on the early presence of environments that are essential for life.

\section{THE FIRST TRUE KUIPER BELT ANALOG}
\label{sec:kuiperbeltanalog}
 
\begin{figure}
    \centering
	\hspace{-0.5cm}
	\includegraphics[width=8.5cm]{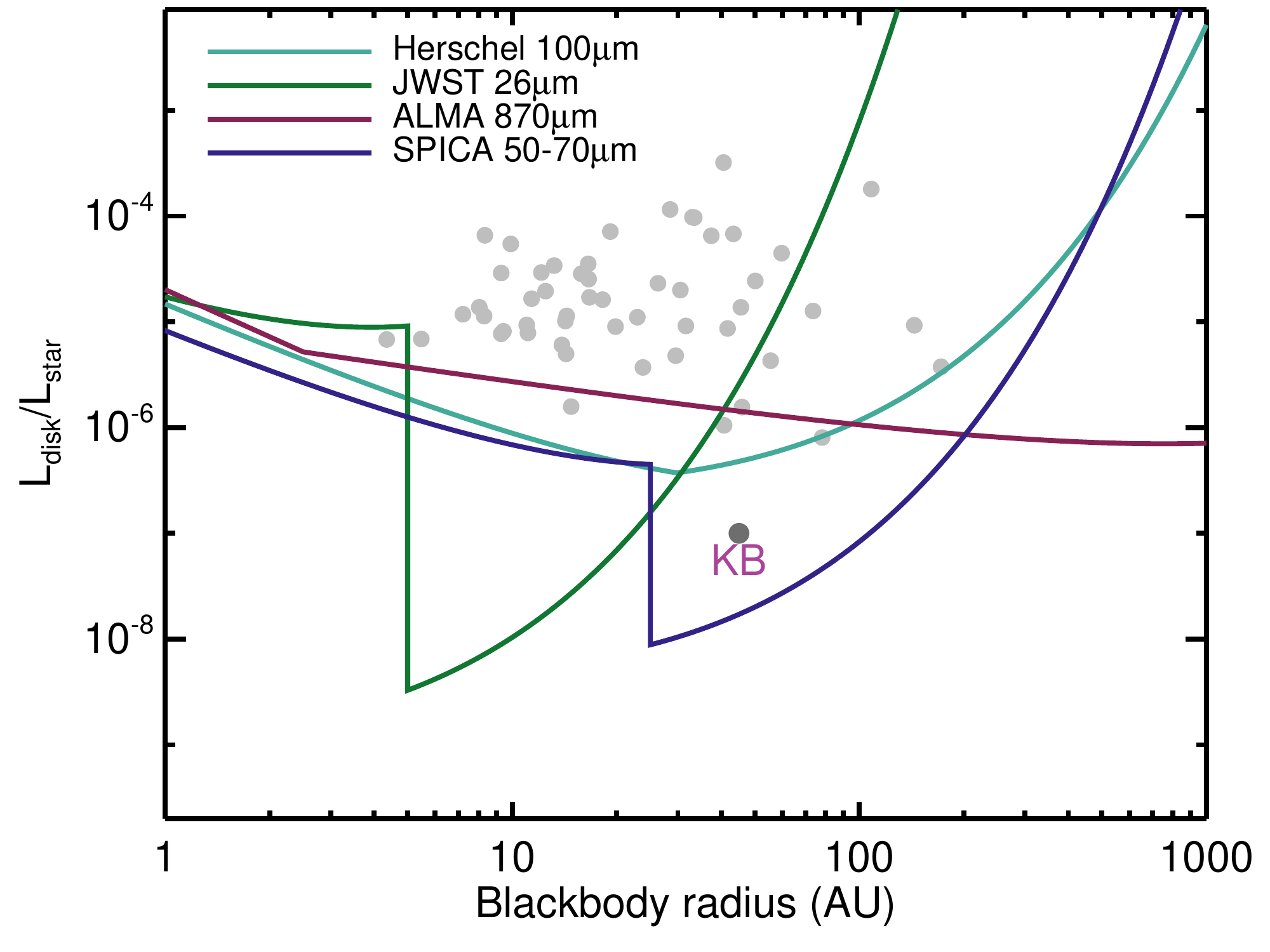}
	\caption{Sensitivity to Kuiper belt analogues at 5\,pc. The blackbody radius is the stellocentric radius obtained from the dust temperature by assuming the dust behaves as a blackbody. Grey dots are known disks around nearby stars. Disks that lie above a given instrument’s line are detectable. The {SPICA} sensitivity at 50-70\,$\mu$m is very similar, so only one line is shown. Thus, only {SPICA} can detect debris disks at Kuiper belt (and fainter) levels.}\label{fig:kb_sens}
\end{figure}

While all stars presumably host debris disks at some level, only the thermal continuum of the brightest 20\,\% are currently detectable, and we know neither our rank in the remaining 80\,\%, nor how this rank is related to the Solar System’s history or planetary architecture. A decade from now we will have detected or set stringent limits on planets around most nearby stars, but the limits on small body populations will be as poor as they are now (see Fig.~\ref{fig:kb_sens}). {SPICA} provides our only opportunity to detect true Kuiper belt analogues.
 
 A true Kuiper belt (KB) analogue is defined here as a debris disk with a fractional luminosity $L_{\rm disk}/L_{\star}\!\sim\!10^{-7}$ and a radius of $\sim\!45\,L_\star^{0.2}$\,au \citep[i.e., scaling with $L_\star$ as suggested by mm-wave imaging of bright debris disks,][]{Matra2019}. With temperatures of $\sim\!50$\,K, KB analogues emit essentially all of their energy at far-IR wavelengths, but their low luminosities imply that this emission is at most 1\,\% of the stellar flux. To detect a debris disk requires: (1) sensitivity sufficient to detect the disk flux, (2) unresolved disks to be brighter than some fraction of the stellar flux due to calibration uncertainties, (3) resolved disks to have surface brightness greater than the local uncertainty in the stellar PSF, and (4) disks to be brighter than the confusion limit. The typical calibration limit (2) is $\sim\!10$\,\%, explaining why it is not possible to detect unresolved KB analogues (e.g., with Spitzer or Herschel), while the sensitivity and PSF limits (1) and (3) preclude the detection of resolved KB analogues with ALMA, JWST, or in scattered light. These points are illustrated in Fig.~\ref{fig:kb_sens} which shows the detectability of debris disks with different radii and fractional luminosities for a Sun-like star at 5\,pc. Disks can be detected above the lines indicated for a given instrument, and essentially all known disks lie above the Herschel 100\,$\mu$m limit, with no current or planned instruments able to detect KB analogues, except {SPICA}.

\begin{figure*}
    \centering
	\hspace{-0.5cm}
	\includegraphics[width=8.5cm]{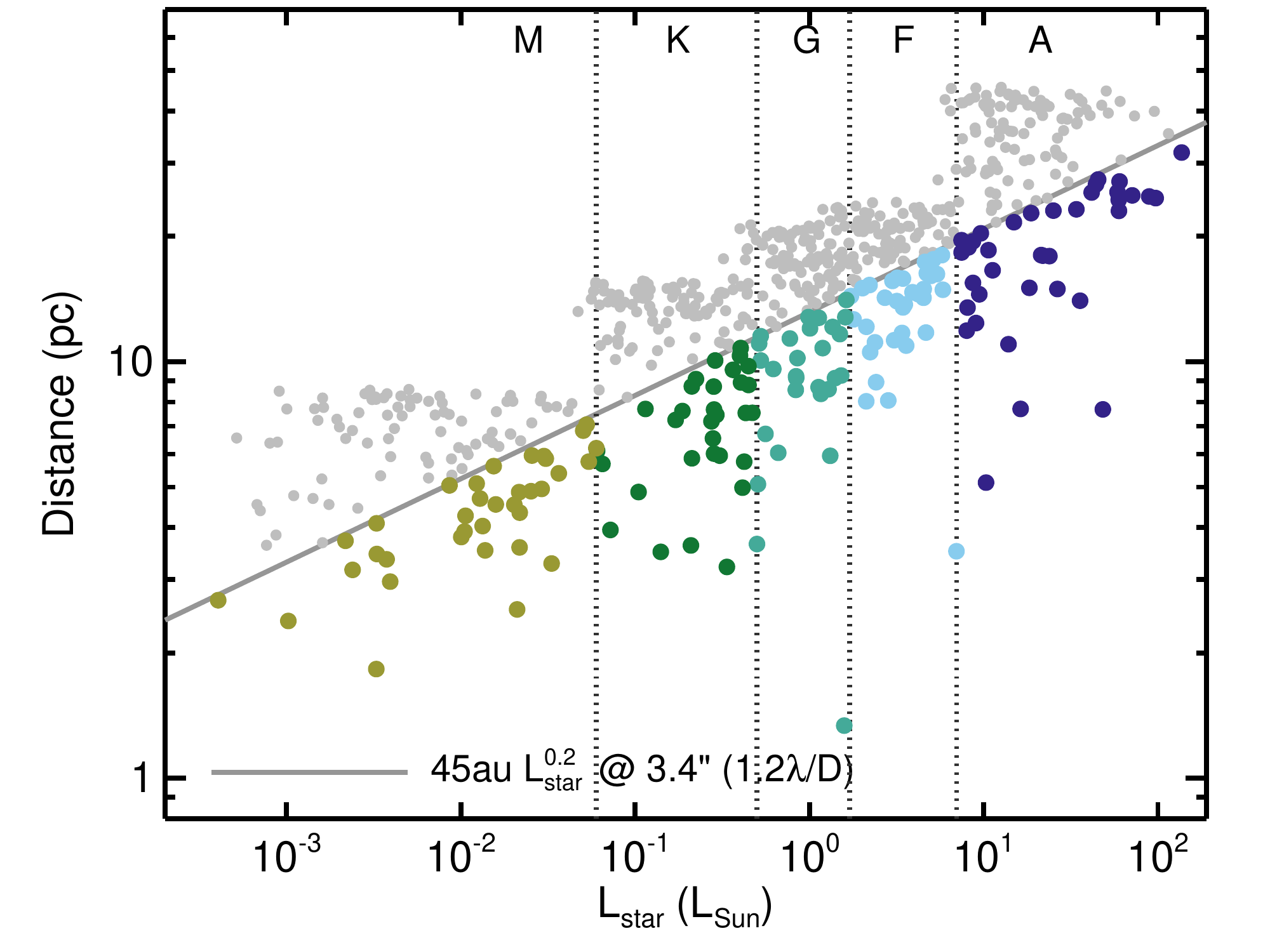}
	\hspace{0.cm}
	\includegraphics[width=8.5cm]{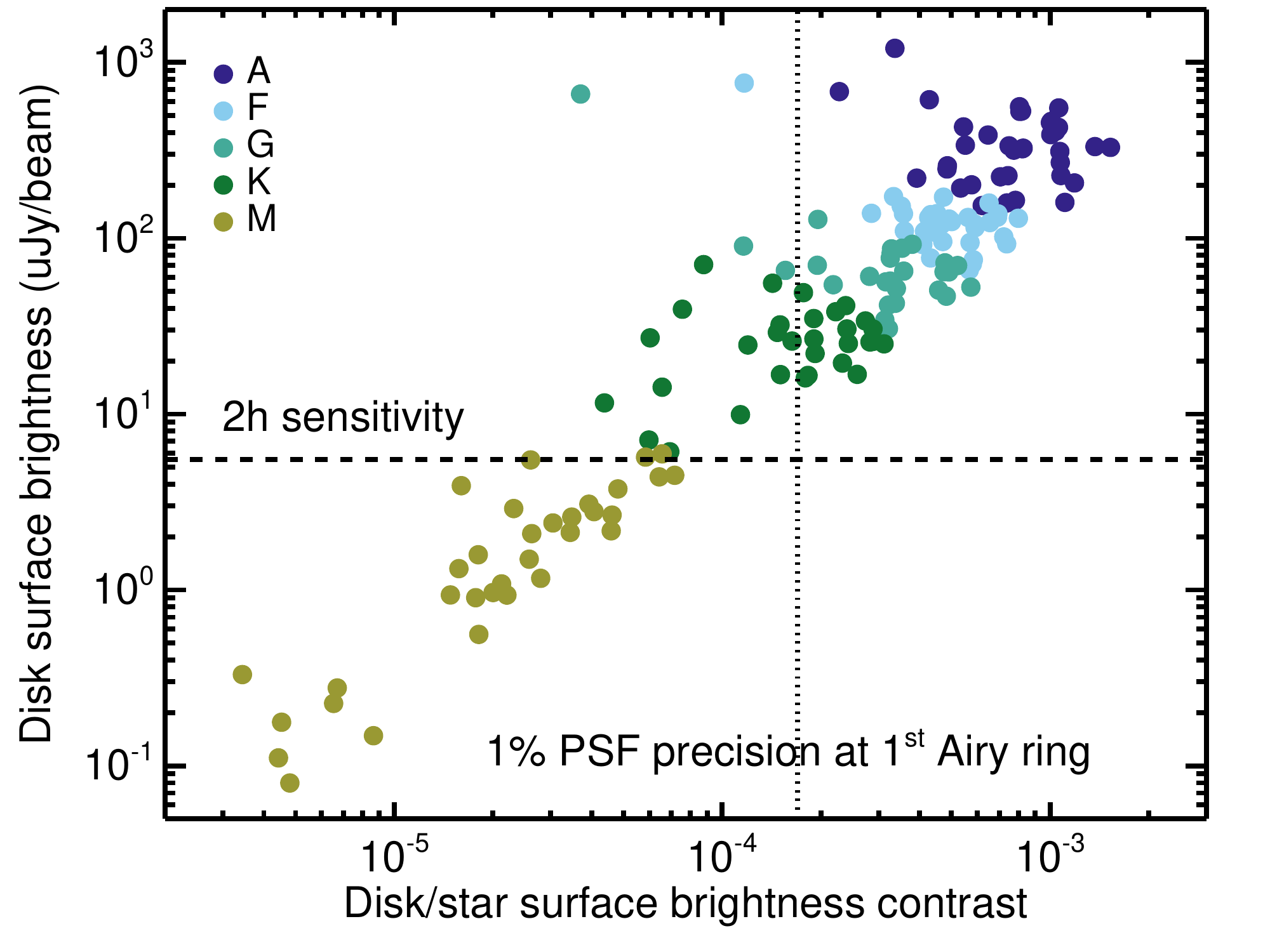}\\
	\vspace{0.3cm}
	\hspace{-0.5cm}
	\includegraphics[width=8.5cm]{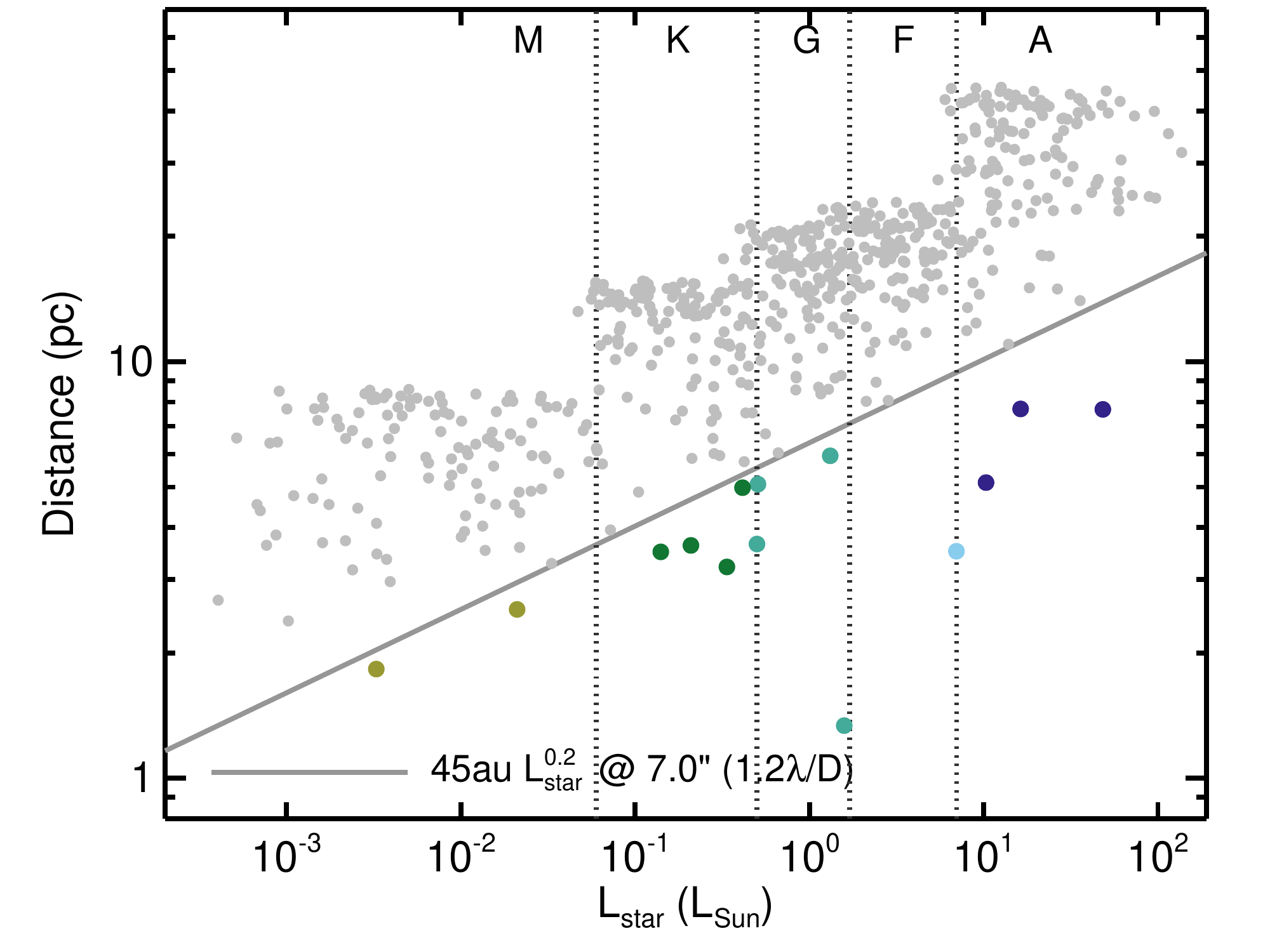}
	\hspace{0.cm}
	\includegraphics[width=8.5cm]{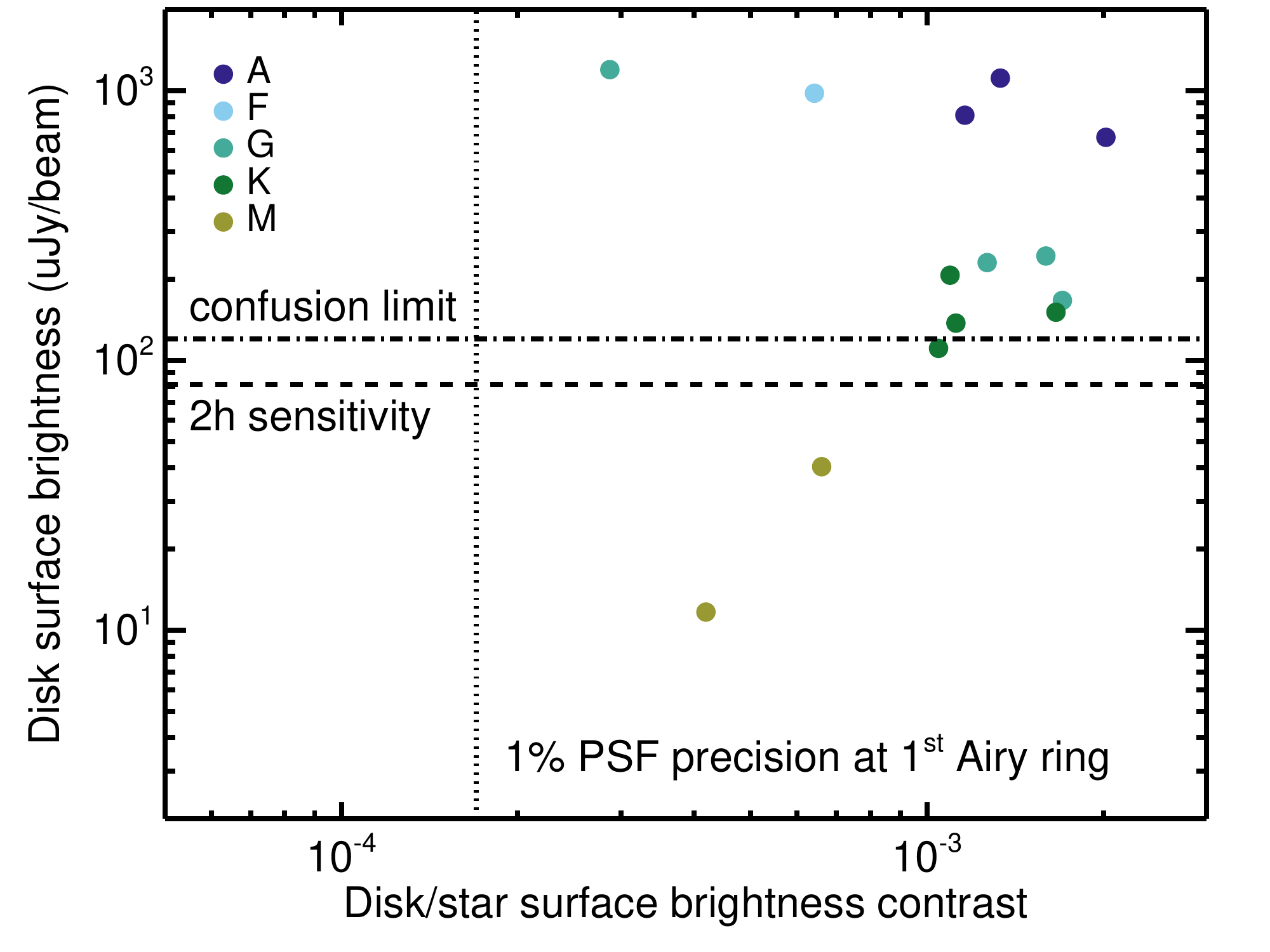}\\
	\caption{Kuiper Belt sample selection. The top row shows the 34\,$\mu$m SMI survey, and the bottom row shows the 70$\mu$m B-BOP survey. (Left panels:) Volume-limited sample of nearby stars (Phillips et al. 2010). The KBs of targets falling below the solid line would be resolvable by {SPICA}. (Right panels:) The surface brightness of resolvable KB analogues is used to determine a further selection based on a 3\,$\sigma$ detection in a 2\,h observation assuming the sensitivity (dashed lines) and confusion limits (dot-dashed lines), and disk to star contrast assuming the PSF is known at 1\,\% accuracy near the first Airy ring (dotted lines). Stars with detectable KBs lie above and right of all lines. Here disks have been assumed to be face-on; inclined disks would move up and to the right on this plot. Detection for the top left target in the upper right panel, $\alpha$ Cen, should still be possible as the angular size of a KB for this target is well beyond 1.2\,$\lambda/D$. KBs around M-type stars cannot be detected at 34\,$\mu$m unless the contrast is better than $1.7 \times 10^{-4}$, or at 70\,$\mu$m because their surface brightness is below the confusion limit (unless stellar proper motion is used to subtract the background).}\label{fig:kb_sample}
\end{figure*}

This unique capability of {SPICA} is possible because of its combination of photometric sensitivity and resolution at wavelengths around the peak KB emission. Constraint (2) can be avoided by observing stars close enough that a KB analogue would be resolved (see Fig.~\ref{fig:kb_sample}, left panels), and constraint (4) is partially or wholly avoided by observing at $\leq\!70\,\mu$m, rather than at 100\,$\mu$m where a KB analogue would be well below the confusion limit. This means that, once the resolution constraint is met, sensitivity (1) and PSF stability (3) constraints on {SPICA's} imaging performance determine the number of stars for which it will be feasible to detect true KB analogues, as shown in the right panels of Fig.~\ref{fig:kb_sample}. Here we consider surveys at 34\,$\mu$m (upper panels) and 70\,$\mu$m (lower panels). Both require that the SPICA PSF be very well understood, but the shorter wavelength is preferred as the higher spatial resolution means that more stars can be observed (compare for example the number of stars closer than the resolution limit in the left panels of Fig.~\ref{fig:kb_sample}).

The sensitivity limit (1) is shown on the right panels of Fig.~\ref{fig:kb_sample} as horizontal dashed lines assuming a 2\,h observation per star, and using the nominal sensitivities (e.g., 10\,$\mu$Jy, $5\,\sigma$~1\,h for SMI). A longer observation is not warranted at 70\,$\mu$m because this already reaches the confusion limit (horizontal dash-dot line), while at 34\,$\mu$m the sample is more constrained by the PSF stability. For the PSF stability limit (3), consider that the surface brightness of a star's first Airy ring is 1.7\,\% of the star. The right panels of Fig.~\ref{fig:kb_sample} show that at both wavelengths face-on KB analogues are 10-100 times fainter than this at a comparable location (though these could be in the minima). Thus to detect such disks the surface brightness of the PSF wings must be known to $\sim\!1$\,\% accuracy, resulting in a PSF stability requirement of $1.7 \times 10^{-4}$ of the peak near the first Airy ring. This is the vertical dotted line shown on the right panels of Fig.~\ref{fig:kb_sample}. This PSF stability constraint can be relaxed by a factor of a few at 70\,$\mu$m because the star is fainter but the disk similarly bright compared with the shorter wavelength (meaning that sensitivity, or more accurately confusion is the limiting factor). Assuming that this PSF requirement can be met, surveys at each of the wavelengths would be able to detect KB analogues around the stars in the top right quadrants of the right panels of Fig.~\ref{fig:kb_sample}. These limits are somewhat pessimistic in the sense that an inclined KB analogue would have higher surface brightness, and higher disk to star contrast, than assumed.

The frequency of Kuiper belt analogues is unknown, but might be expected to be several tens of percent. Deriving the incidence with a precision of a few percent, comparable with that for brighter disks, would allow characterisation of the entire brightness distribution down to sub-KB levels, thus placing the Solar System’s debris disk in context for the first time. This level of precision requires observing several tens of stars. Fig.~\ref{fig:kb_sample} left shows that the KBs around $>\!100$ nearby stars would have a large enough angular size to be resolved by {SPICA} at 34\,$\mu$m, and 14 at 70\,$\mu$m. Of these, Fig.~\ref{fig:kb_sample} right shows that KB analogues around M-types would be too faint given the sensitivity, confusion, and/or PSF stability limits, leaving an unbiased sample of $\sim\!100$ targets at 34\,$\mu$m and 11 at 70\,$\mu$m. Furthermore, $\sim\!7$ from the 34\,$\mu$m sample and 4 from the 70\,$\mu$m sample are already known to host detected disks (i.e., they are in the 20\,\% of the population that have KBs bright enough to be detected by photometry). Thus, while a survey at 70\,$\mu$m could achieve the first detection of a true KB analogue, a survey at 34\,$\mu$m would be better suited to determining the frequency of dust at KB analogue levels.

To summarise, the proposed observing strategy is to image systems at 34 and/or 70\,$\mu$m with SMI or B-BOP (if the latter includes a 70\,$\mu$m detector), but the goal of detecting true KB analogues would only be feasible if the above PSF stability constraint can be met.

\section{A REFERENCE OBSERVING PROGRAM}
\label{sec:reference-mission}

In the context of the ESA M5 process, we studied the feasibility of achieving the science goals outlined in this 
paper within the typical lifetime of a space mission. 
We rely here on the detailed instrument descriptions, wavelength ranges, spectral resolution and sensitivities, outlined in \citet{PRR2018, PRR2020a} for the mid- and far-IR instruments SMI \citep{SMI2020} and SAFARI \citep{PRR2020b} and briefly summarized in the preface.

Table~\ref{tab:reference-mission} summarizes a potential reference observing program consisting of six surveys with strongly overlapping samples. Details of how the surveys are compiled can be found in Appendix~A. These surveys would provide the data required to answer the key science questions outlined in this white paper. The total exposure time required for such a reference program is 1.5~months. If we add typical overheads such as 10\,\% calibration time, slew times (3\,min per source) and assume an overall efficiency of the observatory of 60\,\% (based on Herschel experience), we estimate the full reference observing program to take $\sim\!77$\,d, or $\sim\!2.5$\,months. This is eminently feasible within a mission lifetime of 3\,years.

\begin{table*}[t]
\caption{Summary of the reference mission. Surveys in brackets use the data acquired by the main survey and do not require a different sample and/or longer exposure times. Exposure time estimates are derived using SAFARI simulator version 1.2 issue 6.2 and SMI simulator version 4.}
    \begin{center}
    \vspace{-5mm}
    \begin{tabular}{lllll}
    \hline\hline
    Name of survey & Number  & Instrument & Time  & Observing strategy\\
                   & of targets &         & request$^{\rm a}$ [h]  &  \\
    \hline
    HD survey                & 403 & SAFARI/HR & 653 & HD\,$J\!=\!2-1$ line\\
    (Water survey/CO ladder) &     & & & \\[3mm]
    Mineralogy survey        & 324 & SAFARI/LR & 94 & Forsterite 69\,$\mu$m\\
    (Ice survey)             & & & & \\
                             & 403 & SMI/LR & 51 & Forsterite 33\,$\mu$m\\[3mm]
    Snowline survey          & 507 & SMI/HR & 36 & 17.75\,$\mu$m water line\\
    (Mass loss survey)       & & & & H$_2$ lines \\[3mm]
    Isolated debris disks gas survey & 80 & SAFARI/HR & 55$^{\rm b}$ & [O\,I]\,63, [C\,II]\,158\,$\mu$m lines\\[3mm]
    Isolated debris disks ice survey & 140 & SAFARI/LR & 94 & ice features at 45 and 65\,$\mu$m\\
    \hline
    \multicolumn{3}{c}{} & \multicolumn{2}{l}{{\bf Total:} 983 h (41 d)}\\
    \hline\hline
    \end{tabular}
    \end{center}
    $^{\rm a}$ This is the exposure time without instrument and observatory overheads.\\
    $^{\rm b}$ These times were derived using issue 3.0 of the SAFARI time estimator (w/o background).
    \label{tab:reference-mission}
\end{table*}

\section{BEYOND SPICA}
\label{sec:beyond}

We now let loose the mission design of {SPICA} and comment in the following subsections on a couple of interesting capabilities that should be considered for future infrared instrumentation development.

\subsection{Mid-IR spectral resolution}

The spectral resolution of $R\!\sim\!30\,000$ is crucial to resolve line profiles from the inner warm disk and to separate the peaks of the Keplerian profiles (see Sect.~\ref{app:snowlinesurvey}). Increasing the resolution further can help disentangling more complex lines profiles (more than two peaks), but the true gain arises once line profiles can fully resolve disk wind kinematic features. Fig.~\ref{fig:visir_ne} shows that the spectral resolution of $R\!\sim\!30\,000$ is enough to detect a shift of the wind profile with respect to the disk rest frame. However, a full study of the line asymmetries, including accurate peak determination, arising from various disk dispersal processes warrants a resolution of $R\!\sim\!100\,000$, or $\sim\!3$\,km/s, such as available on the ground with VLT CRIRES. 

\subsection{Far-IR spectral resolution}

The limited SAFARI spectral resolution will make it difficult to kinematically separate the individual components contributing to the emission lines, and thus to isolate the emission within the disk from that in the disk wind and outflow. In practice, it will only be possible to distinguish atomic emission in fast jets, i.e.\ $v_{\rm jet}\!>50$\,km/s, contributing to the [O\,I] line at 63\,$\mu$m, where the resolution is $R\!>\!5000$. To properly discriminate between the various components potentially contributing to the far-IR lines (disk, wind), we require a spectral resolution of $30\,000$, close to that which SMI offers at shorter wavelengths.

\subsection{Wavelength coverage}

The only water line excited by cold gas in the SAFARI wavelength range is the 179.5\,$\mu$m ortho-H$_2$O line ($T_{\rm up}\!=\!114.4$\,K). However, going to slightly longer wavelengths than presently envisioned for SAFARI, the 269.27\,$\mu$m para-water line ($T_{\rm up}\!=\!53.43$\,K), one of the three ground state lines of water could be included. Furthermore, due to this increase in spectral coverage, the gap in the CO ladder would close with respect to the ground-based sub-millimetre facilities. This would be beneficial for measuring the radial temperature gradients in disks, especially to gain accuracy at the spatial scales of a few to a few 10 au.

\subsection{Mirror size}

Finally, a larger primary mirror would undoubtedly increase the sensitivity and diminish confusion at far-IR wavelengths. This increase in angular resolution is very interesting for detecting a true Kuiper Belt analogue (Sect.~\ref{sec:kuiperbeltanalog}). The higher the spatial resolution, the larger the 'volume' and hence potential sample size available for searching for true Kuiper Belt analogues (Fig.~\ref{fig:kb_sample}). The most stringent constraint here is in fact the level of PSF stability for far-IR imaging that is required to detect the faint disk emission with sufficient confidence (see Sect.~\ref{sec:kuiperbeltanalog}).

\section{CONCLUSIONS}

This white paper describes the unique science questions that a spectroscopic infrared space mission such as {SPICA} can answer. Mid- to far-IR infrared spectra will uniquely measure the gas masses of disks (HD, CO ladder) and the water/ice content and determine how these evolve during the planet forming period. The key measurements of infrared lines (e.g.\ HD, OH, water, mid- to far-IR fine structure lines) will pinpoint the crucial transition when disks exhaust their primordial gas and further planet formation may use secondary gas produced from planetesimals. The mid-IR high spectral resolution is also unique in determining indirectly from the water line profiles the location of the snowline dividing the rocky and icy mass reservoir in the disk and how this evolves during the build-up of planetary systems. Infrared spectroscopy (mid- to far-IR) of key solid state bands (e.g.\ forsterite, enstatite, hydrosilicates, calcite) is crucial is assessing whether extensive radial mixing, which is part of our Solar System history, is a general process occurring in all planetary systems. The combined gas, dust and ice observations of debris disks will assess whether extrasolar planetesimals are similar to our Solar System comets/asteroids.

With {SPICA} being taken out of the M5 competition, there is no other facility within the next 20 years that will be able to measure these fundamental quantities in statistically relevant samples of planet forming disks. Among the facilities that the infrared community is constantly pushing forward are the Origins Space Telescope\footnote{https://origins.ipac.caltech.edu} proposed as part of the US Decadal Survey and a potential new instrument on the Stratospheric Observatory for Infrared Astronomy (SOFIA) filling the niche of medium- to high-resolution spectroscopy in the 30-120\,$\mu$m wavelength range. This latter instrument would have less sensitivity compared to {SPICA}, but could serve as a key pathfinder for the brighter objects.

\begin{acknowledgements}
I.K.\ acknowledges funding from 'Grants-in-Aid for Scientific Research 25108004' for a work visit to Japan for discussing the {SPICA} project and relevant science. M.H. was supported by JSPS KAKENHI Grant Numbers JP17H01103, JP18H05441.
M.A.\ acknowledges support from the Prodex Experiment Arrangement C4000128332 for the Swiss contribution to {SPICA}. 
S.N.\ is grateful for support from JSPS (Japan Society for the Promotion of Science) Overseas Research Fellowships when he belonged to Leiden University, and RIKEN Special Postdoctoral Research Fellowship, and he is supported by MEXT/JSPS Grants-in-Aid for Scientific Research (KAKENHI) 16J06887 and 20K22376. W.R.M.R. acknowledges the support by the European Research Council (ERC) under the European Union's Horizon 2020 research and innovation program through ERC Consolidator Grant ``S4F'' (grant agreement No~646908). B.N. and D.F. acknowledge financial support by the Agenzia Spaziale Italiana (ASI) under the research contract 2018-31-HH.0.
D.J is supported by the National Research Council of Canada and by an NSERC Discovery Grant. SPL acknowledges grants from the Ministry of
Science and Technology of Taiwan 106-2119-M-007-021-MY3 and 109-2112-M-007-010-MY3.
\end{acknowledgements}

\section*{affiliations}

{\footnotesize
{$^8$Department of Astronomy, University of Tokyo, 113-0033, Tokyo, Japan}\\
{$^9$Department of Physics, Texas State University, 749 N Comanche Street, San Marcos, TX 78666, USA}\\
{$^{10}$School of Physics and Astronomy, Cardiff University, Queens Buildings, The Parade, Cardiff CF24 3AA, UK}\\
{$^{12}$Anton Pannekoek Institute for Astronomy, University of Amsterdam, Science Park 904, 1098XH Amsterdam, The Netherlands}\\
{$^{14}$Universit{\'e} Paris-Saclay, CNRS, Institut d'Astrophysique Spatiale, 91405, Orsay, France}\\
{$^{15}$Institute of Space and Astronautical Science, Japan Aerospace Exploration Agency, 3-1-1 Yoshinodai, Chuo-ku, Sagamihara, 252-5210 Kanagawa, Japan}\\
{$^{16}$NRC Herzberg Astronomy and Astrophysics, 5071 West Saanich Rd, Victoria, BC, V9E 2E7, Canada}\\
{$^{17}$Department of Physics, University of Warwick, Gibbet Hill Road, Coventry, CV4 7AL, UK}\\
{$^{18}$Institute of Theoretical Physics and Astrophysics, University of Kiel, Leibnizstra\ss e 15, 24118 Kiel, Germany}\\
{$^{19}$LESIA, Observatoire de Paris, Universit{\'e} PSL, CNRS, Sorbonne Universit{\'e}, Univ. Paris Diderot,\\ Sorbonne Paris Cit{\'e}, 5 place Jules Janssen, 92195 Meudon, France}\\
{$^{20}$Institute of Astronomy and Department of Physics, National Tsing Hua University, Hsinchu 30013, Taiwan}
{$^{21}$Stockholm University, AlbaNova University Center, Department of Astronomy, SE-106 91 Stockholm, Sweden}\\
{$^{22}$Leiden Observatory, Leiden University, PO Box 9513, 2300 RA Leiden, The Netherlands}\\
{$^{23}$ESO, Garching, Germany}\\
{$^{24}$College of Science, Ibaraki University, Bunkyo 2-1-1, Mito, Ibaraki 310-8512, Japan}\\
{$^{25}$Institute for Space Imaging Science, Department of Physics and Astronomy, University of Lethbridge, Alberta, T1K 3M4, Canada}\\
{$^{26}$INAF, Osservatorio Astronomico di Roma, Via di Frascati 33, 00078 Monte Porzio Catone (RM)}\\
{$^{27}$Star and Planet Formation Laboratory, RIKEN Cluster for Pioneering Research, 2-1 Hirosawa, Wako, Saitama 351-0198, Japan}\\
{$^{28}$Department of Physics, Faculty of Science and Engineering, Meisei University, 2-1-1 Hodokubo, Hino, Tokyo 191-8506, Japan}\\
{$^{29}$CEA, Centre d'Etudes de Saclay, France}\\
{$^{30}$Observatorio Astron\'omico Nacional (OAN,IGN), Calle Alfonso XII, 3, 28014 Madrid, Spain}\\
{$^{31}$Niels Bohr Institute \& Centre for Star and Planet Formation, University of Copenhagen, {\O}ster Voldgade 5-7, DK-1350 Copenhagen K., Denmark}\\
{$^{32}$SRON Netherlands Institute for Space Research, Postbus 800, 9700 AV Groningen, The Netherlands}\\
{$^{35}$Institute of Astronomy and Astrophysics, Academia Sinica, 11F of Astronomy-Mathematics Building, No.1, Sec. 4, Roosevelt Rd, Taipei 10617, Taiwan}\\
{$^{37}$Institute of Astronomy, University of Cambridge, Madingley Road, Cambridge CB3 0HA, UK}\\
{$^{38}$Space Research Institute of the Austrian Academy of Sciences, Schmiedlstra\ss e 6, 8042 Graz, Austria}
}
\vspace*{1cm}

\begin{appendix}

\section{REFERENCE OBSERVING PROGRAM}

The following subsections describe a potential observing program that would provide the data to answer the research questions raised in the main paper. The source lists have been compiled using the current knowledge and only serve to demonstrate that (a) statistical samples can be compiled and (b) the required observing time remains feasible within a typical mission lifetime of a few years. The disk gas mass case (HD) has been identified as leading and thus the HD survey is largely driving the sample selection also for the mineralogy and snowline case.

We report here observing times without overheads that have been calculated using the instrument simulators of the SAFARI (version 1.2, issue 6.2) and SMI instruments (version 4). Based on the Herschel experience, we could expect an efficiency of 60\,\% and typical slew times of 3\,minutes. We also assume that the calibration time on average amounts to 10\,\% of the observation time.

\begin{table*}[!t]
    \caption{Properties of the star forming regions to be targeted. We propose to select the best 100 targets in each of the five age bins based on the availability of ancillary data and characterization in the next decade.}
    \centering
    \vspace*{2mm}
    \begin{tabular}{lllll}
    \hline\hline
        Region & Age & Distance & $\#$ sources & Ref. for membership\\
               & [Myr] & [pc] &    \\
    \hline\\[-3mm]
    \multicolumn{5}{c}{$0.1-1$\,Myr} \\
    \hline\\[-3mm]
        Perseus & $0.1-1$ & $275-300$ & 42 &  Young et al. (2015) \\
        Ophiuchus & $0.5-1$ & 125 & 66 &  Padgett et al. (2008);\\ 
        & & & & Rebollido et al. (2015) \\
        \hline\\[-3mm]
            \multicolumn{5}{c}{$1-2$\,Myr} \\
        \hline\\[-3mm]
        Taurus-Auriga & $<\!1-2$ & 165 & 198 & Howard et al. (2013);\\
         & & & & Rebull et al. (2010) \\
        Lupus & 2 & 160 & 123 & Mer\'in et al. (2008);\\
         & & & & Bustamente et al. (2015) \\
        \hline\\[-3mm]
            \multicolumn{5}{c}{$2-3$\,Myr} \\
        \hline\\[-3mm]
        Cha & $2-3$ & 190 & 145 & Ribas et al. (2013);\\
         & & & & Luhman et al. (2008);\\
         &  &   &  & Young et al. (2005) \\
        \hline\\[-3mm]
            \multicolumn{5}{c}{$>\!3$\,Myr} \\
        \hline\\[-3mm]
        Upper Sco & $5-10$ & 145 & 106 & Mathews et al. (2013);\\
         & & & & Bahrenfeld et al. (2016) \\
        $\eta$ Cha & $5-9$  & 100 & 14  & Riviere et al. (2015) \\
        Moving Groups & $5-40$  & $20-50$ & 53  & Rebull et al. (2008);\\
         & & & & Donaldson et al. (2012) \\
         &  & &  & Riviere et al. (2013) \\
        \hline\hline
    \end{tabular}
    \label{tab:regions}
\end{table*}

\subsection{HD Survey}
\label{app:HDsurvey}

Experience from past Spitzer and Herschel surveys \citep[e.g.][]{Evans2009, sturm20102010A&A, Dent2013} shows that statistically relevant sample sizes consist of at least 100 sources per age bin (star forming region). The reason behind this is that even at a specific age/within a SFR, there is a large scatter in terms of stellar (e.g., spectral type, mass accretion rate, X-ray luminosity) and disk (e.g., mass, gas-to-dust mass ratio, temperature, surface density) properties. 

In order to address the disk gas mass evolution as function of time, we need to observe SFRs with a wide spread of ages and well characterized YSO populations (Spitzer and Herschel data) covering class\,I to class\,III sources. These are shown in Table~\ref{tab:regions} with their ages, distances, sample sizes and respective references.

We group them in four age bins: $0.1-1$\,Myr, $1-2$\,Myr, $2-3$\,Myr and older than 3\,Myr. These regions will be surveyed in addition by ALMA and JWST in the next decade. ALMA will provide a full census of disk sizes (gas and dust), inclinations and for a few cases in depth characterization of non-axisymmetric structures (and protoplanets); JWST will provide an inventory of water, ices and organic molecules inside $\sim\!10$\,au. 

The final selection of appropriate SFRs should be driven by the largest ancillary datasets (Spitzer, WISE, Herschel, ALMA and JWST data) and the need for sample diversity. Our strategy here is to cover the entirety of these SFRs and Moving Groups with sample sizes of 100 objects per age bin selected across a stellar mass range. We aim, for 10 bins in stellar mass (from M-dwarfs to F-type stars, logarithmically spaced) and 10 objects per mass bin to cover a representative range of mass accretion rates). This provides the basis for determining the gas dispersal timescale. Within a star forming region, we probe different evolutionary stages (class\,I, II, III, transitional disks, disks with substructure from SPHERE, ALMA), mass accretion rates and stellar spectral types; however, the main focus will be on the class\,II disks. Within the selected SFRs, we can show which secondary factors such as spectral type, disk size, mass accretion rates play a role next to SED class and age in setting the gas evolution.

We will select $\sim\!20$ class\,I and II sources in Perseus to capture the earliest stages. From each age bin (see Table~\ref{tab:regions}), we eventually select a total of 100 targets, leading to 400 targets for the reference sample. We note that this is the minimum required to address the gas mass evolution during the planet formation epoch. We expect many additional open time programs to build and expand on this. Examples could be to extend this survey to higher stellar masses (intermediate mass T\,Tau stars and Herbig Ae's), to add very deep surveys at the low mass end (brown dwarfs), or to add more distant high mass star forming regions such as Orion.

The reference program thus contains $400$ objects (out of the ones listed in Table~\ref{tab:regions}) to be targeted in order to answer the key science questions: (1) How does the disk gas mass evolve through the planet forming epoch and (2) Which other factors besides age play a role in this evolution.

\begin{figure}[t]
    \centering
    \includegraphics[width=8cm]{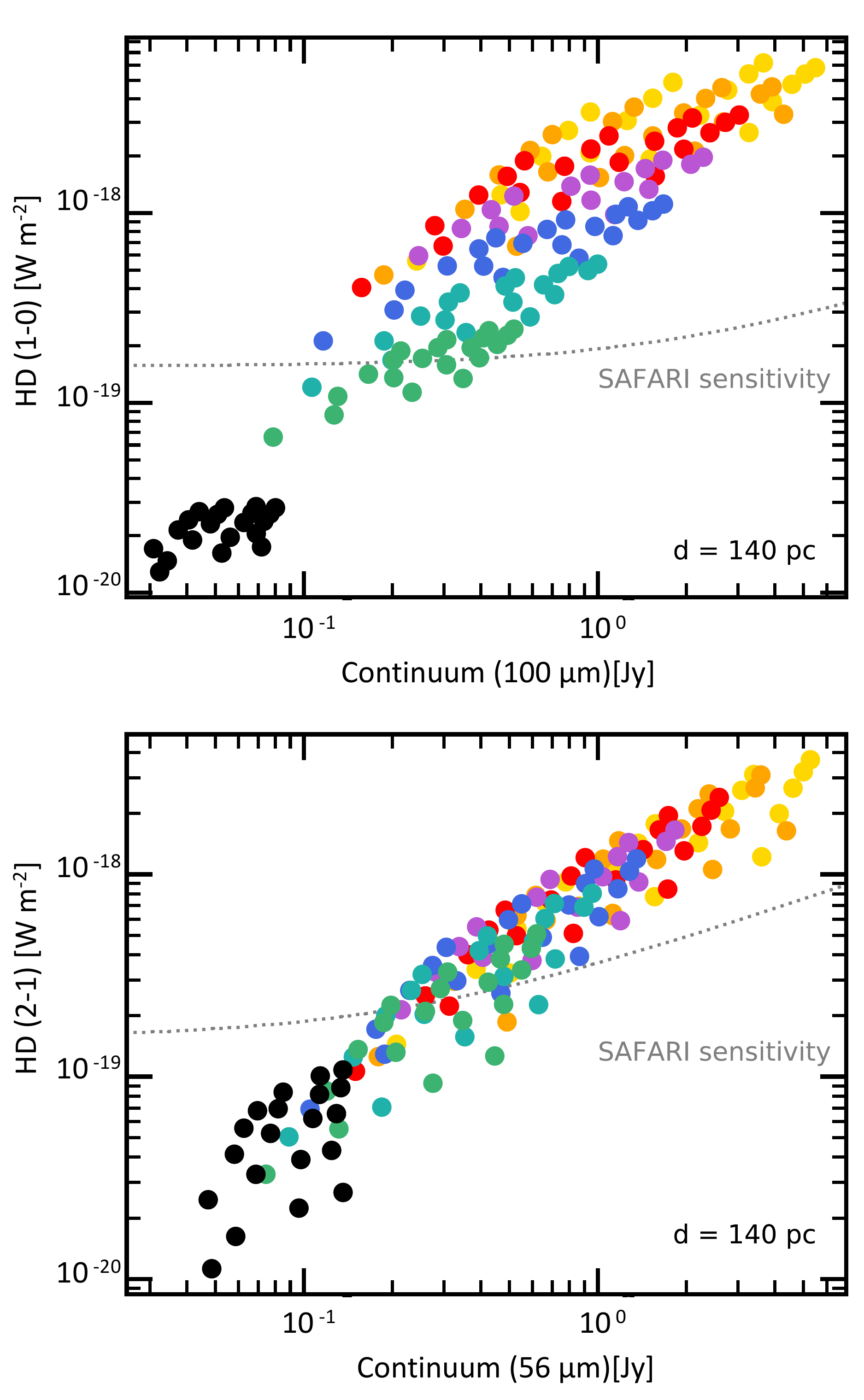}    
    \vspace*{-3mm}
    \caption{HD line fluxes versus continuum for the grid of models taken from \citet{Trapman2017} assuming a distance of 140\,pc. The SAFARI sensitivity curve is displayed as a dashed line for a 1\,h integration. The color scale is the same as in Fig.~\ref{fig:HD-to-diskmass}.}
    \label{fig:HDline-cont}
\end{figure}

The goal is line detection; the lines do not need to be spectrally resolved in the SAFARI wavelength range. However, since molecular lines have a very low line-to-continuum ratio ($\lesssim\!0.005$), i.e.\ weak narrow gas lines (unresolved) on a strong continuum, we require the highest possible spectral resolution to maximize the line detection rate. The low excitation HD lines ($J\!=\!1-0, 2-1$ are expected to have typical widths of $5-20$\,km/s (depending also on inclination). In the SAFARI wavelength range, a resolving power of $R\!\gtrsim\!3000$ is needed. At a resolution of 2800, the HD lines are not blended and line fluxes can be extracted in a straightforward manner. Another key transition is the ortho-H$_2$O $2_{12}-1_{01}$ line at 179.533\,$\mu$m that is very close to the CH$^+$ $J\!=\!2-1$ transition at 179.601\,$\mu$m. In order to deblend the two lines, a resolving power of $R\!\gtrsim\!2800$ ($\Delta \lambda \lesssim\!0.036\,\mu$m) is needed. A smaller resolving power would not permit to deblend these two lines and to measure the flux of the H$_2$O $2_{12}-1_{01}$. This line is very important because the 179.533\,$\mu$m line is the only transition (in the SAFARI spectral range) that traces the cold water gas reservoir. 

At a distance of 140\,pc, the expected disk integrated flux of the HD $J\!=\!1-0$ line ranges from $\sim\!1\times 10^{-19}\,$W\,m$^{-2}$ for $M_{\rm disk}\!\sim\!2\times 10^{-4}\,$M$_{\odot}$ to $\sim\!1\times 10^{-18}\,$W\,m$^{-2}$ for $M_{\rm disk}\!=\!10^{-2}\,$M$_{\odot}$ \citep{Trapman2017}. The {\it SPICA} detection limit of $\sim\!1.3\times 10^{-19}\,$W\,m$^{-2}$ (1\,h, 5\,$\sigma$, point-source mode), therefore, will allow us to study the evolution of disks at least down to disk gas masses of $\sim\!3\times 10^{-4}\,$M$_{\odot}$. Disk models and observations indicate a clear correlation between line and continuum fluxes (Fig.~\ref{fig:HDline-cont}), so that we could expect for stronger continuum sources also on average higher line fluxes. This enables an efficient observing strategy, where the limiting sensitivity for HD is informed by the predicted continuum (from Spitzer, Herschel, ALMA data) at 60 and 100\,$\mu$m. 

Also, for the more distant star forming region, Perseus (factor 2 more distant compared to above estimates), we require a factor 16 longer integration times to reach the $M_{\rm disk}\!\sim\!2 \times 10^{-4}\,$M$_{\odot}$ mass limit. However, none of the class\,II targets in that region will be brighter than 0.1\,Jy and thus we exploit the full SAFARI sensitivity.  

Based on the strategy outlined above and a 'current best' sub-selection of targets from Table~\ref{tab:regions}, a deep survey for the gas mass estimates will take a total observing time of 928\,h (38.7\,d) (integration time of 612.9\,h or 25.5\,d). In this subselection (see Table~\ref{tab:HD-classes}), class\,II disks dominate (244), but we retain a statistically significant sample of class\,I sources (61) for tracking differences in SED evolutionary stage and also a significant sample of class\,III sources (97) for assessing the transition from primordial to secondary gas, a crucial point in the mode of planet formation; this class\,III sample also constitutes potentially a fraction of 'young' debris disks for which we can study the detailed gas composition (see Sect.\ref{sec:debrisdisks}).

\begin{table}[h]
\caption{Distribution of classes within the subselected sample for the HD mass survey.}
    \centering
    \vspace*{2mm}
\begin{tabular}{lrrrr}
\hline\hline
Region & class\,I & class\,II & class\,III & total \\
\hline
 \multicolumn{5}{c}{$0.1-1$\,Myr} \\
\hline
Perseus       & 29  & 13 &  & \\
Ophiuchus     & 14  & 50 & 2 & \\
              &     &    &   & 108 \\
\hline
 \multicolumn{5}{c}{$1-2$\,Myr} \\
\hline
Taurus-Auriga & 13  & 65 & 22 & \\
              &     &    &   & 100 \\
\hline
 \multicolumn{5}{c}{$2-3$\,Myr} \\
\hline
Cha           & 5  & 75 & 20 & \\
              &     &    &   & 100 \\
\hline
 \multicolumn{5}{c}{$>\!3$\,Myr} \\
\hline
Upper Sco     & 0  & 33 & 12 & \\
$\eta$ Cha    &  0 & 5 & 9 & \\
MGs           &  0 & 3 & 33 & \\
              &     &    &   & 95\\
\hline\hline
\end{tabular}
\label{tab:HD-classes}
\end{table}

\begin{table*}[t]
 \caption{Mineral resonances with wavelengths positions above 20\,$\mu$m}
 \centering
    \begin{tabular}{llc}\hline\hline
    Mineral        & composition & Peak Position ($\mu$m) \\ \hline
    Forsterite     & Mg$_2$SiO$_4$  &  23.5, 33.5, 49.3, 68.9\\ 
    Enstatite (Clino)  & MgSiO$_3$  &  43, 65.9 \\ 
    Enstatite (Ortho)  & MgSiO$_3$  &  43, 49.1, 51.5, 68.5, 72.5 \\
    Calcite        & CaCO$_3$       &  73, 90  \\    
    Dolomite       & CaMg(CO$_3$)$_2$ &  51, 62.5  \\
    Montmorillinite & $\rm(Na,Ca)_{0.33}(Al,Mg)_2Si_4O_{10}(OH)_2 \cdot nH_2O$ & 29, 49 \\
    Serpentine     & $\rm(Mg,Fe)_3Si_2O_5(OH)_4$  & 28, 33, 43, 77  \\ \hline\hline
    \end{tabular}
    \label{tab:mineralfeaturelist}
\end{table*}

\subsection{Mineralogy/Ice Survey}
\label{app:mineralogysurvey}

The mid- and far-IR spectral region contains many bands of abundant minerals and of water ice (see Fig.~\ref{fig:dustopa}). Table~\ref{tab:mineralfeaturelist} shows the key mid-/far-IR bands covered by SAFARI and SMI. In order to get the full inventory also the mid-IR bands (e.g.\ forsterite at 23.5 and 33.5\,$\mu$m) should be observed with SMI.  

A Herschel far-IR spectroscopic study was done with 32 Herbig Ae/Be stars, which is a too small sample to disentangle multiple parameters which could drive changes in the mineralogy. We thus want to increase the number of targets to $\sim\!400$ to enable a comprehensive statistical study of the evolutionary trend with {\it SPICA}. The source sample will be shared for large parts with the HD survey which acquires SAFARI/HR spectroscopy. In addition, $\sim\!100$ debris disk sources should also be observed to see the continuous evolutionary trend from pre-main-sequence stars to main sequence stars. This latter sample is shared with the debris disk gas/ice surveys.

Most bands are a few $\mu$m to several tens of $\mu$m wide. However the Forsterite 69\,$\mu$m band shape and peak position carry valuable diagnostic value of Fe content in the olivine structures. For example, 1\,\% inclusion will result in $\sim\!0.3\,\mu$m peak shift to longer wavelengths. To distinguish this shift, it requires a spectral resolution $R\!>\!230$. A spectral resolution of $>\!230$ is also important because in some spectral ranges vibrational resonances of dust species may blend, such as the enstatite complex near 40-43\,$\mu$m, or the serpentine band near 21\,$\mu$m that may blend with crystalline forsterite and amorphous silicate (SMI wavelength range). In addition it is important to have a reliable calibration of the relative spectral response function over a wavelength range allowing the detection of broad (5-10\,$\mu$m) and weak (1-2\,\% above the local continuum) bands. An additional requirement is the calibration of SMI with respect to SAFARI, because dust species have vibrational resonances covering the wavelength ranges of SMI and SAFARI. The relative strengths of the resonances carry information about e.g.\ the temperature distribution, chemical gradients and abundance gradients. Band strength ratios should be determined with an accuracy of 5\,\%. 

This survey will provide also the data to study water ice in disks through the two prominent features at 45\,$\mu$m and 60\,$\mu$m. Observations of both are important as the relative flux of the two gives insights on the formation temperature of the ice. Moreover, the 45\,$\mu$m feature is sensitive to the structure of the ice (Fig.~\ref{fig:fig_5.4}). It is very pronounced and narrow in crystalline ices, with an additional band at 60\,$\mu$m. On the other hand, amorphous water ice has a broad structure around 45\,$\mu$m and no 60\,$\mu$m band. The ice features are rather broad, in particular the 60\,$\mu$m one. Based on our experience with PACS, the baseline stability is very important. The poor baseline stability of PACS (largely due to pointing problems of the telescope) hampered the detection of water ice. To be able to detect the ice emission and to determine the shape of the band a relative flux calibration (i.e., between consecutive spectral bins) of 1\,\% (or better) is needed. Another strong requirement is the simultaneous coverage of the two water ice peaks from (at least) $40-80\,\mu$m and to avoid overlap of orders at the positions of the ice features.

An abundance of a few per cent of crystalline forsterite in the outer disk regions is expected on the basis the detection of the 69\,$\mu$m band in Herbig Ae/Be disks and in $\beta$~Pictoris with Herschel. Model simulations show that a T\,Tauri disk at 150\,pc with a continuum flux of 1\,Jy and a constant forsterite abundance of 5\,\% shows a 69\,$\mu$m forsterite band strength of 10\,mJy above the local continuum (1\,\%). To detect this weak 69\,$\mu$m feature with at least 3\,$\sigma$, we need at least a S/N ratio of 300 or more at 69\,$\mu$m for the total continuum spectra. Thanks to the high sensitivity of SAFARI, this will be achieved in relatively short observing time (less than 1 h for most sources).
Based on the HD survey sample, we estimate that we can make SAFARI/LR spectroscopic observations of 324 targets in 94\,h ($\sim\!3.9$\,d).

\begin{figure*}[th]
    \centering
    \includegraphics[width=7cm]{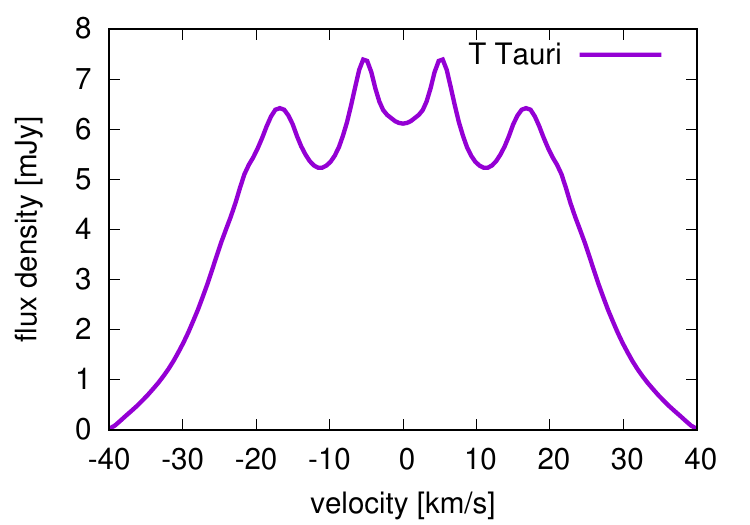}
    \includegraphics[width=7cm]{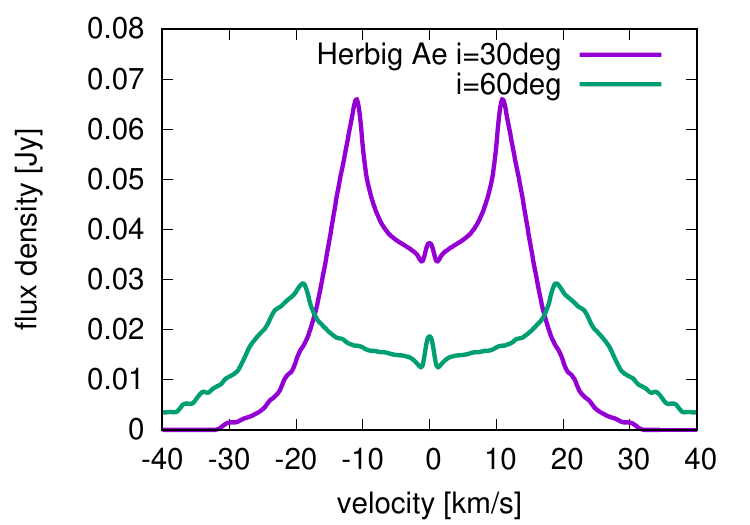}
    \caption{The 17.75\,$\mu$m water line flux densities obtained by model calculations of (left) a T\,Tauri disk located at 300\,pc and (right) a Herbig Ae disk located at 420\,pc. The dust continuum is subtracted. References for the T\,Tauri models: \citet{Antonellini2015,Antonellini2016}, for the Herbig Ae models: \citet{Notsu2016,Notsu2017,Notsu2018}}
    \label{fig:18um_waterlineflux}
\end{figure*}

\begin{figure*}[h]
    \centering
    \includegraphics[width=5.7cm]{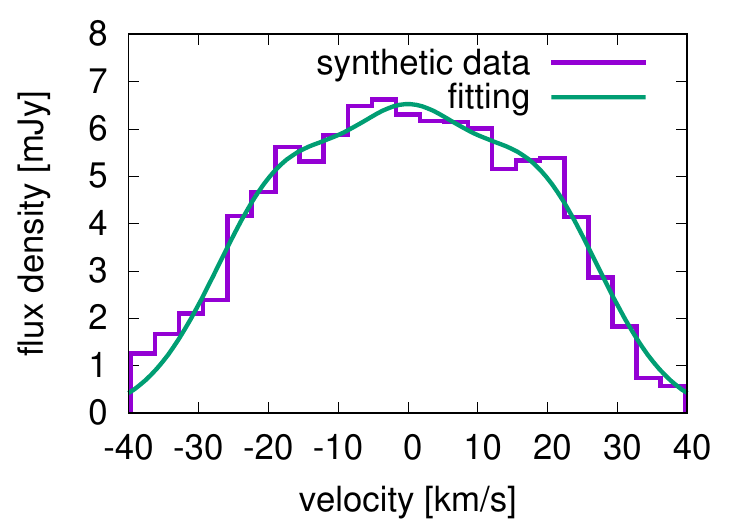}
    \includegraphics[width=5.7cm]{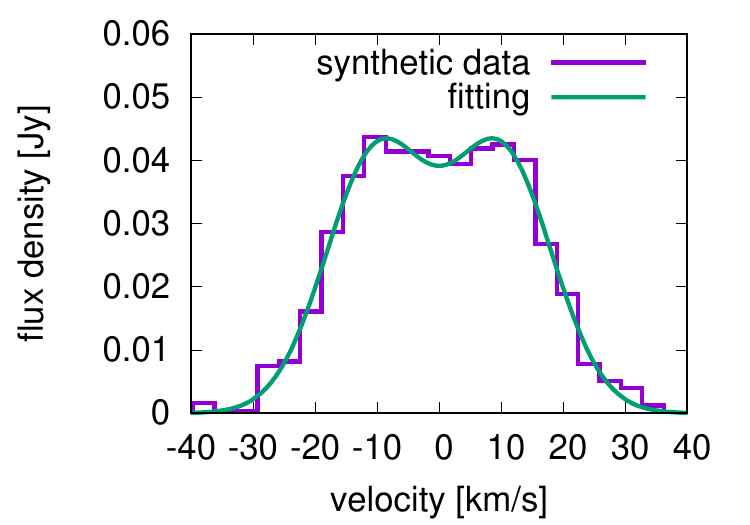}
    \includegraphics[width=5.7cm]{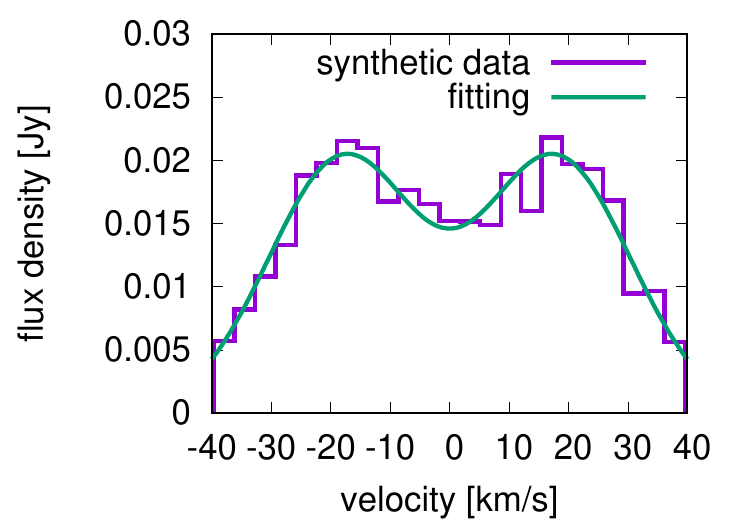}
    \caption{The 17.75\,$\mu$m water line profiles deconvolved with spectral resolution of $R\!=\!29,000$ and sampling of $N\!=\!1.5$ for a T\,Tauri disk located at 300\,pc (left) and a Herbig Ae disk located at 420\,pc with inclination angle of 30 degree (center) and 60 degree (right). Noise levels of 0.28\,mJy and 1.4\,mJy are used for a T\,Tauri disk model and Herbig Ae disk model, respectively. The synthetic data can be fitted with a double peak profile. The spectral resolution of $R\!\sim\!30,000$ is required in order to resolve the position of double peaks.
    }
    \label{fig:18um_waterlineresolution}
\end{figure*}

In addition, SMI spectroscopy is required to cover the warmer (inner disk) dust counterpart and address the mixing between inner and outer disk; depending on the JWST data collection, this part of the survey can be adjusted to ensure that we have the full short wavelength counterpart of our SAFARI sample. Based on the HD survey sample and assuming a priori no JWST data, we estimate that we can make SMI/LR spectroscopic observations of 403 targets in 51\,h ($\sim\!2.1$\,d).

Furthermore, to see the evolutionary trend, we will cover $\sim\!140$ isolated older debris disks. Based on the debris disk survey sample selected for the ice studies (sample A, B, C), we estimated that we can obtain SAFARI/LR spectroscopic observations of 140 targets with 94\,h ($\sim\!3.9$\,d).

In total, this results in a program of 239\,h ($\sim\!10$\,d) observing time is estimated for this program.

\subsection{Snowline Survey}
\label{app:snowlinesurvey}

The source sample for this survey again largely overlaps with the HD survey in nearby star forming regions ($\leq\!300$\,pc, Appendix~\ref{app:HDsurvey}), which includes mostly T\,Tauri disks. In addition, since the 17.75\,$\mu$m water line is expected to be relatively strong for earlier type stars, we include a homogeneous (coeval) sample of Herbig Ae/Be disks in the Orion A/B giant molecular clouds at the distance of $\sim\!420$\,pc. 

In order to assess the evolution and physics of the water snowline, we cover parameters such as ages, gas accretion rates (e.g., H\,$\alpha$ line equivalent widths), grain size evolution (e.g., SED slopes of dust emission), and gas-to-dust mass ratios. To reach significant statistics in each of these parameters, we will select $\sim\!300$ targets from nearby young and intermediate-aged star clusters, as well as $\sim\!200$ Herbig Ae/Be disks from the Orion A/B star forming region. The characterization of substructures (such as inner holes/gaps/spirals), obtained from submillimeter observations with ALMA and infrared observations with 8-10\,m class ground-based telescopes, will provide additional information for the evolution of the sample.

Inclination angles will be known from other molecular line observations such as CO submillimeter emission lines by ALMA. In addition, even though JWST cannot spectrally resolve the mid-IR lines ($R\lesssim\!3\,000$), it will provide total line fluxes for the key water line at 17.75\,$\mu$m; this will be especially useful for the final target selection. 

Our model calculations show that the line strength of the 17.75\,$\mu$m water emission from inside the snowline of a fiducial T\,Tauri disk located at 300\,pc (at Serpens) is $\sim\!1\times 10^{-19}$\,W\,m$^{-2}$ and that of a fiducial Herbig Ae disk located at 420\,pc (at Orion A/B) is $\sim\!1\times 10^{-18}$\,W\,m$^{-2}$ after subtracting dust continuum emission (Fig.~\ref{fig:18um_waterlineflux}). The flux and flux density sensitivities of SMI/HR are $\sim\!1\times 10^{-20}$\,W\,m$^{-2}$ and $\sim\!1.4$\,mJy, respectively, for 5\,$\sigma$ in one hour integration time. The synthetic analysis of the line profiles with spectral resolution of $R\!=\!29,000$ and sampling of $N\!=\!1.5$ shows that the noise level is low enough so that the synthetic data can be fitted with a double peak profile within one hour integration time for
a T\,Tauri disk located at 300\,pc (at Serpens) and within ten minutes integration time for a Herbig Ae disk located at 420\,pc (at Orion A/B) (Fig.~\ref{fig:18um_waterlineresolution}). Thus, in order to recover the water line profile (total line flux of $8.5\times 10^{-20}$\,W\,m$^{-2}$ at 300\,pc, FWHM$\sim\!40-60$\,km/s, T\,Tauri disk), we need a noise level of 0.28\,mJy for a continuum of 10\,mJy. We then scale the noise level with the continuum using the 12\,$\mu$m fluxes from WISE.

\begin{figure}
    \centering
    \includegraphics[width=6.2cm]{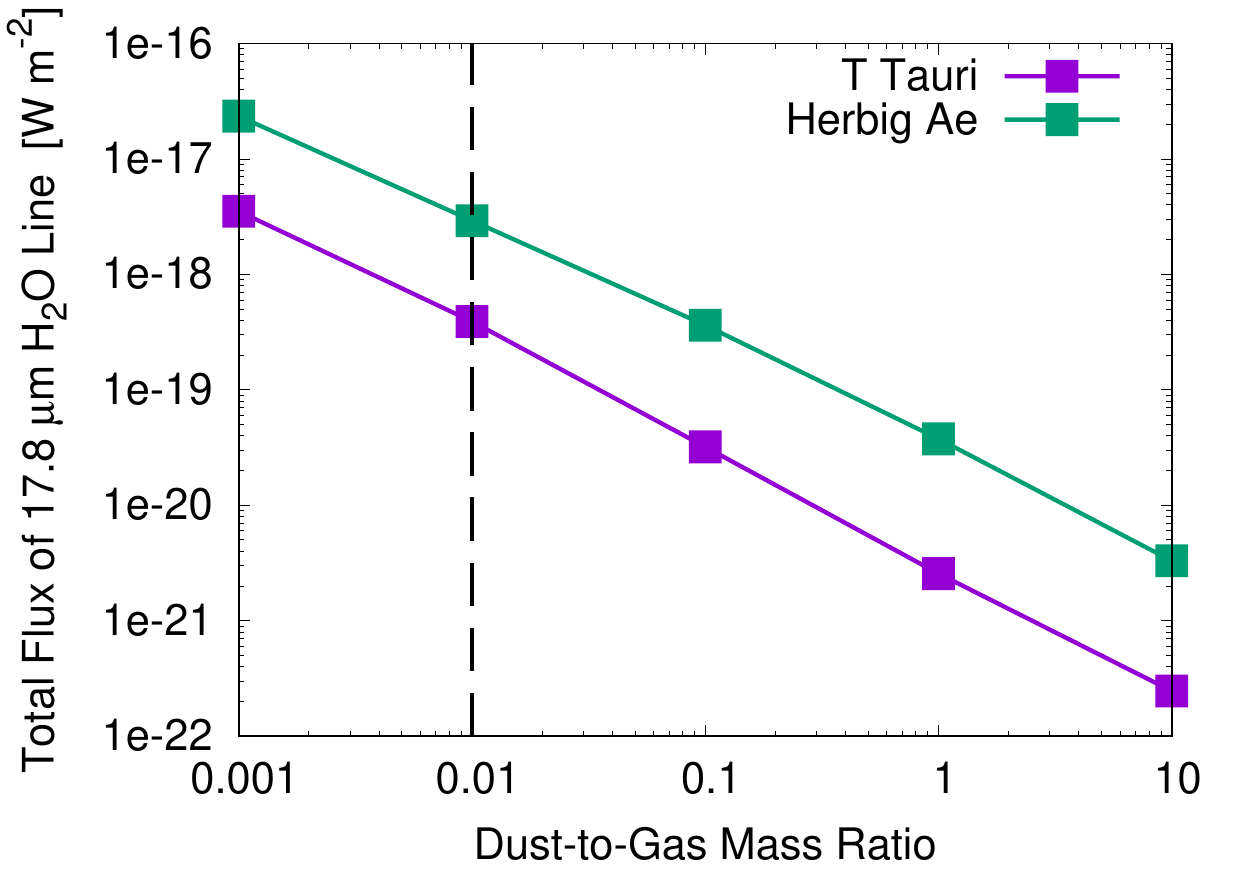}
    \includegraphics[width=6.2cm]{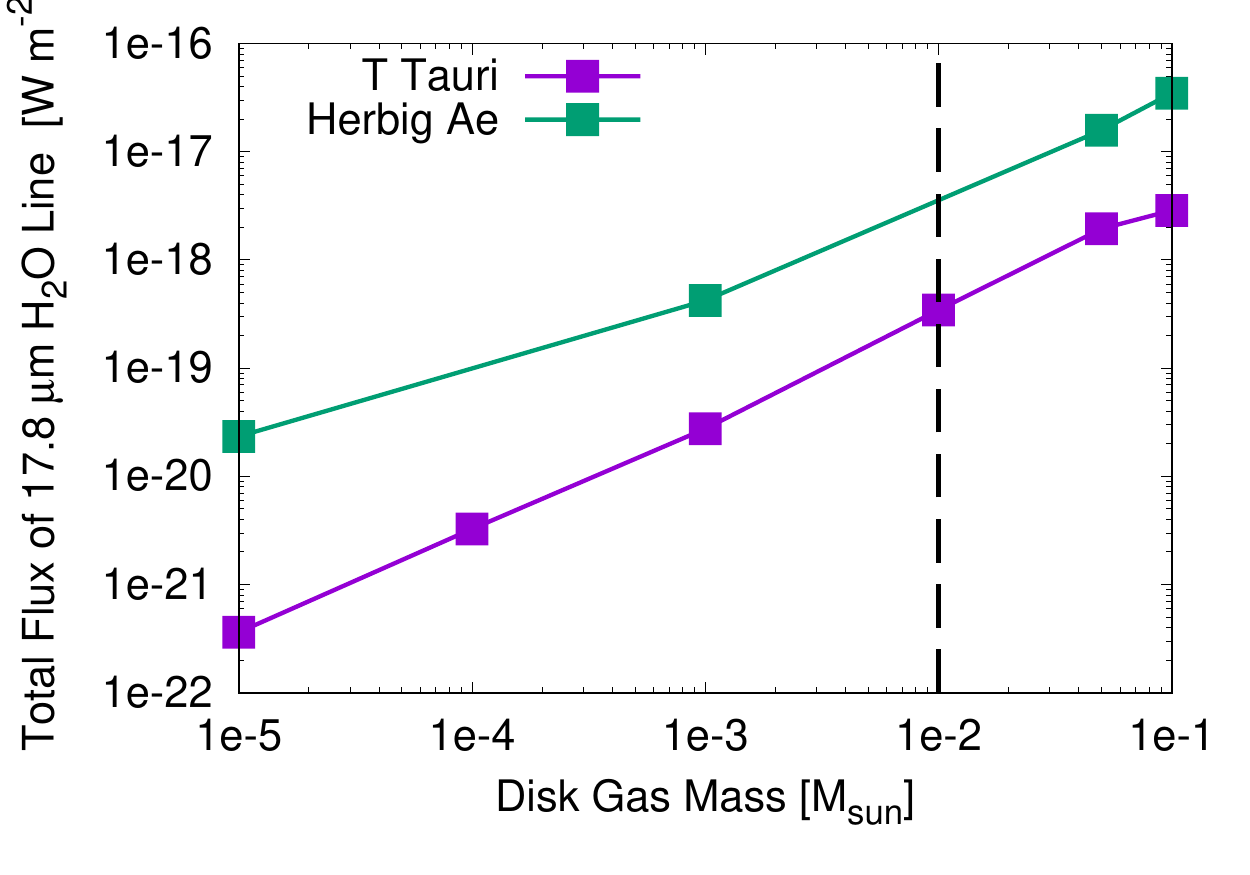}
    \includegraphics[width=6.2cm]{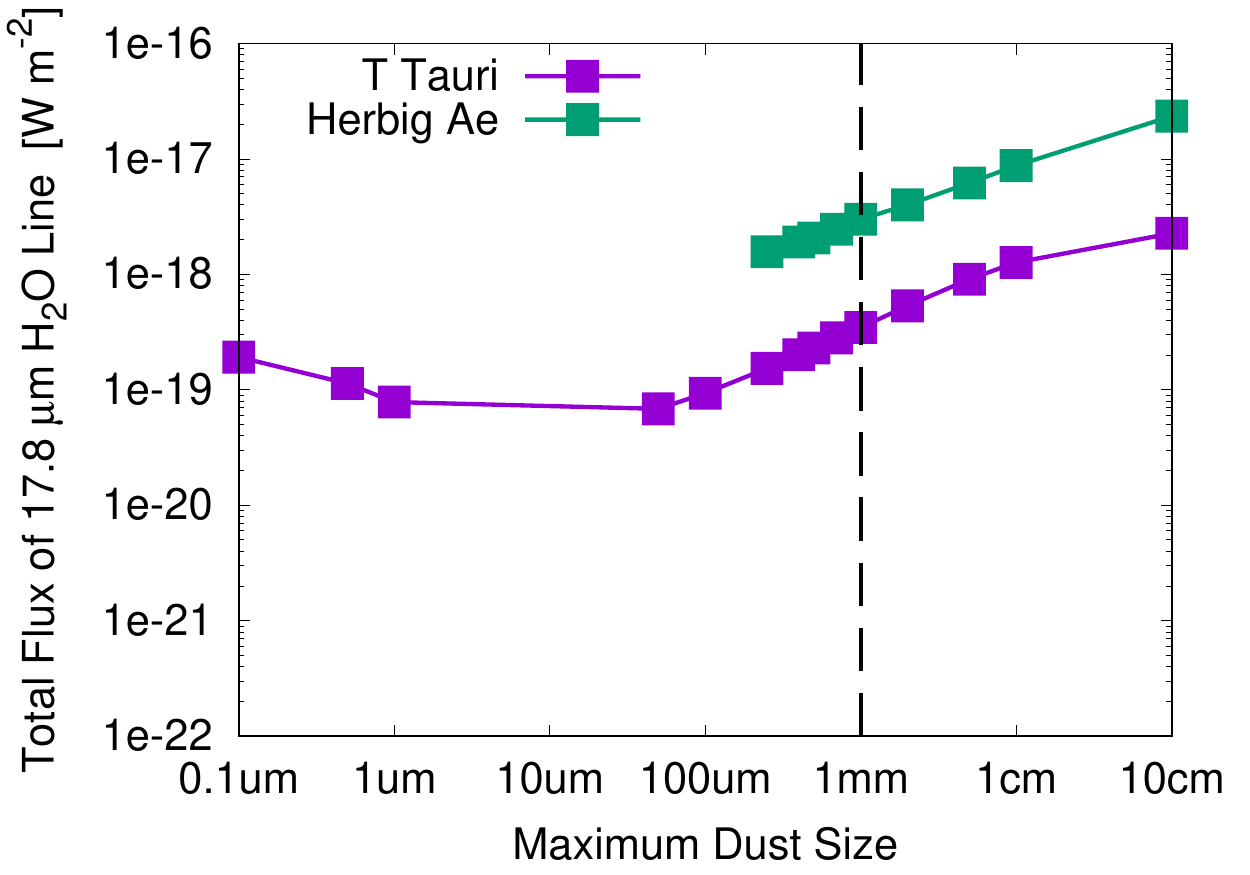}
    \includegraphics[width=6.2cm]{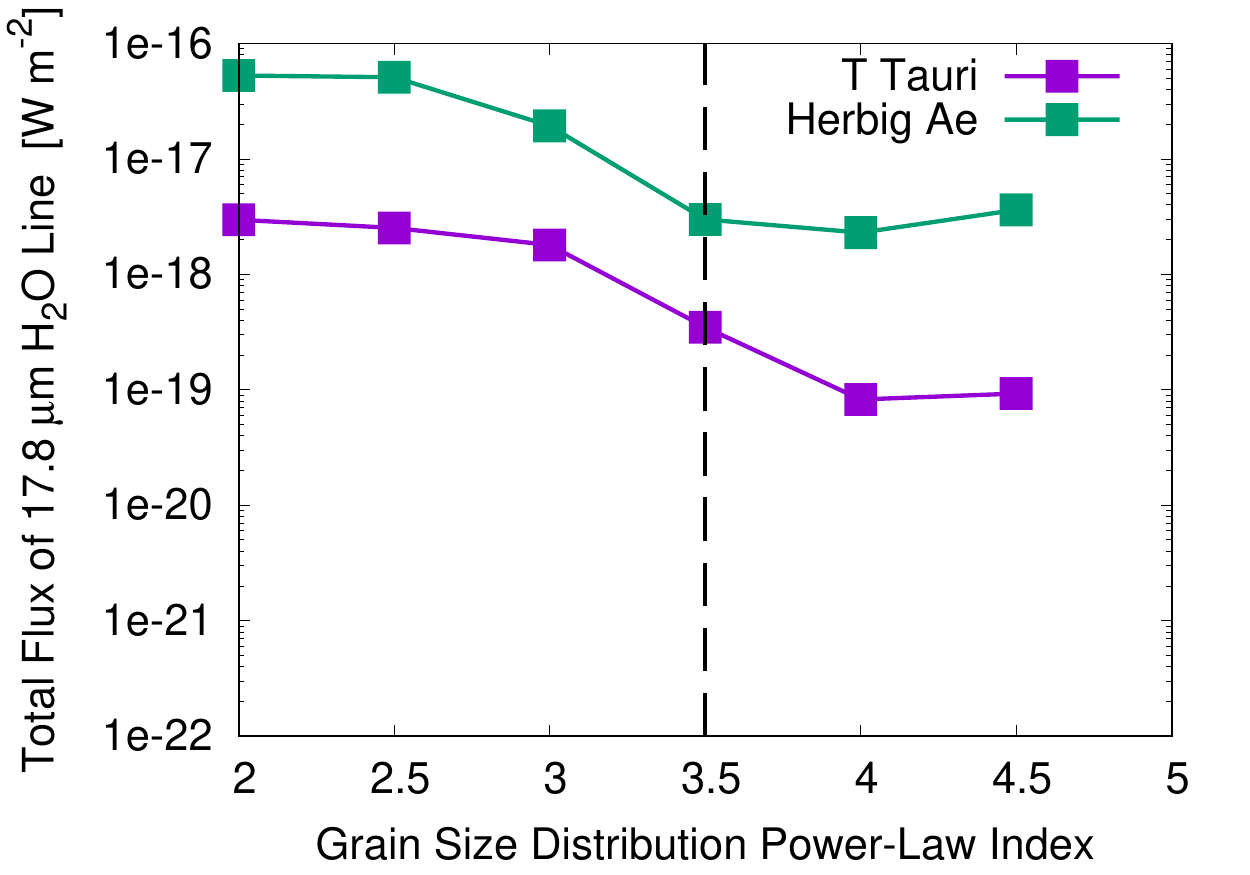}
    \includegraphics[width=6.2cm]{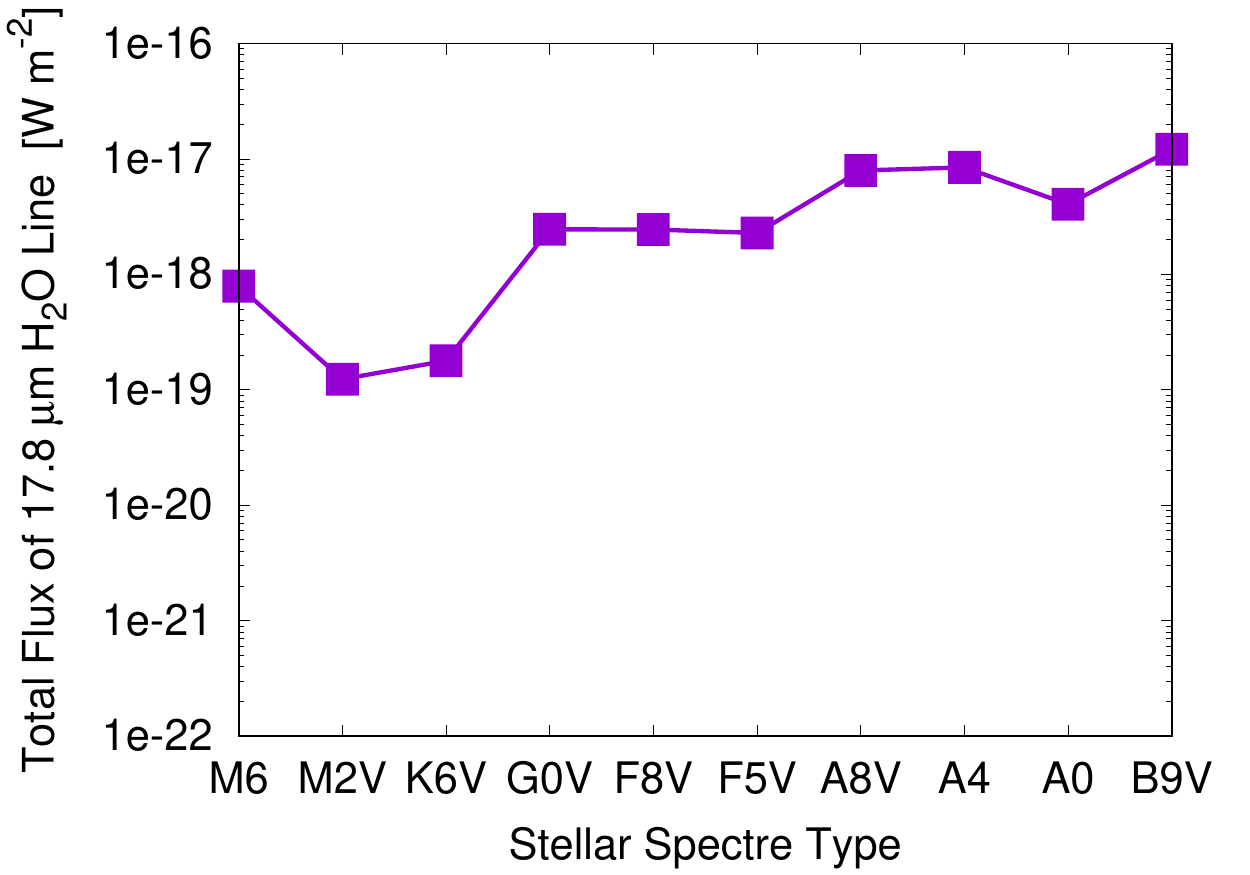}
    \caption{The 17.75\,$\mu$m water line fluxes for various disk parameters around T\,Tauri and Herbig Ae stars. The dashed lines indicate the fiducial models \citep{Antonellini2015, Antonellini2016} }
    \label{fig:18um_waterline_parameters}
\end{figure}

Since the water line flux depends on the various parameters of the disks (Fig.~\ref{fig:18um_waterline_parameters}),
the target objects will be selected more specifically based on the forthcoming JWST observations of water lines with lower spectral resolution and ALMA observations of CO lines for determining inclination angles of the disks.

We note that this snowline survey will also provide the data for the disk dispersal case (Sect.~\ref{sec:dispersal}) since the water line is systematically weaker than the H$_2$ and fine structure lines targeted there. Also, SMI/HR covers the wavelengths of various molecular lines which enable us to carry out additional science, such as measuring the C/O ratio in the inner disks using various mid-IR lines and measuring gas masses at the last stage of gas dispersal from the disks using H$_2$ lines.

\subsection{Isolated Debris Disks Gas/Ice survey}

The goal of this program is to determine the composition of planetesimals in debris disks using a combination of gas line detections with SAFARI/HR to probe their volatile content that is released in collisions, and solid state features with SAFARI/LR to probe the composition of the dust. While an observation of even a single debris system would be significant, to answer the broader question of how the chemical make-up correlates with the stellar type and the presence of planets, and to ultimately link chemical composition to planet formation, a sample size on the order of 100 systems is desired. This will allow the sample to be divided by various properties to allow to search for trends. That is, it is not sufficient to simply derive a volatile content of a few disks, rather the aim is to make a predictive model that can be used to constrain protoplanetary disk evolution models, and to determine how this composition affects the evolution of any planetary system including its habitability. Thus it is important to ascertain how volatile content depends on planetesimal belt radius, stellar luminosity and age, or other properties of the planetary system. There are also several parameters that need to be constrained for the model to make accurate predictions, such as the level of viscosity in the gas disk and how that depends on ionisation. These will provide fundamental insights into the underlying physics that are not the focus of this science case, but again argue for a large sample for success.

The targets that are most amenable to such a study are already known, since the sample of nearby debris disks with detectable dust emission is relatively complete based on IRAS, Spitzer and Herschel surveys. Thus we first compiled a sample of 486 such disks from which the model in \citet{Kral2017} was used to predict the gas line fluxes. These line fluxes were calculated for two assumptions: gas stemming from Solar System like comet/planetesimal composition, and gas from four times drier material. We also used SED modelling to predict the contribution of stellar and dust emission at the 44\,$\mu$m of the ice feature from which to assess its detectability.

We propose a 3 tier strategy (three samples consisting of different objects):
\begin{itemize}
    \item Sample A: Observe $\sim\!40$ disks in LR and HR to a depth such that [O\,I] and [C\,II] can be detected if planetesimals are 4 times drier than Solar System comets.
    \item Sample B: Observe $\sim\!40$ disks in LR and HR to a depth such that [O\,I] and [C\,II] can be detected if planetesimals are similar to Solar System comets.
    \item Sample C: Observe $\sim\!60$ disks in LR only to detect the water ice feature.
\end{itemize}
The motivation for this is that Sample A will say how dry planetesimals are typically and determine dependence on spectral type. Sample B will expand this to determine volatile dependence on planetesimal belt radius, age, presence of planets etc. Sample C will use ice features as a proxy for volatile content (using relations calibrated with samples A and B) allowing us to probe parameter space not covered in the other samples due to the prohibitively long integration times that would be required to detect gas (later spectral types and lower fractional luminosity). This leads to a total sample of 80 objects (sample A and B) for which we request HR observations and a total sample of 140 targets (sample A, B and C) for which we request LR observations. For the latter, we aim for a S/N ratio of 5 on the predicted photospheric continuum level.

We expect that the samples A and B will already have CO gas detections with ALMA within the next decade. On their own these CO detections would not be able to make predictions for the volatile content of the planetesimals (in particular of water ice), but in combination with the {\it SPICA} gas lines such accurate predictions would be possible. There are already $\sim\!20$ debris disks with CO detected by ALMA, and the detection thresholds predicted by \citet{Kral2017} imply that $\sim\!100$ CO and [C\,I] detections can be expected with ALMA in the next decade, covering a similar sample to that proposed for {\it SPICA} observations of [C\,II] and [O\,I].

The total integration time using the SAFARI estimator version 1.2 (issue 3.0; this version did not yet include the impact of the source background) is 54.5\,h for sample A and B for the high resolution spectra. In addition, we request 94\,h for sample A, B and C for the SAFARI/LR mode (version 1.2, issue 6.2).

This isolated debris disk sample complements the class\,III disks in SFRs from the HD survey. The isolated debris disks are typically much older than 100\,Myr, while the class\,III disks in SFRs and MGs present the younger counterpart (1-100\,Myr). For those, we target primarily the fainter HD lines, ensuring that the [O\,I]\,63\,$\mu$m and [C\,II]\,158\,$\mu$m for all targets will be detected. 

\end{appendix}

\bibliographystyle{pasa-mnras}
\bibliography{references}

\end{document}